%% file: fermion_consistent_DM.tex
\documentclass[11pt,a4paper]{article}

\usepackage{diagbox}

\usepackage{jheppub}
\usepackage{booktabs}
\usepackage{subfigure}
\usepackage{multirow}
\usepackage{subfigure}
\usepackage{graphicx}
\usepackage{epsfig}
\usepackage{lscape}
\usepackage{rotating}
\usepackage{epsf}
\usepackage{pstricks}
\usepackage{bm}
\usepackage{slashed}
\usepackage{dsfont}
\usepackage{amsmath}
\usepackage{color}
\usepackage{float}
\usepackage{braket}

\usepackage{stackengine,xcolor}

\usepackage[normalem]{ulem}

\newcommand{\beq}{\begin{eqnarray}}
\newcommand{\eeq}{\end{eqnarray}}

\title{Minimal Consistent Dark Matter models for systematic experimental characterisation: Fermion Dark Matter}

\author[a,b]{Alexander Belyaev,}
\author[c,d]{Giacomo Cacciapaglia,}
\author[a]{Daniel Locke,}
\author[a,e]{Alexander Pukhov}

\affiliation[a]{School of Physics and Astronomy, University of Southampton, Highfield, Southampton SO17 1BJ, UK}
\affiliation[b]{Particle Physics Department, Rutherford Appleton Laboratory, Chilton, Didcot, Oxon OX11 0QX, UK}
\affiliation[c]{Institut de Physique des Deux Infinis (IP2I) de Lyon, CNRS/IN2P3, UMR5822, F-69622 Villeurbanne Cedex, France}
\affiliation[d]{Universit\'e de Lyon; Universit\'e Claude Bernard Lyon 1, F-69001 Lyon, France}
\affiliation[e]{Skobeltsyn Inst. of Nuclear Physics, Moscow State Univ., Moscow 119992, Russia}

\abstract{The search for a Dark Matter particle is the new grail and hard-sought nirvana of the particle physics community. From the theoretical side, it is the main challenge to provide a consistent and model-independent tool for comparing the bounds and reach of the diverse experiments. We propose a first complete classification of minimal consistent Dark Matter models, which provides the missing link between experiments and top-down models. Consistency is achieved by imposing renormalisability and invariance under the full Standard Model symmetries. We apply this paradigm to fermionic Dark multiplets with up to one mediator. We also reconsider the one-loop contributions to direct detection, including the relevant effect of (small) mass splits in the Dark multiplet. Our work highlights the presence of unexplored viable models, and paves the way for the ultimate {systematic} hunt for the Dark Matter particle.}

\preprint{}
\makeindex

\begin{document}
\maketitle
\newpage

\input{01_intro.tex}

\input{02_classification.tex}

\input{03_one_DM_mult.tex}

\input{04-DM+onemore_multiplet.tex}

\input{05_pheno.tex}

\input{06_conclusions.tex}

\input{Appendix.tex}

\newpage
\bibliography{bib}
\bibliographystyle{JHEP}
\end{document}

%% file: 01_intro.tex

\section{Introduction}

Dark Matter (DM) exploration is becoming an increasingly appealing subject at present \cite{Bertone:2018krk}, 
{particularly} when the Large Hadron Collider (LHC) experiments like ATLAS and CMS, as well as other non-collider experiments, do not indicate any clear signal Beyond the Standard Model (BSM).  The evidence for DM provides, arguably, the strongest  experimental indication of BSM physics. Thanks to the great advances in precision cosmology and astrophysics, it is well-established from several independent  observations the presence of a source of mass in the Universe, not accounted for in the Standard Model (SM). The observations include galactic rotation curves, cosmic microwave background fits of WMAP \cite{WMAP:2010qai} and PLANCK \cite{Adam:2015rua} data, gravitational lensing, large scale structure formation in the Universe, as well as the existence of so-called bullet clusters. All of this data points towards the presence of BSM matter, roughly $5$ times more abundant than ordinary baryonic matter \cite{Planck:2018vyg} in the present day Universe.

While evidence for the presence of DM in the Universe has become more convincing, our knowledge of its nature remains veiled; there are many particle candidates, however no experiment so far was able to probe their properties.
The mass of DM candidates covers a vast range, from sub-eV (axion-like) to astrophysical masses (primordial black holes).  Here we will be interested in masses in the GeV--TeV range, so that the
DM particles can be probed at colliders like the LHC by measuring their production in particle collisions, at direct detection underground experiments~\cite{Aprile:2018dbl,PandaX-4T:2021bab,Akerib:2018lyp} that are sensitive to elastic scattering of the DM particles in the local galactic halo off target nuclei, and finally at indirect detection experiments that measure the products from DM annihilation and/or decay in the Universe {constituting} positron, gamma-ray and anti-proton fluxes.
The fact that such DM candidates can be probed by a large array of experiments, of different nature, made the interest in DM rapidly increase in the particle physics community, especially after the discovery of the Higgs boson at the LHC. 

One of the most important issues behind DM searches is related to how to combine the results of experimental searches, so different in nature, in a consistent and yet model-independent and general way.
Starting from Ref.~\cite{Goodman:2010ku}, an Effective Field Theory (EFT) approach has been adopted in collider and direct detection searches. Since then, the level of sophistication in DM exploration at the LHC and in direct detection has been constantly increasing.
Although many ATLAS and CMS papers have been using EFTs in Run 1 data analysis and interpretation~\cite{Aad:2012fw,Aad:2013oja,Aad:2014vka,ATLAS:2014wra,Khachatryan:2014tva,Khachatryan:2015nua}, the limitations of this approach soon became clear. 
In EFTs, contact interactions are used to model the couplings of the DM candidate to ordinary matter: this approximation works well in direct detection, where the energy of the collision is very low, corresponding to the velocities of the Earth and of Dark particles in the local halo, while it fails at energy scales close to or above the mass of the mediator generating the effective contact interactions. Eventually, this invalidates the comparison between direct detection results and the LHC searches at ATLAS and CMS. 

At the next step beyond EFTs, the exploration of collider DM phenomenology adopted {\it simplified models}, where the Dark sector is characterised by the DM candidate and a mediator that makes the connection with the SM particles \cite{Buchmueller:2013dya,Cheung:2013dua,Dutta:2014kia,Busoni:2014sya,Papucci:2014iwa,Bai:2014osa,Berlin:2014cfa,Hamaguchi:2014pja,Busoni:2014haa,Balazs:2014jla,Buchmueller:2014yoa,Abdallah:2014hon,Harris:2014hga,Racco:2015dxa,Jacques:2015zha}.
Some of these models have been used in recent ATLAS and CMS experimental interpretations of Run~1~\cite{Khachatryan:2014rra,Aad:2014vea,Khachatryan:2014rwa,Aad:2014tda,Aad:2015zva}
and Run~2~\cite{ATLAS:2021fjm,ATLAS:2020uiq,CMS:2020ulv,CMS:2019ykj,CMS:2018ysw,CMS:2018nlv}
LHC  data.
In simplified models, the mass of the mediator and, potentially, its width are non-trivial parameters of the model. However, one remains agnostic about the theory behind the Dark sector and tries to parametrise the interactions in the simplest terms: this often leads to writing interactions which are not invariant under the full SM gauge symmetry but only under the unbroken colour SU(3) and electromagnetic U(1). Nevertheless, one still needs to know if it is possible to construct viable models that lead to a given simplified scenario, consistent with the full symmetries of the SM \cite{Kahlhoefer:2015bea,Goncalves:2016iyg}.
The latter point is particularly important at the LHC, a machine which is probing energies well above the electroweak (EW) symmetry breaking scale, so that for many events the full weak SU(2)$\times$U(1) is a good symmetry.
For instance, if a mediator or DM candidate comes in a multiplet of the weak Isospin SU(2), its charged partners may play an important role in the LHC phenomenology often being more important than the neutral state itself. This is the case for charginos in supersymmetry.
In addition, simplified models often violate gauge invariance at high scales \cite{Kahlhoefer:2015bea}, which is a crucial principle for building a consistent BSM model that incorporates the SM together with new physics. For example, considering simplified models with a new heavy gauge vector boson mediating DM interactions, one should also introduce a
mechanism responsible for generating the mediator mass and ensuring 
gauge invariance for the model \cite{Kahlhoefer:2015bea}. Eventually, this necessarily requires introducing an additional sector into the model that may affect the DM phenomenology \cite{Kahlhoefer:2015bea,Duerr:2016tmh}. 

These drawbacks strongly indicate the next step in the evolution of the DM investigation, based on building
 {\it Minimal Consistent Dark Matter} (MCDM) models. MCDM models can be still understood as  toy models that, however, take in full account the consistency with the symmetries of the SM.
In our approach, MCDM models consists of one DM multiplet and at most one mediator multiplet. Furthermore, a particular MCDM model can be easily incorporated into a bigger, more complete and fundamental, BSM model and be explored via complementary constraints 
from collider and direct/indirect DM search experiments as well as relic density constraints. The exploration of complementarity of the collider and non-collider  constraints within the complete models such as MCDM ones is very appealing especially now as we have a large amount of data from the LHC. Combining searches may shed light on the BSM physics in the form of DM, which can be near the corner of the combined collider and non-collider limits. Another attractive feature of the MCDM approach is the minimal but self-consistent parameter space that can be potentially mapped to the parameter space of known (and completely new) BSM models.

Many implementations of MCDM models have been studied in the literature~\cite{Deshpande:1977rw,Cirelli:2005uq,Hambye:2009pw,Papucci:2014iwa,Harris:2014hga,Berlin:2014cfa,Bai:2014osa}, however
there has been no attempt on their systematic classification yet. This is precisely the aim of the present work.
In this study we shall: 
\begin{itemize}
\item[a)] perform a complete classification of MCDM models, with at most one mediator and including only renormalisable interactions (with some notable exceptions); 
\item[b)] present the main features for each class of MCDMs constructed using the main building principles we state below.
\end{itemize}
We believe that this classification, and the MCDM approach in general, will create a solid framework for the consistent exploration of DM models at collider and non-collider experiments for the complementary probe of Dark sectors.

The paper is organised as follows: after articulating the main principles behind the MCDM approach in Section~\ref{sec:2}, we summarise the main properties of models with only a DM candidate in Section~\ref{sec:DMmult}. Here we also present a detailed calculation of the one-loop cross section for direct detection, which includes for the first time the mass split between components of the DM electroweak multiplet. In Section~\ref{sec:mediator} we classify and characterise models with a single mediator. Finally, in Section~\ref{sec:new-model-pheno} we study in detail a new model that emerges from the classification, featuring a Dirac fermionic DM candidate and a CP-odd scalar mediator. In some regions of the parameter space, the scalar mediator can be accidentally stable and contribute to the relic density. We offer our conclusions and outlook in Section~\ref{sec:conclusions}.

%% file: 02_classification.tex

\section{Classification of MCDM models} \label{sec:2}

The building blocks we use to construct MCDM models are {vector-like} multiplets defined in terms of their spin and electroweak quantum numbers. We will only consider spin-0 ($S$), spin-1/2 ($F$ for Dirac and $M$ for Majorana \footnote{Here, we call `Majorana' a multiplet with zero $U(1)$ charges and in a real representation of the non-abelian gauge symmetries, SU(2), such that  $\Psi^C = \Psi$.}), and spin-1 ($V$). For models with higher spin, we refer the reader, for instance, to Refs~\cite{Ding:2012sm,Khojali:2016pvu,Khojali:2017tuv,Asorey:2010zz}.
The electroweak quantum numbers will be encoded in the weak Isospin, $I$, and the hypercharge, $Y$, of the multiplet. Furthermore, we will denote with a tilde the multiplets that belong to the Dark sector, i.e. they cannot decay into purely SM final states. The multiplets we consider, therefore, read:
$$
\widetilde{S}^I_Y\,, \quad \widetilde{F}^I_Y\,, \quad \widetilde{M}^I_0\,, \quad \widetilde{V}^I_Y\,, \qquad S^I_Y\,, \quad F^I_Y\,, \quad M^I_0\,, \quad V^I_Y\,.
$$
As some mediator multiplets may carry QCD quantum numbers, we will use a superscript ${}^c$ to label this feature.

To construct consistent minimal models, we follow these main building principles:
\begin{itemize}
\item[A)] We add one Dark multiplet (including the singlet case) and all its renormalisable interactions to SM fields, 
	excluding the ones that trigger the decays of the multiplet, which is therefore stable {by construction}. The models will automatically 
	include a Dark symmetry, being $\mathbb{Z}_2$ or $U(1)$ depending on the multiplet. The weak Isospin and 
	hypercharge are constrained by the need for a neutral component, therefore we will have the following two 
	cases:
		\begin{itemize}
		\item[-] for integer isospin $I=n$, $n \in \mathbb{N}$, then $Y = 0, 1 \dots n$;
		\item[-] for semi-integer isospin $I = (2n+1)/2$, $n \in \mathbb{N}$, then $Y = 1/2, 3/2 \dots (2n+1)/2$.
		\end{itemize}
		Note that the case of negative hypercharge can be obtained by considering the charge conjugate field, thus 
		the sign of $Y$ is effectively redundant, and we will consider $Y \geq 0$.

\item[B)] We consider models where only one Dark multiplet is present, and mediators are SM fields.~\footnote{Note that 
	this model building approach has been used in~\cite{Cirelli:2005uq} to construct models of so-called Minimal Dark 
	Matter, so some of the results we present here can be found in this reference. However, our approach has some 
	differences: in Ref.~\cite{Cirelli:2005uq}, the symmetry making the DM candidate stable or long lived emerged  
	at low energy, at the level of renormalisable interactions, while decays could be induced by higher dimensional 
	couplings to the Higgs multiplets. In our case, we assume that a parity or global $U(1)$ symmetry is also respected 
	by higher dimensional operators. {Henceforth, we do not take into account  any constrains on the isospin of the multiple.}} While our principle is to be limited to renormalisable interactions, under the assumption 
	that higher order ones are suppressed by a large enough scale to make them irrelevant for the DM properties, in some 
	cases we will consider dimension-5 operators.

\item[C)]  In additional to point B), we consider adding just one mediator multiplet, characterised by the respective weak 
	Isospin, $I'$, and hypercharge, $Y'$.	The mediator multiplet can be odd or even with respect to the Dark symmetry, and
	its quantum numbers are limited to cases where renormalisable couplings to the Dark multiplet and to the SM are allowed. 
	This leaves open the possibility of multiplets carrying QCD charges, which we label with a superscript ${}^c$. The mediators
	are labeled as following:
		\begin{itemize}
		\item  ${S}_{Y'}^{I'}$, ${F}_{Y'}^{I'}$, ${M}_{0}^{I'}$ and ${V}_{Y'}^{I'}$ for even mediator multiplets;
		\item  $\widetilde{S}_{Y'}^{I'(c)}$, $\widetilde{F}_{Y'}^{I'(c)}$, $\widetilde{M}_{0}^{I'(c)}$ and $\widetilde{V}_{Y'}^{I'(c)}$ for odd mediator multiplets.
		\end{itemize}
	{The odd mediator multiplets can also contain a DM candidate if a neutral component is present.}

\item[D)] We consider all renormalisable interactions allowed by the symmetries of quantum field theory. Our basic assumption for MCDM models is that higher-order 
	operators are suppressed by a scale high enough that the LHC is unable to resolve the physics generating the operators. 
	The effect on the DM properties is also considered negligible (except for dim-5 operators generating mass splits).

\item[E)] We ensure cancellation of triangle anomalies, so that the MCDM models entails consistent gauge symmetries.

\end{itemize}


\begin{table}[htb]
\begin{center}
\begin{tabular}{||l||c|c|c||}
%
\hline
\diagbox[]{Spin of\\ Mediator}{Spin of \\Dark \\Matter}&
0&1/2&1
\\
\hline
\hline
&&&\\
no mediator& 
$\widetilde{S}_Y^I$
& 
$\widetilde{F}_Y^I$
&
$\widetilde{V}_Y^I$
\\&&&\\
\hline
&&&\\
spin 0 even mediator& 
$\widetilde{S}_Y^I{S}_{Y'}^{I'}$
& 
$\widetilde{F}_Y^I S_0^{I'}$
&
$\widetilde{V}_Y^I{S}_{Y'}^{I'}$
\\&&&\\
spin 0 odd mediator& 
$\widetilde{S}_Y^I\widetilde{S}_{Y'}^{I'}$
& 
$\widetilde{F}_Y^I \widetilde{S}_{Y'}^{I'}$  \ \ \  $\widetilde{F}_Y^I \widetilde{S}_{Y'}^{I'c}$
&
$\widetilde{V}_Y^I\widetilde{S}_{Y'}^{I'}$
\\
&&&\\
\hline
&&&\\
spin $1/2$ even mediator & --
& 
(via dim-6 operators)
&
--
\\&&&\\
spin $1/2$ odd mediator &
$\widetilde{S}_Y^I \widetilde{F}_{Y'}^{I'} \ \ \ \widetilde{S}_Y^I \widetilde{F}_{Y'}^{I'c}$
& 
$\widetilde{F}_Y^I \widetilde{F}_{Y\pm 1/2}^{I\pm 1/2}$
&
$\widetilde{V}_Y^I \widetilde{F}_{Y'}^{I'} \ \ \ \widetilde{V}_Y^I \widetilde{F}_{Y'}^{I'c}$
\\
&&&\\
\hline
&&&\\
spin $1$ even mediator & 
$\widetilde{S}_Y^I{V}_{0}^{I'}$
& 
$\widetilde{F}_Y^I V_0^{I'}$
&
$ \widetilde{V}_Y^I {V}_{Y'}^{I'}$
\\
&&& 
\\
spin $1$ odd mediator &
$\widetilde{S}_Y^I \widetilde{V}_{Y'}^{I'}$
& 
$\widetilde{F}_Y^I \widetilde{V}_{Y'}^{I'}$ \ \ \ $\widetilde{F}_Y^I \widetilde{V}_{Y'}^{I'c}$
&
$\widetilde{V}_Y^I\widetilde{V}_{Y'}^{I'}$
\\
&&&\\
\hline
\end{tabular}
\caption{\label{tab:MCDM}Classification of MCDM models
in Spin(DM)-Spin(mediator) space. When possible, the Dirac fermion can be replaced by a Majorana one, $F \to M$.}
\end{center}
\end{table}

With the notations above, following the precepts A) to E), we can classify all MCDM models with up to one mediator multiplet 
using a 2-dimensional grid in Spin(DM)-Spin(mediator) space, as presented in Table~\ref{tab:MCDM}.
Each specific DM model is denoted by a one- or two-symbol notation, indicating the DM multiplet first, followed by the mediator multiplet.
{In general, the interactions of the DM candidate to the SM are mediated by SM particles (e.g. by the EW gauge bosons and the Higgs) and other components of the DM multiplet, besides the components of the mediator multiplet. Hence, highly non-trivial interference effects can arise. Furthermore, some couplings entail flavour structure, which need care as they may incur very strong bounds.}
Eventually, the case with no mediator multiplet is denoted by just one symbol labelling the DM multiplet.
In this case the role of mediators can only be played by SM particles and members of the DM multiplet.

In the remainder of this paper, we will focus on spin-1/2 DM multiplets, leaving the other two cases for a future publication.

%% file: 03_one_DM_mult.tex

\section{Case of one  DM multiplet: $\tilde{F}^I_Y$ and $\tilde{M}^I_0$ models} \label{sec:DMmult}

Models where the DM belongs to a single EW multiplet, while no other light states are present, have been studied in great detail, starting from the seminal paper in Ref.~\cite{Cirelli:2005uq}. In this section we briefly review the main properties of these minimal models, and add a detailed discussion of the following novel aspects:

\begin{itemize}
\item[i)] We provide an improved formula for the mass split induced by EW loops, which is numerically more stable than the one given in Ref.~\cite{Cirelli:2005uq}.

\item[ii)] We discuss in great detail the effect of couplings to the Higgs boson arising as dimension-5 operators. While going beyond  renormalisability principles, they are generated by integrating out a single mediator (thus, they can be considered as a limiting case from some of the models discussed in Section~\ref{sec:mediator}). Furthermore, a class of these operators have special phenomenological relevance as they  help salvage some of the minimal models with non-zero hypercharge.

\item[iii)] We provide a detailed and up-to date discussion of direct detection bounds at one-loop level. We include for the first time the effect of mass splits within the DM multiplet, and show their relevance. 

\item[iv)] We discuss the impact of nuclear uncertainties and of the variation of the gluon contribution due to the mass splits. Both generate comparable uncertainties in the total spin-independent cross sections, which emerge as an uncertainty in the DM mass limits of hundreds of GeVs.

\end{itemize}
This section also serves to fix the notation we will adopt in the rest of the paper. When writing Lagrangians and interactions we will consistently use $\Psi = \Psi_L + \Psi_R$ for  Dirac DM multiplet, with $\Psi_R = \Psi_L^C$ for the Majorana case (where ${}^C$ indicates the charge conjugate field), 
$\psi^i$ for the components of a Dirac multiplet and $\chi^i$ for the components of a Majorana multiplet. Furthermore, we only consider $Y \geq 0$, as the case of negative hypercharge is straightforwardly analogous to the corresponding positive value case.
{We will use $M_{DM}$ to denote the mass of the neutral component that serves as DM candidate.}

In the ``stand alone" case, only gauge interactions of the EW gauge bosons, $W^\pm$, $Z$ and photon, are allowed at renormalisable level. 
This simple class of models has well established properties~\cite{Cirelli:2005uq}, which we list below:
\begin{itemize}
\item[-] A gauge coupling ${\Large g_{Z \bar{\psi}_0 \psi_0}}$ is always present for Dirac multiplets with $Y \neq 0$, which are thus excluded by direct detection even for under-abundant points (for $M_{DM} < m_Z/2$ the invisible width of the $Z$ also excludes the model).
On the contrary, when $Y=0$,  the coupling ${\Large g_{Z \bar{\psi}_0 \psi_0}}$
always vanishes.

\item[-] Due to the absence of couplings to the Higgs field, the mass split between the neutral and charged components of the DM multiplet are generated by EW loop corrections and are always small (below a few hundred MeVs, with the precise values depending on the hypercharge of the multiplets). This leads to long lived particles, especially at high mass. The lightest component is not always guaranteed to be neutral: this only occurs for multiplets with $Y=0$ and maximal hypercharge, $Y=I$.
\item[-] For $Y \geq 1$ and isospin $I \neq Y$ (hence, $I \geq 2$), the mass range with the neutral component being the lightest is excluded by the $Z$ width. Hence, these multiplets in isolation cannot provide a DM candidate.
\item[-] For $Y = 1/2$ and $I \geq 3/2$, the lightest component is neutral for $M_{DM} \lesssim 570$~GeV. Above this threshold, the charge $-1$ state becomes the lightest in absence of Higgs couplings.
\item[-] For $Y=1/2$, a dim-5 operator with the Higgs boson generates a mass that splits the neutral component in two Majorana mass eigenstates (pseudo-Dirac case). This salvages the models from exclusion via the $Z$ interactions. 
\item[-] Taking into account loop-induced mass splits, the loop-induced cross sections ensures that current and future direct detection experiments can probe multiplets with $I \geq 1$, where $I \geq 2$ can be completely ruled out, while the case of a doublet $I=1/2$ is always below detection. Uncertainties in the nuclear form factors and mass splits for the gluon contribution generate uncertainties of hundreds of GeV in the DM mass limit.

\end{itemize}
We should finally note that, for DM multiplets with $\{I, Y\}  = \{0,0\}$, $\{1/2, 1/2\}$, $\{1,0\}$ and $\{1,1\}$, a linear Yukawa coupling with the SM leptons is allowed by gauge symmetries, while larger isospin multiplets are automatically protected at renormalisable level. However, higher order couplings involving the Higgs can always generate decays of the DM multiplets, and it has been the main motivation of Ref.~\cite{Cirelli:2005uq} to find multiplets that are long-lived enough to be Cosmologically stable, thus pointing towards multiplets with $I=2$. In this work we will be more pragmatic and allow for any multiplet by forbidding implicitly all operators that could mediate the decays of the DM candidate. The origin of such a symmetry is to be searched in the more complete model containing the DM multiplet. Moreover, as the MCDM models are to be considered effective low energy descriptions of the DM phenomenology, we do not consider the upper limit on the isospin value coming from the absence of Landau poles in the renormalisation group running of the EW gauge couplings below the Planck mass.

After reviewing the properties of Dirac and Majorana multiplets in Sec.~\ref{sec:DMmult_Dirac} and~\ref{sec:DMmult_Majorana} respectively, in Sec.~\ref{sec:DMmult_Higgs} we study in detail the effect of dim-5 couplings to the Higgs field. In Sec.~\ref{sec:DDminimal} we provide novel detailed results on one-loop cross sections for direct detection, including for the first time the mass split in the multiplet, and present current exclusion limits and future projections. We also show that, due to delicate cancellations among various amplitudes, both the mass split and nuclear uncertainties have sizeable impact on the cross sections and on the DM mass limits.

\subsection{Dirac multiplets ($\tilde{F}^I_Y$)} \label{sec:DMmult_Dirac}

In the case of Dirac multiplets, i.e. when both chiralities are present, the lowest order Lagrangian, to be added to the SM one, reads
\beq \label{eq:LagrDirac0}
\Delta\mathcal{L}_{\text{Dirac}} = i \bar{\Psi} \gamma^\mu D_\mu \Psi - m_D \bar{\Psi} \Psi\,,
\eeq
where the covariant derivative includes the EW gauge bosons. It is invariant under a global U(1)$_{\rm DM}$ symmetry, thus an asymmetric contribution to the relic abundance may be present if the complete model preserves this symmetry. 

\begin{figure}[tb]
\begin{center}
\includegraphics[width=7cm]{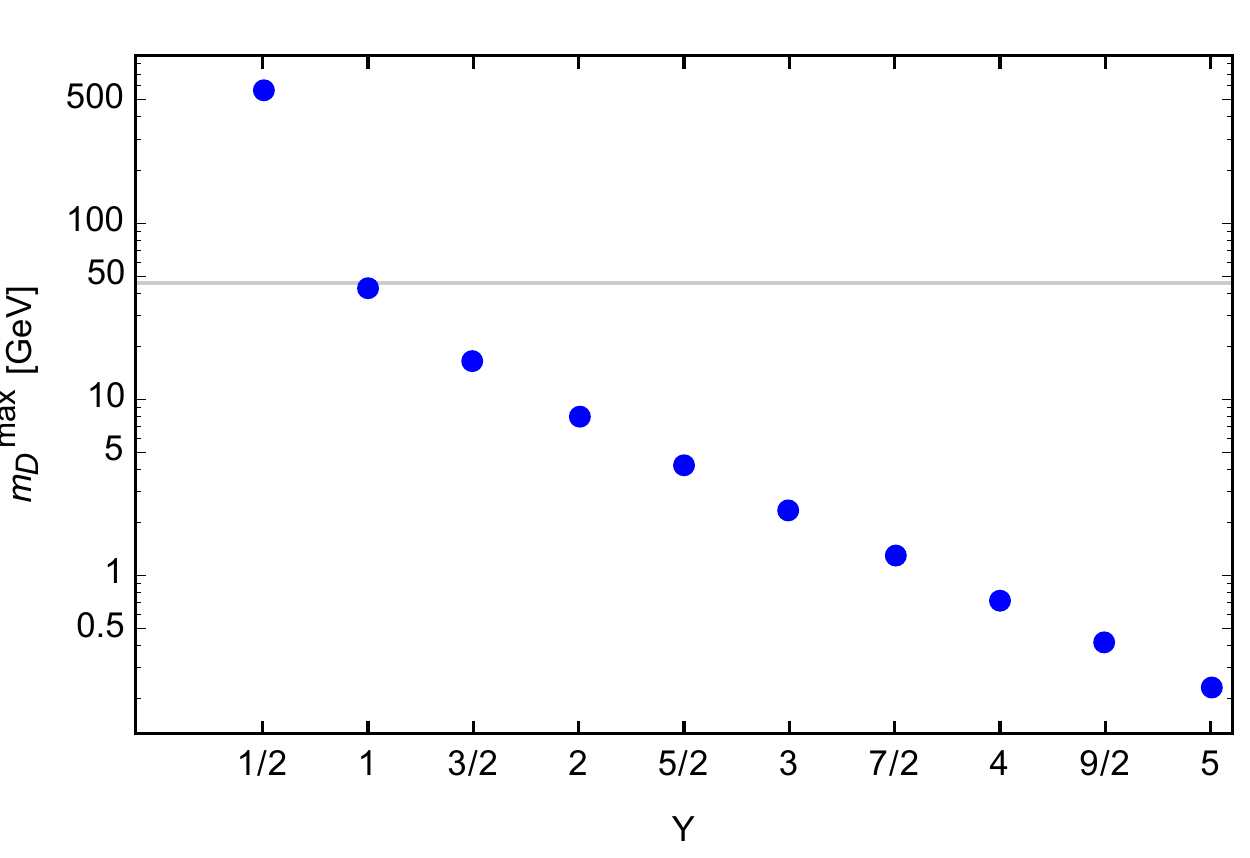} \qquad
\includegraphics[width=7cm]{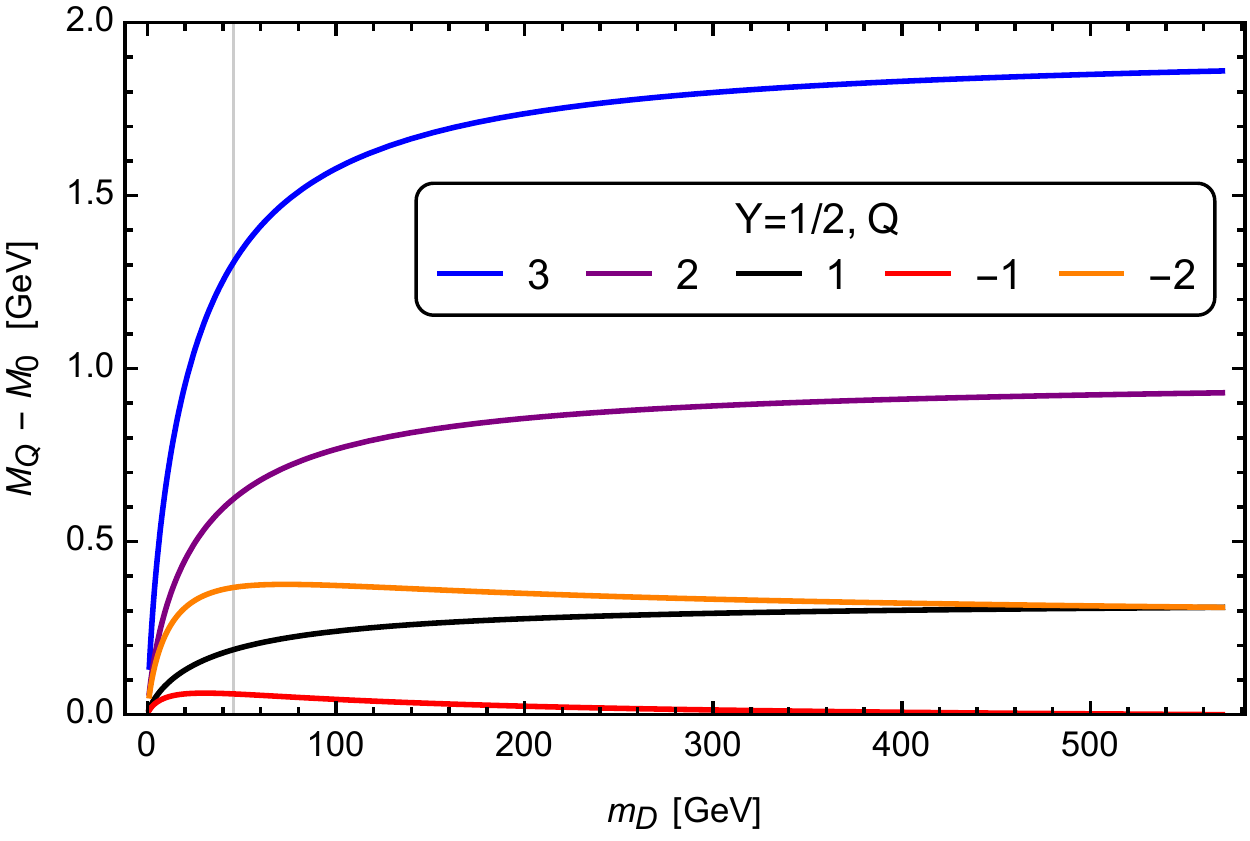}
\end{center}
\caption{{\it Left}: maximum value of $m_D$ above which the lightest component has charge $Q=-1$ for various values of $Y$. The horizontal line indicates $m_Z/2$, below which decays of the $Z$ exclude the model. {\it Right}: spectrum for a generic multiplet with $Y=1/2$, with $m_D < 570$~GeV. The vertical line shows $m_D \approx m_Z/2$, below which the model is excluded by the $Z$ decays.} \label{fig:deltaM_F}
\end{figure}

Except for the singlet case $\tilde{F}^0_0$, the multiplet contains charged states:
\beq
\Psi = \left( \begin{array}{c}
\psi^{n+} \\
\vdots \\
\psi^+ \\
\psi_0 \\
\psi^- \\
\vdots \\
\psi^{m-}
\end{array} \right)\,, \qquad \mbox{with} \quad n = I+Y\,, \;\; \mbox{and}\;\; m=I-Y\,. 
\label{eq:Dirac}
\eeq
The Dirac mass term in Eq.~\eqref{eq:LagrDirac0} gives equal mass to all components of the multiplet. This degeneracy can only be lifted by radiative corrections due to the EW gauge bosons. This contribution has been first computed in Ref.~\cite{Cirelli:2005uq}, and can be written as
\begin{multline}
M_Q - M_{Q'} = \frac{\alpha m_D}{4 \pi s_W^2} \left[ (Q^2-{Q'}^2) \left( f_F (x_W) - c_W^2 f_F (x_Z) - s_W^2 f_F (x_\gamma) \right) + \right. \\
\left. 2 Y (Q-Q') \left(  f_F(x_Z)- f_F (x_W) \right) \right]\,, \label{eq:DMf}
\end{multline}
where $f_F (x)$ is a loop function and $x_V = m_V/m_D$. This expression explicitly shows that the mass differences vanish in the limit of equal masses for $W$, $Z$ and photon. For the loop function, we found an alternative form that is numerically more stable than the one given in Ref.~\cite{Cirelli:2005uq} (see Appendix~\ref{app:radcorr} for more details). The result, which is exact, reads
\beq
f_F (x) = \frac{x}{2} \left[ 2 x^3 \ln x - 2 x - \sqrt{x^2-4} (x^2+2) \ln \frac{x^2-2+x \sqrt{x^2-4}}{2} \right]\,.
\label{eq:DMffx}
\eeq
This function has been defined in such a way that $f_F (x_\gamma) \equiv f_F (0) = 0$.
It is instructive to study how the mass split looks in the limit of DM mass small and large compared to the $W$ and $Z$ masses.
For light DM, $M_{DM} \approx m_D \ll m_W$, the leading contribution reads
\beq
\left. M_Q - M_{Q'} \right|_{m_D \ll m_W} \approx \frac{3 \alpha}{2 \pi} (Q^2 - {Q'}^2) m_D \left( \log \frac{m_W}{m_D} + \frac{1}{4} \right)\,. 
\eeq
This mass split tends to zero for vanishing DM mass and is proportional to the difference in squared charges, as an indication that it is dominated by the photon exchange. Furthermore, in this limit the lightest component of the multiplet is always the neural one.
In the opposite limit, $m_D \gg m_W$, the leading term in the expansion reads
\beq
\left. M_Q - M_{Q'} \right|_{m_D \gg m_W} \approx \frac{\alpha m_W}{2 (1+c_W)} \left[ (Q^2 - {Q'}^2 ) + \frac{2 Y (Q-Q')}{c_W} \right]\,.
\eeq
For $Y=0$, the charged states are always heavier than the neutral one as the surviving term is proportional to the difference of squared charges. On the contrary, for $Y \neq 0$ the second term, which depends on the sign of the charges (we chose $Y>0$ without loss of generality), does not guarantee that the $Q=0$ state is always the lightest one. In particular, the state $Q=-1$ is always lighter than the $Q=0$ one in this limit, for any value of $Y \neq 0$ and of the isospin of the multiplet.
Thus, there exists an upper limit on $m_D$, above which the lightest state in the multiplet is charged, and this value is determined by the $Q=-1$ state. The values of the mass upper bounds for various $Y$ are shown in the left panel of Fig.~\ref{fig:deltaM_F}: 
the highest value is achieved for $Y=1/2$ which gives $m_D^{\rm max} \approx 570$~GeV (we recall that for $Y=0$ there is no limit), while for $Y=1$ we find $m_D^{\rm max} \approx 42$~GeV, which is already below $m_Z/2$. Hence, multiplets with $Y \geq 1$ are excluded by the $Z$-width measurement in the region where the lightest state is neutral, as long as a $Q=-1$ state exists in the multiplet. In fact, this upper limit is removed for multiplets with maximal hypercharge, $Y=I$, for which only states with positive charge are present.
In the right panel of Fig.~\ref{fig:deltaM_F} we show the mass splits for various charges and for $Y=1/2$ as a function of the DM mass, i.e. the mass of the neutral component. This shows that the $Q=-1$ state is always the lightest above the neutral one for $m_{D} \lesssim 570$~GeV, with a mass split always smaller than $100$~MeV.

The analysis of the loop induced mass split, therefore, shows that only 4 classes of models are potentially interesting: 
\begin{itemize}
\item[a)] the singlet $\tilde{F}_0^0$;
\item[b)] multiplets with maximal hypercharge $\tilde{F}^I_I$, including the doublet $\tilde{F}_{1/2}^{1/2}$;
\item[c)] multiplets $\tilde{F}_{1/2}^I$ ($I$ semi-integer with $Y=1/2$), with $m_D \leq 570$~GeV for $I \geq 3/2$;
\item[d)] multiplets $\tilde{F}_{0}^I$ ($I$ non-zero integer with $Y=0$).
\end{itemize}
As already mentioned, all models with $Y\neq 0$, i.e. b) and c), are excluded by direct detection via the $Z$ exchange. As we will see, however, a dim-5 couplings to the Higgs can salvage the models with $Y=1/2$ (see Section~\ref{sec:DwithMhiggs}).

\subsubsection{Pseudo-Dirac multiplets}

For completeness, we recall that Dirac multiplets with $Y=0$ can be split in two Majorana multiplets $\tilde{M}^I_0$. This can be effectively described by the addition of a new mass term to the Lagrangian in Eq.~\eqref{eq:LagrDirac0}:
\beq \label{eq:LagrDirac0M}
\Delta \mathcal{L}_{\text{p--Dirac}} = i \bar{\Psi} \gamma^\mu D_\mu \Psi - m_D \bar{\Psi} \Psi - \frac{1}{2} \left( \delta m \; \bar{\Psi}^C \Psi\, + \mbox{h.c.} \right).
\eeq
Without loss of generality, we consider $\delta m$ to be real and positive.~\footnote{In principle $\delta m$ can be complex, however the phase can always be removed by a redefinition of $\Psi$. A physical phase appears in couplings of $\Psi$ that are not invariant under the phase redefinition.}
The Lagrangian above effectively describes two Majorana multiplets (see Section~\ref{sec:DMmult_Majorana}) with masses
\beq
M_{1,2} = m_D \pm \delta m\,.
\eeq
We highlighted the mass term $\delta m$ as it breaks the U(1)$_{\rm DM}$ to a $\mathbb{Z}_2$, hence it may be a small perturbation depending on how this breaking is implemented in the UV completion of the model. Note also that this term is not generated radiatively as long as it is not generated by the complete model. Hence, it may be natural to have a small mass split between the two Majorana multiplets, which leads to a model with two DM candidates, with the relic density dominated by the lighter one for large mass split. 
We recall that in all pseudo-Dirac models the lightest component is guaranteed to be neutral.

\subsection{Majorana multiplets ($\tilde{M}^I_0$)} \label{sec:DMmult_Majorana}
In the case of a Majorana multiplet, $\tilde{M}^I_0$, the Lagrangian to be added to the SM one reads:
\beq \label{eq:LagrMaj0}
\Delta \mathcal{L}_{\rm Majorana} = i \frac{1}{2}\bar{\Psi} \gamma^\mu D_\mu \Psi  - \frac{1}{2}  m_M \; \bar{\Psi} \Psi\,,
\eeq
where $\Psi^C = \Psi$ and the multiplet can be written in terms of a Weyl spinor $\Psi = \left( \begin{array}{c} \chi \\ \bar{\chi} \end{array} \right)$ with components
\beq
\chi = \left( \begin{array}{c}
\chi^{n+} \\
\vdots \\
\chi^+ \\
\chi_0 \\
(\chi^+)^C \\
\vdots \\
(\chi^{n+})^C
\end{array} \right)\,, \qquad \mbox{with} \quad n = I\,,
\eeq
so that the Majorana DM candidate $\chi_0$ is accompanied by $n=I$ Dirac charged partners.
The phenomenology of this multiplet is in large part the same as for a $\tilde{F}^I_0$ Dirac multiplet, in particular the mass split between the various components is given by the same formula given by Eqs.~\eqref{eq:DMf} and \eqref{eq:DMffx}.
Hence, the lightest component is always the neutral one.

\subsection{Mass split from dim--5 Higgs couplings} \label{sec:DMmult_Higgs}

In this section we consider minimal couplings to the Higgs field, which can arise at the level of dim--5 operators. While being suppressed by a UV scale, they are relevant because they can induce a mass split between the components of the DM multiplet, potentially competitive with the EW loops, and change drastically the phenomenology of the multiplet. Hence, while they are not renormalisable couplings, we will consider them here as minimal extensions of the single multiplet models.
Furthermore, as we shall see in Section~\ref{sec:mediator}, they  arise by integrating out a heavier fermion or scalar mediator.

\subsubsection{Basic case for Dirac and Majorana multiplets}

The Brout-Englert-Higgs doublet $\phi_H$, which has $I=1/2$ and $Y=1/2$, can only couple to the DM multiplet via higher dimensional operators.
The lowest order operators have mass--dimension 5 (dim--5) and read:
\beq \label{eq:dim5}
\Delta \mathcal{L}_{\text{dim-5}} \supset - \frac{\kappa}{\Lambda}\ \phi_H^\dagger  T^a_{1/2} \phi_H\ \bar{\Psi} T^a_I \Psi - \frac{\kappa'}{\Lambda}\ \phi_H^\dagger \phi_H\ \bar{\Psi} \Psi \,,
\eeq
where $T^a_I$ are the three SU(2)$_L$ generators for the multiplet with isospin $I$, and $\Lambda$ is a new scale that we assume being beyond the LHC reach to resolve. {For Majorana multiplets, however, the first term is absent as it vanishes identically.}
The second term generates a common mass contribution for all components, thus it simply shifts the mass of the multiplet
\beq
m'_D = m_D + \kappa' \frac{v^2}{2 \Lambda}\,, \qquad m'_M = m_M + \kappa' \frac{v^2}{\Lambda}\,,
\eeq
and generates a coupling to the Higgs, $- \frac{\kappa' v}{\Lambda} h\ \bar{\Psi} \Psi$, that contributes to direct detection, where $v=246$~GeV.

\begin{figure}[tb]
\begin{center}
\includegraphics[width=\textwidth]{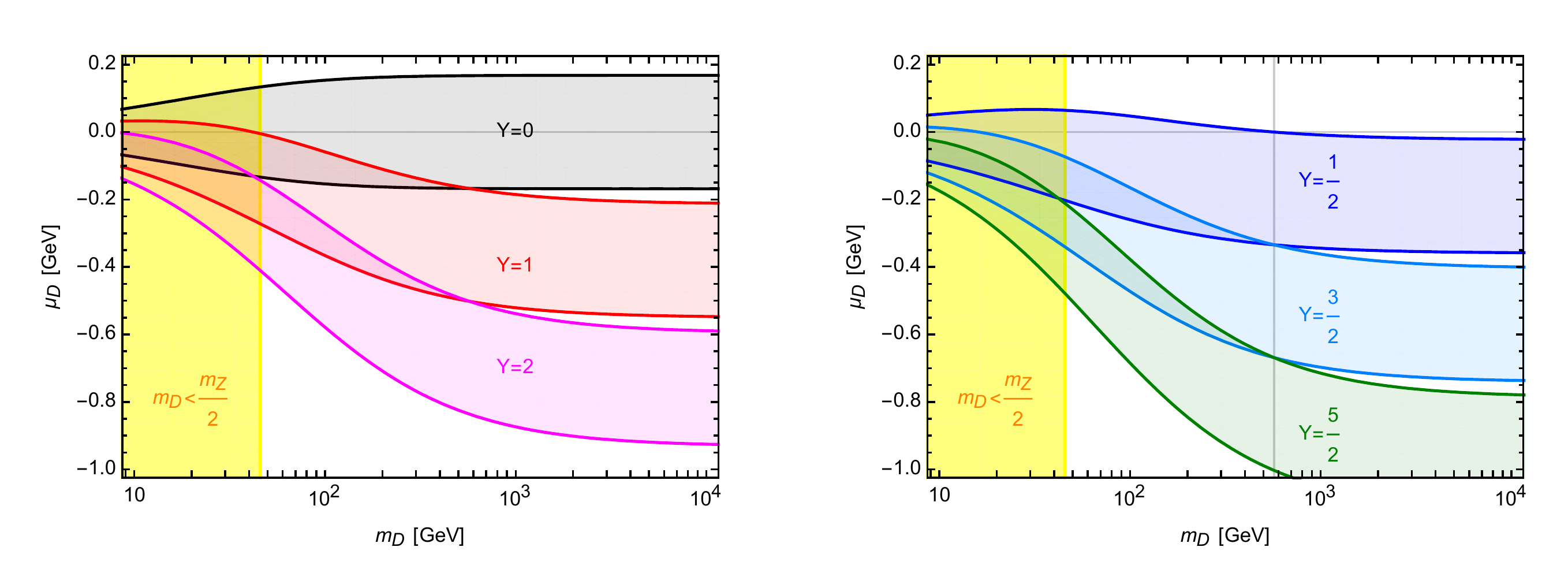}
\end{center}
\caption{The feasible region in the $m_D$--$\mu_D$ parameter space, where the lightest state is neutral, lies within the two lines, where the upper one comes from the $Q=-1$ state and the lower one from $Q=1$. For multiplets with maximal hypercharge, $Y=I$, the region above the upper line is also allowed. The yellow shaded region is excluded by $Z$ decays into the DM multiplet components. {The vertical line in the right hand plot shows the $m_D \lesssim 570$~GeV limit for $Y=1/2$ and $\mu_D =0$.}} \label{fig:HiggsOps}
\end{figure}

The first one, instead, induces a mass split among the various components, thus it may affect the conclusions about the spectrum {of Dirac multiplets} we reached in the previous section.
We recall that the form of the SU(2) generators for a generic isospin $I$  is
\beq
T^3_I = \begin{pmatrix}
I & 0 & \dots & \dots  & 0 \\
0 & I-1 & \dots & \dots & 0 \\
\vdots & \vdots & \ddots & & \vdots \\
\vdots & \vdots & & -I+1 & 0 \\
0 & 0 & \dots & 0 & -I
\end{pmatrix} \,, \quad T^+_I = \frac{1}{\sqrt{2}} \begin{pmatrix}
0 & c_1 & 0 & \dots & 0 \\
0  & 0 & c_2 & \dots & 0  \\
\vdots & \vdots & \ddots &  & \vdots \\
    &    & &   0 & c_{{2I}} \\
 0 & 0 & \dots & 0 & 0  
  \end{pmatrix}\,, \quad
  T^-_I = (T_I^+)^\dagger\,,
\eeq
with
\beq
{c_k = \sqrt{k(2I + 1 - k)}\,, \;\; k = 1, \dots 2I, \;\; \mbox{and} \;\; c_{2I+1-k} = c_k\,. }
\eeq
Once the Higgs field develops its VEV, the only non-vanishing component is 
\beq
\phi_H^\dagger T^3_{1/2} \phi_H = - \frac{1}{2} \varphi_0^\ast \varphi_0 = - \frac{1}{4} (v+h)^2\,,
\eeq
which couples to $\bar{\Psi} T^3_{I} \Psi$. The resulting term in the Lagrangian reads {(C.f. Eq.~\eqref{eq:Dirac} for a characterisation of the components)}
\begin{multline}
\mathcal{L}_{\kappa} = - \mu_D \left( 1 + \frac{h}{v} \right)^2 \left( I\ \bar{\psi}^{n+} \psi^{n+} + (I-1)\ \bar{\psi}^{(n-1)+} \psi^{(n-1)+} + \dots \right. \\
\left. - Y\ \bar{\psi}_0 \psi_0 - \dots - I\ \bar{\psi}^{m-} \psi^{m-} \right)\,,
\end{multline}
where $\mu_D = - \frac{\kappa v^2}{4 \Lambda}$ and we have used the relation $T^3 = - Y$ for the neutral component. In terms of mass splitting, these couplings can be expressed as
\beq
\left. M_Q - M_{Q'} \right|_{\rm Higgs} = \mu_D (Q - Q')\,.
\eeq
Together with the EW loops in Eq.~\eqref{eq:DMf}, the master formula for the mass splits reads:
\begin{equation} \label{eq:DMfmuD}
M_Q - M_{DM} = \delta m_{\rm EW}^{(1)} \ Q^2 + \left(2Y\ \delta m_{\rm EW}^{(2)} + \mu_D \right) Q\,,
\end{equation} 
where loop coefficients $\delta m_{\rm EW}^{(i)}$ can be read off Eq.~\eqref{eq:DMf}. The asymptotic values for large multiplet masses read
\begin{equation}
\lim_{m_D \to \infty} \delta m_{\rm EW}^{(1)} = 166~\mbox{MeV} \quad  \mbox{and} \quad \lim_{m_D \to \infty} \delta m_{\rm EW}^{(2)} = \frac{166~\mbox{MeV}}{c_W} = 188~\mbox{MeV}\,.
\end{equation}
Eq.~\eqref{eq:DMfmuD} shows that for too large $|\mu_D|$, either the $Q=1$ or $Q=-1$ state becomes lighter than the neutral one. The model, therefore, features a feasible DM candidate only if
\begin{equation}
\begin{array}{c} 
-\delta m_{\rm EW}^{(1)} - 2 Y\ \delta m_{\rm EW}^{(2)} < \mu_D < \delta m_{\rm EW}^{(1)} - 2 Y\ \delta m_{\rm EW}^{(2)}\,, \\
\\
\mbox{or} \qquad  \mu_D > -\delta m_{\rm EW}^{(1)} - (2 Y)\ \delta m_{\rm EW}^{(2)} \qquad \mbox{for} \;\; Y = I.
\end{array}
\end{equation} 
The condition for maximal hypercharge stems from the fact that the $Q=-1$ state is absent. The allowed ranges of $\mu_D$ as a function of $m_D$ are shown in Fig.~\ref{fig:HiggsOps} for various values of integer and semi-integer $Y$, where the upper limit should be removed for multiplets with maximal hypercharge. Hence, $\mu_D$ allows to salvage multiplets with $Y > 1/2$. It remains the issue of exclusion by direct detection via the $Z$ coupling: to elude it, one needs to generate a mass split in the neutral state that we discuss in the next subsection.

To connect the feasible values of $\mu_D$ with the scale at which this interaction is generated, it is useful to compare it with the asymptotic value of the EW loops:
 \beq
|\mu_D| < 166~\mbox{MeV} \quad \Leftrightarrow  \quad \frac{\Lambda}{|\kappa|} > 90~\mbox{TeV}\,.
\eeq
This corresponds to the range of $\mu_D$ allowed asymptotically in the $Y=0$ case, and gives a reference  for the scale of new physics $\Lambda$.

\subsubsection{Dirac multiplets with Majorana coupling: case $Y=1/2$} \label{sec:DwithMhiggs}

Models with $Y\neq 0$ are excluded by direct detection via the $Z$ coupling. It is well known that this bound can be avoided if the neutral state is split into two Majorana mass eigenstates via a coupling to the Higgs field. For $Y=1/2$, this occurs at dim--5 level via the operator:
\beq
\Delta \mathcal{L}_{\text{dim--5}} = -\frac{1}{2}  \frac{\kappa_M}{\Lambda}\ \phi_H  T^a_{1/2} \phi_H\ \bar{\Psi} T^a_I \Psi^C + \mbox{h.c.} 
\label{eq:dim5Maj}
\eeq
The operator above is similar in nature to the Weinberg operator in the SM~\cite{Weinberg:1979sa} that gives a Majorana mass to the left-handed neutrinos. Note also that it preserves a $\mathbb{Z}_2$ symmetry on the DM candidate, but breaks the U(1)$_{\rm DM}$.
Its most important effect is to split the neutral Dirac state into two Majorana mass states: the $Z$ boson can only couple the two states to each other, without any diagonal couplings. As long as the heavier Majorana state is not Cosmologically stable, the DM candidate is the lightest one and elastic scattering off nuclei mediated by the $Z$ is absent. The price to pay is a new coupling to the Higgs boson, which also contributes to direct detection. As a fist step, we need to determine what is the effect of the new coupling $\kappa_M$ on the mass ordering inside the multiplet.

In the operator~\eqref{eq:dim5Maj}, the only non-vanishing component of the Higgs current is
\beq
\phi_H  T^+_{1/2} \phi_H = \frac{1}{\sqrt{2}} \varphi_0^2 = \frac{1}{2\sqrt{2}} (v+h)^2\,,
\eeq
which couples to $\bar{\Psi} T_I^- \Psi^C$. The resulting Lagrangian for a generic semi-integer isospin $I$ reads
\begin{multline}
\Delta \mathcal{L}_{\text{dim--5}} =- \frac{1}{2} \mu_M \left( 1 + \frac{h}{v} \right)^2 \left(c_1 \bar{\psi}^{(n-1)+} (\psi^{(n-1)-})^C + \dots + c_k \bar{\psi}^{(n-k)+} (\psi^{(n-k)-})^C + \right. \\
\left. \dots + c_{I+1/2} \bar{\psi}^0 (\psi^0)^C + \dots + c_k \bar{\psi}^{(n-k)-} (\psi^{(n-k)+})^C + \dots \right. \\
\left. + c_1 \bar{\psi}^{(n-1)-} (\psi^{(n-1)+})^C \right) + \mbox{h.c.}
\end{multline}
where $\mu_M = \frac{\kappa_M v^2}{4 \Lambda}$, and we recall that the neutral state corresponds to $k = n = I+1/2$. All states receive a mass correction except the one with the largest electric charge, $\psi^{n+}$. 

For the neutral state, the mass matrix can be written in a Majorana form as follows:
\beq
- \frac{1}{2} \begin{pmatrix} (\bar{\psi}^0)^C & \bar{\psi}^0 \end{pmatrix} \begin{pmatrix}
\tilde{m}_D^0 - 1/2 \mu_D & c_{I+1/2} \mu_M \\
c_{I+1/2} \mu_M & \tilde{m}_D^0 - 1/2 \mu_D
\end{pmatrix} \begin{pmatrix}
(\psi^0)^C \\ \psi^0 \end{pmatrix}\,,
\eeq
where $\tilde{m}_D^0$ includes the one-loop EW corrections. 
 The Majorana mass eigenvalues are
\beq
M_{0,1/2} =\tilde{m}_D^0 - \frac{1}{2} \mu_D \pm c_{I+1/2}\ |\mu_M|\,, \qquad c_{I+1/2} = I+\frac{1}{2}\,.
\eeq
Note that $c_{I+1/2}$ is the largest coefficient in the $T^+_I$ generator and the lightest state always receives a negative contribution to its mass, independently on the sign of $\kappa_M$. Henceforth, this operator always tends to make one neutral state lighter.  For a doublet, $I=1/2$, the charged state does not receive a mass correction from $\kappa_M$, hence the mass split between the charged state and the lightest neutral one can be written as
\beq
\left. M_+ - M_{0,1} \right|_{\tilde{F}_{1/2}^{1/2}} = \delta m_{\rm EW}^{(1)} +  \delta m_{\rm EW}^{(2)} + \mu_D + |\mu_M|\,. 
\eeq
This shows that the presence of a non-zero $\mu_M$ always enlarges the parameter space where the lightest state is neutral, in particular allowing for larger negative values of $\mu_D$ compared to the case with $\mu_M = 0$.

\begin{figure}[tb]
\begin{center}
\includegraphics[width=\textwidth]{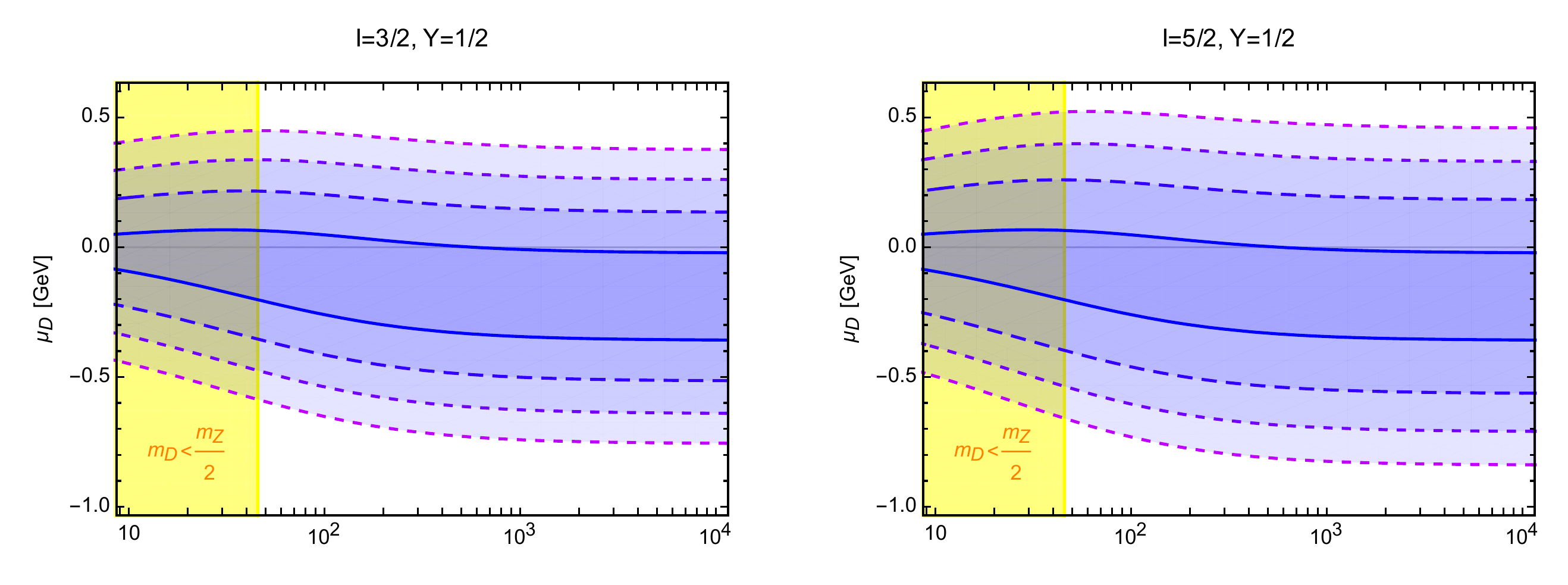}
\end{center}
\caption{Feasible range for $Y=1/2$, $I=3/2$ (Left) and $I=5/2$ (Right) in the presence of a Majorana dim-5 coupling. The lines from solid to dotted correspond to $\mu_M = 0,\ 0.1,\ 0.2,\ 0.3$~GeV. The region with a lightest neutral component lies between the two lines. The yellow shaded region is excluded by the $Z$ decays into the multiplet components. } \label{fig:HiggsOpsM}
\end{figure}

For larger values of the isospin, $I > 1/2$, we need to study the correction to the masses of the charged states, whose mass matrix can be written as
\beq
- \frac{1}{2} \begin{pmatrix} \bar{\psi}^{(n-k)+} & (\bar{\psi}^{(n-k)-})^C \end{pmatrix}
\begin{pmatrix}
\tilde{m}_D^{(n-k)+} + (I-k) \mu_D & c_k \mu_M \\
c_k \mu_M & \tilde{m}_D^{(n-k)-} - (I-k+1) \mu_D
\end{pmatrix} \begin{pmatrix}
\psi^{(n-k)+} \\ (\psi^{(n-k)-})^C \end{pmatrix} 
\eeq
where $\tilde{m}_D^{(n-k)\pm}$ include the one-loop EW corrections. Using the parametrisation adopted in the previous subsection, the mass eigenstates for charge $Q=(n-k)$ states can be written as
\beq
M_{Q,1/2} = \tilde{m}_D^0 + Q^2 \delta m_{\rm EW}^{(1)} - \frac{1}{2} \mu_D \pm \sqrt{Q^2 (\mu_D +\delta m_{\rm EW}^{(2)})^2 + c_k^2 \mu_M^2}\,.
\eeq
The state that receives the potentially largest negative contribution to the mass has charge $Q=1$, for which $c_k \to c_{I-1/2} = c_{I+3/2} = \sqrt{\left(I+\frac{3}{2} \right) \left( I-\frac{1}{2} \right)}$ and the mass difference between the lighter charged and neutral states reads
\beq
M_{+,1} - M_{0,1} = \delta m_{\rm EW}^{(1)} + \left(I+\frac{1}{2} \right) |\mu_M| - \sqrt{(\mu_D +\delta m_{\rm EW}^{(2)})^2 + \left(I+\frac{3}{2} \right) \left( I-\frac{1}{2} \right) \mu_M^2}\,.
\eeq
The lightest state remains the neutral one as long as
\beq
- \delta m_{\rm EW}^{(2)} - \sqrt{X} < \mu_D < - \delta m_{\rm EW}^{(2)} +  \sqrt{X}\,, \qquad X=(\delta m_{\rm EW}^{(1)})^2 + \mu_M^2 + |\mu_M| \ \delta m_{\rm EW}^{(1)} \ (1+2I)\,.
\eeq
This region in the $m_D$--$\mu_D$ parameter space is represented in Fig.~\ref{fig:HiggsOpsM} for $I=3/2$ (Left) and $I=5/2$ (Right), where the curves from solid to dashed correspond to increasing $\mu_M$ from $0$ to $300$~MeV. This plot shows that a non-zero $\mu_M$ always enlarges the allowed band. The same trend occurs for larger isospin values. 
To have a feeling of the scale involved in the generation of $\mu_M$, as a reference the minimal value of $\mu_M$ above which the neutral state is always the lightest for $\mu_D = 0$ and $I=3/2$ is:
\beq
|\mu_M| > 11.5~\mbox{MeV}  \quad \Leftrightarrow  \quad \frac{\Lambda}{|\kappa_M|} < 1300~\mbox{TeV}\,.
\eeq\

A similar splitting can be obtained also for multiplets with hypercharge larger than $1/2$, at the price of higher dimensionality of the operator. For any given semi-integer $Y = N + 1/2$, the operator contains $2 N$ additional $\phi_H$ fields, hence having a mass dimension of $\mbox{dim} = 5+2N = 4 + 2Y$. The main issue with this case is that a sizeable $\mu_M$ would require a relatively low new physics scale:
\beq
|\mu_N| = \frac{\kappa_M}{2 \Lambda^{2Y}} \left(\frac{v^2}{2}\right)^{Y+1/2} > 11.5~\mbox{MeV}  \quad \Leftrightarrow  \quad \Lambda < \kappa_M ^{\frac{1}{2Y}} \left( \frac{v^{2Y+1}}{2^{Y+3/2} (11.5~\mbox{MeV})} \right)^{\frac{1}{2Y}}\,.
\eeq
For $Y=3/2$, this implies $\Lambda < \kappa_M^{1/3}\ 3.4$~TeV, while for $Y=5/2$ we have $\Lambda < \kappa_M^{1/5}\ 1.0$~TeV. Hence, the scale generating these operators is required to be within the range of colliders like the LHC in order for the operator to have sizeable effects.

\begin{figure}[H]
\centering
\includegraphics[width=0.6\textwidth]{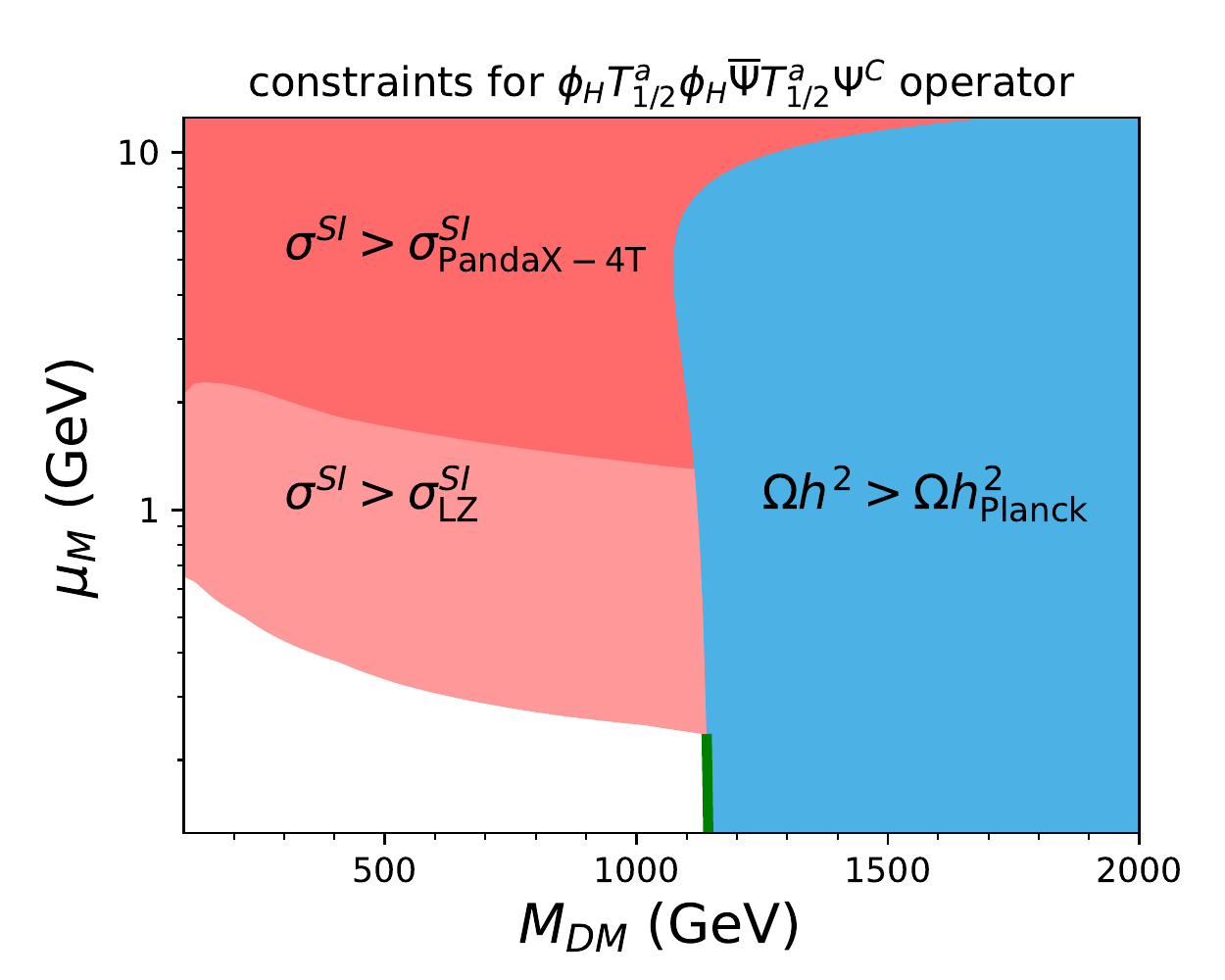}
\caption{
\label{fig:5DEFT}
Exclusion regions for the dim--5 operator in Eq.~\eqref{eq:dim5Maj} in the case of the $\tilde{F}_{1/2}^{1/2}$ model in the $(\mu_M, M_{DM})$ plane: the blue shaded region is excluded by relic density over-abundance; 
the  dark pink region is
excluded by PandaX-4T~\cite{PandaX-4T:2021bab} DM direct detection searches;
the light pink region presents the region that will be probed by future DM direct detection searches with LZ detector~\cite{Akerib:2018lyp}. The narrow green band  indicates
the allowed region with $\Omega h^2=0.12$ that is not accessible to future direct detection experiments.}
\end{figure}

As an example of how the dim--5 operator in Eq.~\eqref{eq:dim5Maj} 
is constrained by relic density and DM direct detection experiments, we show in Fig.~\ref{fig:5DEFT} the 
exclusion regions for a doublet $\tilde{F}_{1/2}^{1/2}$ model in the $(\mu_M, M_{DM})$ plane.
We set $\mu_D = 0$, and recall that the mass splits are given by $\Delta M_0 = 2 |\mu_M|$ between the two Majorana mass states, and $\Delta M_+ = |\mu_M| +$ the EW loops.
The blue shaded region is excluded by relic density over-abundance while the dark pink region is excluded by {current direct detection limits from PandaX-4T~\cite{PandaX-4T:2021bab}}.
The light pink region presents the projected region that future DM direct detection searches with the LUX-ZEPLIN (LZ) detector will be able to probe~\cite{Akerib:2018lyp}. We can see that DM masses 
above $1.1$~TeV are excluded by complementary 
relic density and DM direct detection constraints.
For the relic density, increasing $|\mu_M|$ reduces the co-annihilation via the $W$ and $Z$ gauge bosons, hence requiring a slightly lighter mass, while for $\mu_M\gtrsim 8$~GeV the Higgs couplings start dominating, pushing the DM mass to higher values. However, this region is already excluded by direct detection, as
PandaX-4T excludes  $\mu_M$ above $\sim 2$~GeV. 
The projected LZ limit will probe $\mu_M$ down to $\sim 250$~MeV, a region where the mass split from EW loops becomes relevant. Direct detection due to the EW loops, however, remains too small to be detected, as we will discuss in the text subsection. The narrow green band indicates the region with $\Omega h^2=0.12$ that will not be accessible to direct detection experiments.

\subsection{Loop-induced Direct Detection} \label{sec:DDminimal}

Loop-induced direct detection cross sections in DM models with a single multiplet have been explored in several papers~\cite{Cirelli:2005uq,Essig:2007az,Hisano:2011cs}. In particular, Ref.~\cite{Hisano:2011cs} presents complete results at one-loop (including two-loops for the couplings to gluons via a heavy flavour quark), in the limit where the DM candidate is a Majorana state from a pseudo-Dirac multiplet. Furthermore, the masses of the DM multiplet components are considered to be exactly the same. A cancellation is observed among various amplitudes, leading to a cross section that is significantly smaller than what could be naively expected. 

Motivated by this cancellation, in this section we revisit the one-loop calculation and extend the results to cases where the DM candidate is a Dirac state and for Majorana multiplets. We also included the effect of mass splits in the DM multiplet: while the mass splits are numerically small, these effects can alter the delicate cancellation among the various terms, hence changing dramatically the final result. Furthermore, we will discuss the impact of uncertainties in the nucleon form factors and parton density functions, which can be highly enhanced by the cancellations.

\begin{figure}[htb]
\begin{tabular}{ccc}
\includegraphics[width=0.3\textwidth]{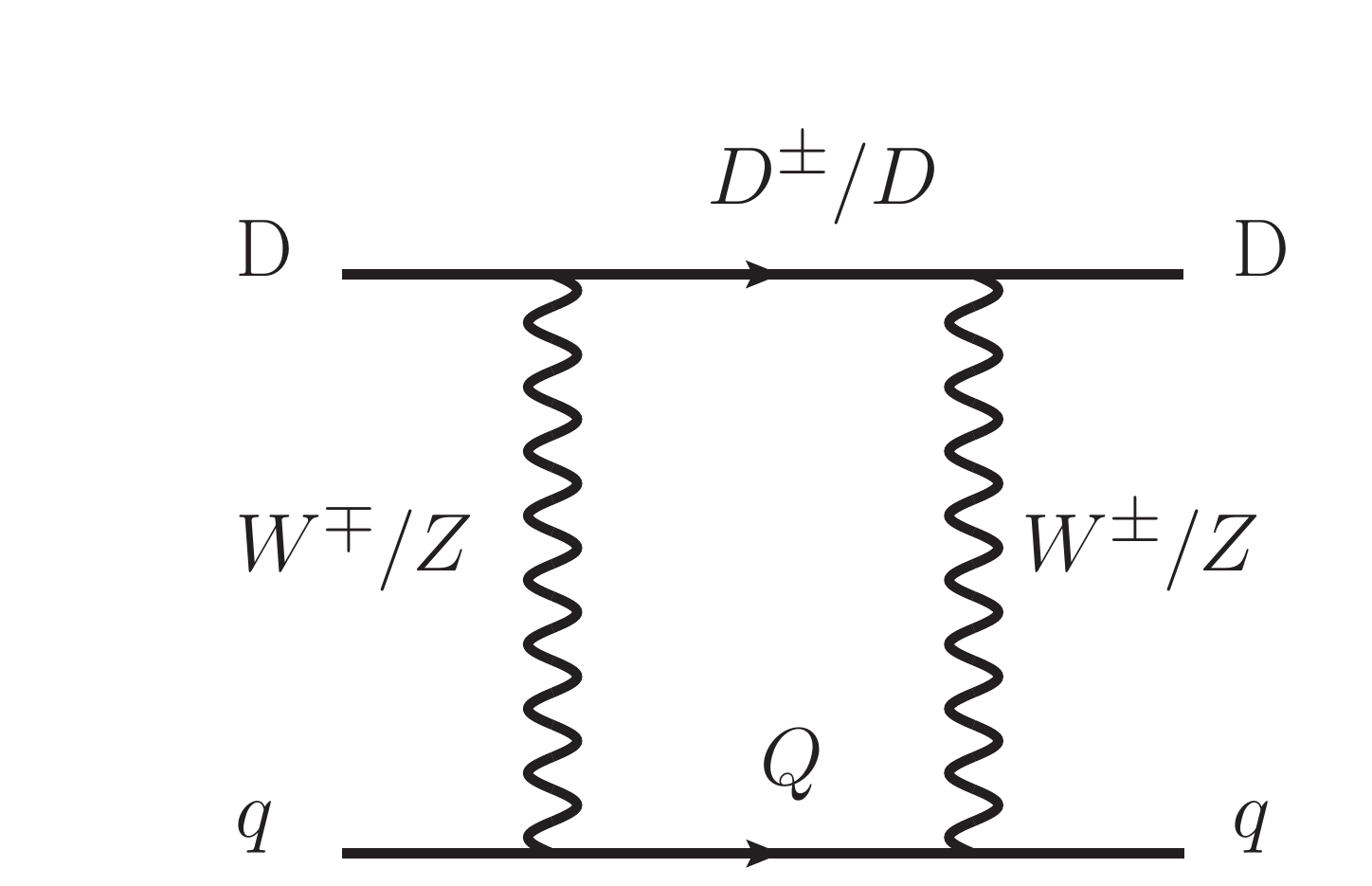} &
\includegraphics[width=0.3\textwidth]{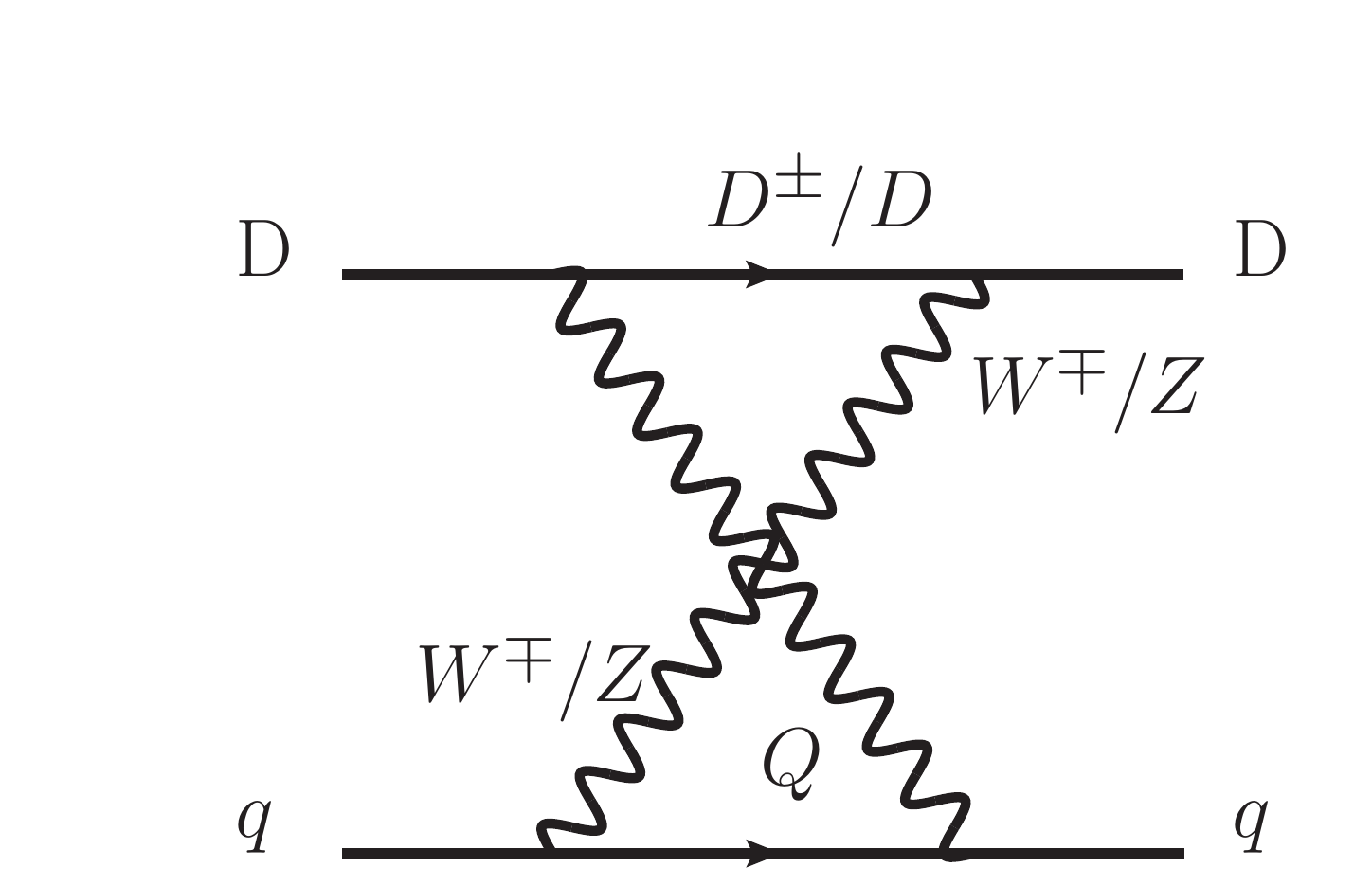} &
\includegraphics[width=0.3\textwidth]{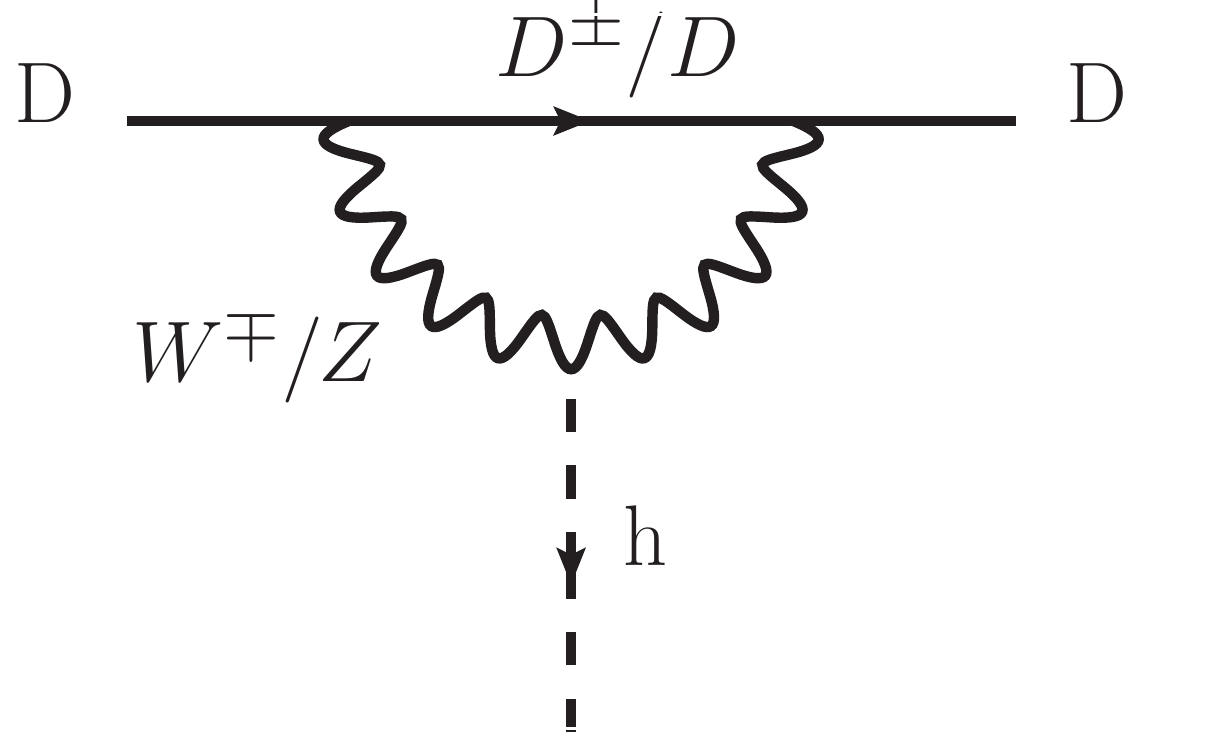} \\
A) & B) & C) 
\end{tabular}
\caption{One-loop diagrams contributing to DM direct detection. If the external quark $q$ is a heavy flavour, it can be connected to the gluons in the nucleons by closing a second loop. \label{fig:triLoop}}
\end{figure}

The one-loop diagrams relevant for direct detection are shown in Figure~\ref{fig:triLoop}, where $q$ ($Q$) are external (internal) SM quarks.  We do not calculate the two-loop diagram resulting from closing the external quark lines for heavy flavours, instead we employ the results of Hisano \emph{et al.}~\cite{Hisano:2011cs}. Furthermore, our calculations are done for spin-independent (SI) cross sections in the limit of zero external momenta and assuming that internal quark masses are comparable to the external ones.~\footnote{This approximation is not valid for the bottom quark with $W$ bosons in the loops, as the top runs inside the loop. However, this contribution is already small, suppressed by the nucleon form factors associated to the bottom. For the external top quark, the mass is fully taken into account in the two-loop coupling calculation to gluons~\cite{Hisano:2011cs}.} 

The amplitudes can be parametrised in terms of the following effective Lagrangian~\cite{Hisano:2011cs}:
\begin{align}\label{eq:DD_Lagrangian}
\mathcal{L}_{eff} &= f_q m_q \bar{D}D\ \bar{q}q
+ \frac{g_q^{(1)}}{M_{DM}} \bar{D} i\partial^\mu \gamma^\nu D\ \mathcal{O}^q_{\mu \nu}
+ \frac{g_q^{(2)}}{M_{DM}^2} \bar{D}(i\partial^\mu )(i\partial^\nu) D\ \mathcal{O}^q_{\mu \nu}
+ f_G \bar{D}D\ G_{\mu \nu}^a G^{a \mu \nu}\,,
\end{align}
where $D$ is the DM fermion, which may be pseudo-Dirac, Majorana or Dirac, 
and the Twist-2 quark current is given by

\beq
\mathcal{O}^q_{\mu \nu} = \frac{i}{2}\ \bar{q} \left( D_\mu \gamma_\nu + D_\nu \gamma_\mu - \frac{1}{2} g_{\mu \nu} \slashed{D} \right) q \,.
\eeq
The first term in Eq.~\eqref{eq:DD_Lagrangian} proportional to $f_q$ is the scalar-scalar (SS) operator, the second and third proportional to $g_q^{(1),(2)}$ are twist-2 operators, and the last one proportional to $f_G$ describes the effectively two-loop coupling to gluons. The coefficients can be explicitly computed (see Appendix~\ref{app:MDM_DD}) and give:
\begin{align}
f_q &= \frac{\alpha^2}{4 m_H^2} \Big[ \frac{(n^2-(2Y+1)^2)}{16 m_W} \kappa_w \Delta_H(w, y_-) + \frac{(n^2-(2Y-1)^2)}{16 m_W} \kappa_w \Delta_H(w, y_+) + \frac{Y^2}{4 c_W^4 m_Z} \kappa_z \Delta_H(z, y_0) \Big] \nonumber \\
& + \frac{\alpha_2^2}{32m_W^3}\Big[(n^2-(2Y+1)^2) \kappa_w \Delta_{S}(w,y_-,a_V^\pm,a_A^\pm) + (n^2-(2Y-1)^2) \kappa_w \Delta_{S}(w,y_+,a_V^\pm,a_A^\pm) \Big] \nonumber \\
& + \frac{\alpha_2^2 Y^2}{4 c_W^4m_Z^3} \kappa_z \Delta_S(z,y_0,a_V^0,a_A^0)\,, \nonumber \\ 
g_q^{(1)} &= \frac{\alpha_2^2}{64m_W^3} \Big[(n^2-(2Y+1)^2) \kappa_w \Delta_{T1}(w,y_-,a_V^\pm,a_A^\pm) + (n^2-(2Y-1)^2) \kappa_w \Delta_{T1}(w,y_+,a_V^\pm,a_A^\pm) \Big]  \nonumber \\
&+\frac{\alpha_2^2 Y^2}{4c_W^4 m_Z^3} \kappa_z \Delta_{T1}(z,y_0, a_V^0,a_A^0)\,, \nonumber \\
g_q^{(1)} &= \frac{\alpha_2^2}{64m_W^3} \Big[(n^2-(2Y+1)^2) \kappa_w \Delta_{T2}(w,y_-,a_V^\pm,a_A^\pm) + (n^2-(2Y-1)^2) \kappa_w \Delta_{T2}(w,y_+,a_V^\pm,a_A^\pm) \Big]  \nonumber \\
&+\frac{\alpha_2^2 Y^2}{4c_W^4 m_Z^3}\kappa_z \Delta_{T2}(z,y_0, a_V^0,a_A^0)\,,
\label{eqn:X_PseudoDirac}
\end{align}
where $w=m_W^2/M_{DM}^2$, $z=m_Z^2/M_{DM}^2$ and $y_i=(M_i-M_{DM})/M_{DM}$. The couplings of the $W$ boson are explicitly given for a multiplet with $n=2I+1$ and hypercharge $Y$. The vector and axial couplings for the quarks are given by $a_V^{0}=\frac{1}{2}T_{3q}-Q_q s^2_W$, $a_A^{0}=-\frac{1}{2}T_{3q}$, $a_V^{\pm}=a_A^{\pm}=\frac{1}{2}$. 
Diagram C in Fig.\ref{fig:triLoop} only contributes to the SS operator and gives rise to the loop function $\Delta_H$, while diagrams A and B generate the loop functions $\Delta_S$, $\Delta_{T1}$, $\Delta_{T2}$. 
{Finally the normalisation factors $\kappa_z$ and $\kappa_w$ depend on the nature of the DM candidate: if D is pseudo-Dirac $(\kappa_z,\kappa_w)=(1,1)$, whereas for Majorana $(\kappa_z,\kappa_w)=(0,1/2)$ and for Dirac $(\kappa_z,\kappa_w)=(4,2)$. }

The mass splits within the DM multiplets are represented by the parameters $y_0$, $y_+$ and $y_-$, which encode the mass split between the two Majorana mass states ($y_0 = 0$ for a Dirac multiplet, while the whole term vanishes for a Majorana multiplet as $Y=0$) and between the charge $Q=\pm 1$ states and the DM state, respectively.
In the limit of zero mass splits, i.e. $y_i \to 0$, our results reproduce the formulas in \cite{Hisano:2011cs}, which we report below for reference:
\begin{align}
f_q &= \frac{\alpha^2}{4 m_H^2} \left[ \frac{(n^2-(4Y^2+1))}{8 m_W} \kappa_w g_H(w) + \frac{Y^2}{4 c_W^4 m_Z} \kappa_z g_H(z) \right] +  \frac{\alpha_2^2 Y^2}{c_W^4m_Z^3} (a_A^2-a_V^2) \kappa_z g_s(z)\,, \nonumber \\
g_q^{(1)} &=  \frac{\alpha_2^2}{8 m_W^3} (n^2-(4Y^2+1)) \kappa_w g_{T1}(w) + \frac{2}{   m_Z^3}\frac{\alpha_2^2 Y^2}{c_W^4} (a_A^2+a_V^2) \kappa_z g_{T1}(z)\,,  \nonumber \\
g_q^{(2)} &= \frac{\alpha_2^2}{8 m_W^3} (n^2-(4Y^2+1))\kappa_w g_{T2}(w) + \frac{2}{ m_Z^3}\frac{\alpha_2^2 Y^2}{c_W^4} (a_A^2+a_V^2) \kappa_z g_{T2}(z)\,.\label{eqn:Xlim}
\end{align}

As already mentioned, for the coupling to gluons we use the two-loop computation presented in Ref.~\cite{Hisano:2011cs} for vanishing mass splits of the DM multiplet. The contribution can be expressed in terms of long-distance (LD, dominated by momenta of the order of the light quark masses) and short-distance (SD, dominated by momenta of the order of the $W/Z$ bosons or of the DM states) contributions, as follows:
\begin{equation}
f_G = \sum_{q=u, d, s, c, b, t} \left. f_G \right|_q^{\rm SD} + \sum_{Q=c,b,t} c_Q \left. f_G \right|_Q^{\rm LD}\,,
\end{equation}
where the LD contribution of the light quarks are taken into account in the SS coefficients $f_q$ and NLO corrections in QCD are embedded in the coefficients $c_Q$. Explicit results can be found in Ref.~\cite{Hisano:2011cs}.

The SI cross section for DM scattering off target nucleon $N$ is expressed as
\begin{equation}
\sigma^{SI}_N = \frac{4}{\pi} \frac{M_{DM}^2 m_N^2}{(M_{DM}+m_N)^2} |f_N|^2\,,
\end{equation}
where 
\begin{equation}
\frac{f_N}{m_N} = \sum_{q=u,d,s} f_{Tq} f_q + \sum_{q=u,d,s,c,b} \frac{3}{4} (q(2)+\bar{q}(2))(g_q^{(1)} + g_q^{(2)}) - \frac{8 \pi}{9 \alpha_s}f_{TG}f_G\,.
\label{eqn:fNovermN}
\end{equation}
Here, $f_{Tq}$ are the proton form factors for the quarks, while $f_{TG} = 1-\sum_{q=u,d,s}f_{Tq}$ applies for gluons (note that charm, bottom and top are considered heavy flavours in this formula, and associated to the gluon form factor), while $q(2)$ and $\bar{q}(2)$ are second moments (evaluated at $\mu=m_Z$) for quarks and anti-quarks respectively.

\begin{figure}[H]
	\centering
	\includegraphics[width=0.8\textwidth]{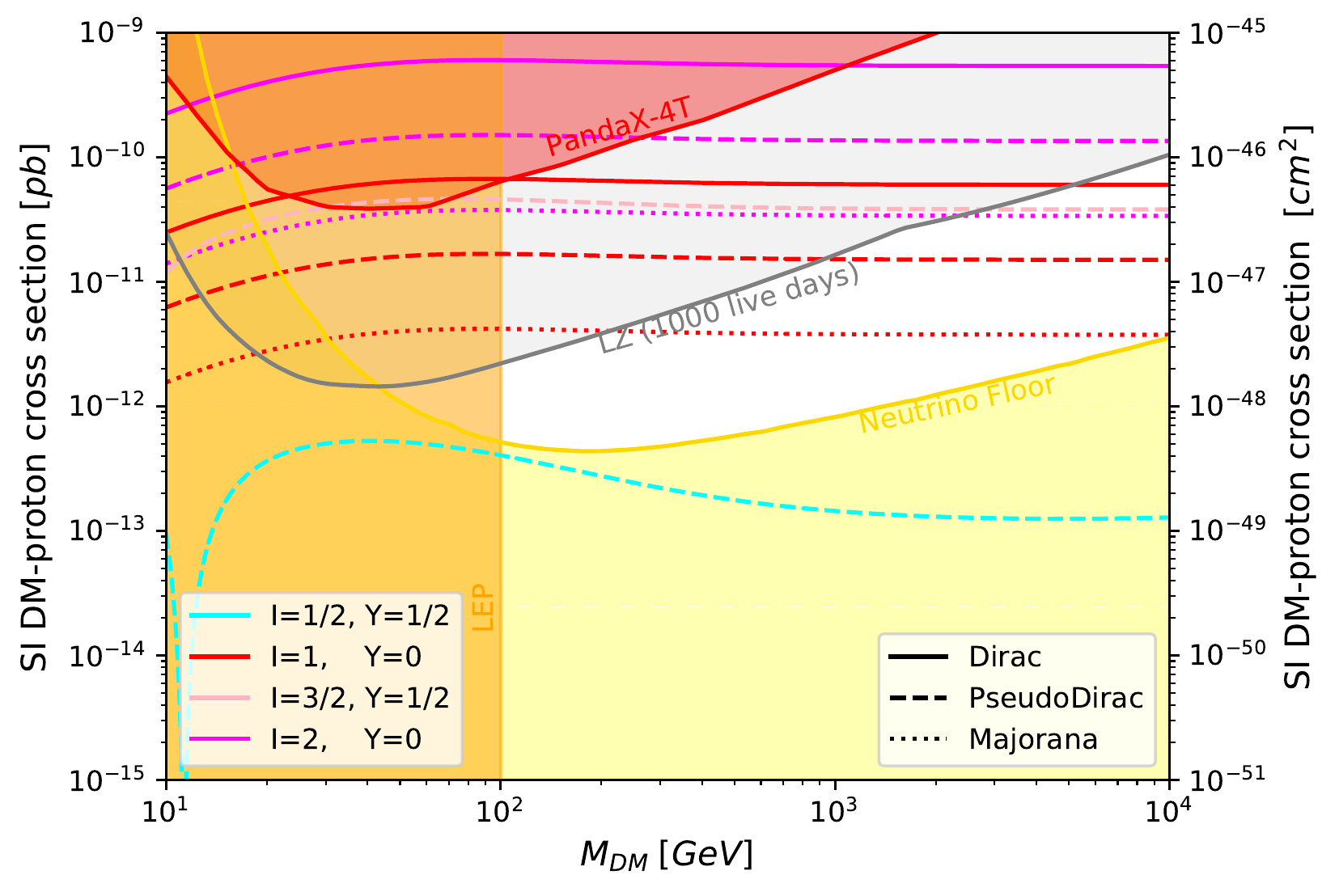}
	\caption{The SI DM-proton cross section for a single fermion multiplet, for surviving cases $I\leq2$ for which the neutral component is the lightest.
		The cases for Dirac multiplet with $Y\ne 0$ are not shown, since they are excluded by direct detection via $Z$-boson exchange.  \label{fig:dmdd-loops} }
\end{figure}

In the zero mass split case, a strong cancellation has been observed between the contribution of the twist-2 operators ($g_q^{(1)}$ and $g_q^{(2)}$) and the gluon ($f_G$) contributions, while the SS one ($f_q$) tends to be smaller \cite{Hisano:2011cs}. This result is shown in Fig.~\ref{fig:dmdd-loops}, were we plot the SI cross sections for various cases compared to the current exclusion from PandaX-4T \cite{PandaX-4T:2021bab} and the projection from the future LZ~\cite{Akerib:2018lyp}. The border of the yellow shaded region labelled ``Neutrino Floor" corresponds to the sensitivityestimate achievable at each DM mass for a one neutrino event exposure at liquid Xenon detectors~\cite{Billard:2013qya}. We have digitised data for the PandaX-4T, LZ and neutrino floor limits, and they are now publicly available on the PhenoData platform~\cite{PhenoData:PandaX-4T, PhenoData:LZ, PhenoData:NeutrinoFloor}.
We would like to note that the neutrino floor limit for one neutrino event can be improved (i.e. lowered) by future experiments with lower energy threshold \cite{Billard:2013qya} potentially by about one order of magnitude.
One can see that only multiplets with $I \geq 2$ can be completely probed by LZ up to masses of $10$~TeV. The observed cancellation, however, is very sensitive to two important effects: the nuclear uncertainties on the form factors and on the second moments, and the mass splits within the DM multiplet. We will discuss both below, starting from the former.

\subsubsection{Impact of uncertainties on nucleon form factors and parton distribution functions (PDFs)}

The proton form factors for light quarks may be calculated~\cite{Ellis:2018dmb} in terms of light quark mass ratios, $m_u/m_d=0.46\pm0.05$ and $m_s/(m_u+m_d)=13.75\pm0.15$, and quantities associated with nucleonic matrix elements, $\Sigma_{\pi N}=46\pm11$~MeV, $\sigma_s=35\pm16$~MeV and $z=1.258\pm0.081$. Explicitly, they are given by
\begin{align}
m_p f_{T_u}^p &= \frac{2m_u}{m_u+m_d} \left[ \frac{z}{1+z} \Sigma_{\pi N} + \frac{m_u+m_d}{2m_s} \frac{1-z}{1+z} \sigma_s \right]\,, \nonumber \\
m_p f_{T_d}^p &= \frac{2m_d}{m_u+m_d} \left[ \frac{1}{1+z} \Sigma_{\pi N} - \frac{m_u+m_d}{2m_s} \frac{1-z}{1+z} \sigma_s \right]\,,\nonumber \\
m_p f_{T_s}^p &= \sigma_s\,. \label{eqn:formFactors} 
\end{align}
In order to combine errors from all sources, we use the Monte Carlo approach whereby we estimate the sampling distribution of the cross section via the generation of points from the sampling distributions of underlying parameters. For the form factors, we sample from a multivariate Gaussian defined by the input parameters given above, assuming that errors are uncorrelated. The distribution of form factor values are computed using Eqs.\eqref{eqn:formFactors}. 

The uncertainties on the second moments $q(2)$ and $\bar{q}(2)$ derive from the uncertainties in the parton distribution functions (PDFs). We take into account both the uncertainties in the PDF fitting procedure, and in the scale variation. In practice, we concurrently sample from the CTEQ18NLO \cite{Hou:2019efy} PDFs, using the Hessian implementation in LHAPDF \cite{Buckley:2014ana}, before numerically integrating these PDFs to generate the second moment values. We probe the variation from PDF scale by sampling from a log-normal distribution for the PDF scale, $\mu$, with central value $\mu=m_Z$ such that the $1\sigma$ bands fall on $\mu=m_Z/2$ and $\mu=2m_Z$ (i.e $\log_2(\mu/m_Z)$ is normally distributed with mean 0 and standard deviation 1).

\begin{figure}[H]
	\centering
	\includegraphics[width=0.8\textwidth]{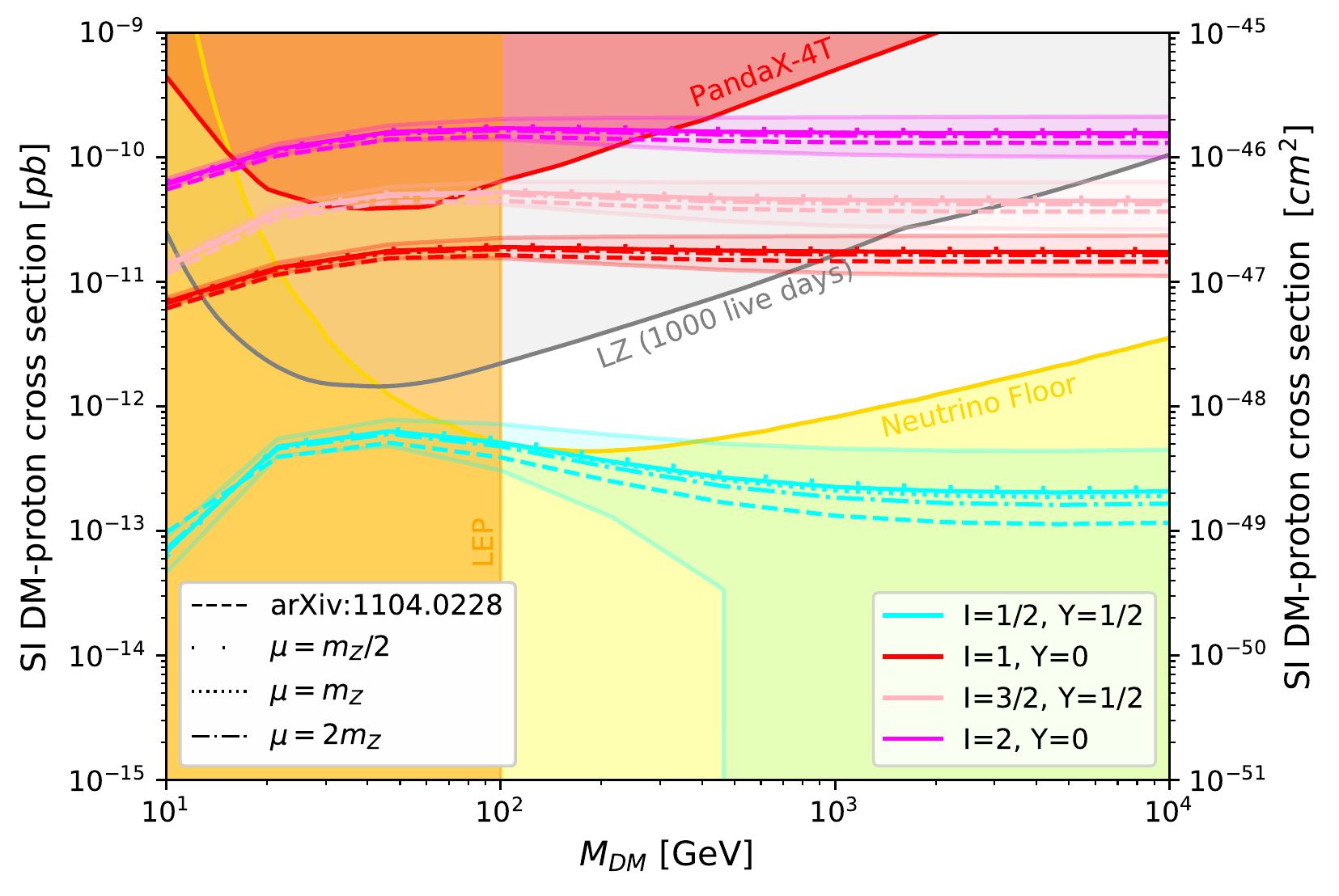}
	\caption{Impact of form factor and PDF uncertainties on the SI cross section for various models with a single DM fermion multiplet. We show results for pseudo-Dirac cases. {For $Y=0$, Majorana and Dirac cases can be obtained by a simple scaling of the cross sections by a factor of $1/4$ and $4$ respectively. We recall that only the pseudo-Dirac case is allowed for $Y\neq 0$.} \label{fig:DD_errors}}
\end{figure}

The uncertainties propagated on the SI cross section are depicted in Fig.~\ref{fig:DD_errors}, where, for each model, the solid line represents the mean and the band signifies the 95\% confidence interval (we show results for pseudo-Dirac case, as the Majorana and Dirac cases can be obtained by a simple numerical scaling). For comparison, the dashed lines show the results of Ref.~\cite{Hisano:2011cs}, where the values used are $f_{Tu}=0.023$, $f_{Td}=0.032$, $f_{Ts}=0.020$ and second moments (evaluated at $\mu=m_Z$) $u(2)(\bar{u}(2))=0.22(0.034)$, $d(2)(\bar{d}(2))=0.11(0.036)$, $s(2)(\bar{s}(2))=0.026(0.026)$, $c(2)(\bar{c}(2))=0.019(0.019)$, $b(2)(\bar{b}(2))=0.012(0.012)$. The difference is due to the fact that we use a different set of PDFs. To study the effect of scale dependence from the PDFs, we also show in various dotted styles the values of the cross sections for $\mu=m_Z/2$, $m_Z$, $2m_Z$ respectively, while keeping all the other parameters fixed. The result shows that the main contribution to the uncertainties derive from the form factors.

For models with $I \geq 1$, the uncertainties in the cross section derives in a sizeable uncertainty in the bound on the DM mass, which amounts to several hundred GeV. For the doublet, $I=1/2$, instead, we observe that a cancellation may occur for $M_{DM} > 450$~GeV, hence making a prediction for the direct detection reach impossible in this mass range. Nevertheless, the cross section always remains below the reach of LZ, and will likely escape detection.


\subsubsection{Impact of mass splits}

As we have shown, the current uncertainties in the nucleon form factors and PDFs produce relevant uncertainties on the SI cross sections, resulting in variations of several hundred GeV in the DM mass bound or producing cancellations in the doublet case. Even if these uncertainties were substantially reduced, the one-loop calculation is sensitive to the mass splits within the DM multiplet, which were not taken into account so far. Here, we extended the loop computation to take into account mass splits, and we re-evaluate the cancellations observed in Ref.~\cite{Hisano:2011cs}. As input parameters, we use the same ones used in \cite{Hisano:2011cs} and recapped at the end of last section.

\begin{figure}[H]
\centering
	\includegraphics[width=1.0\textwidth]{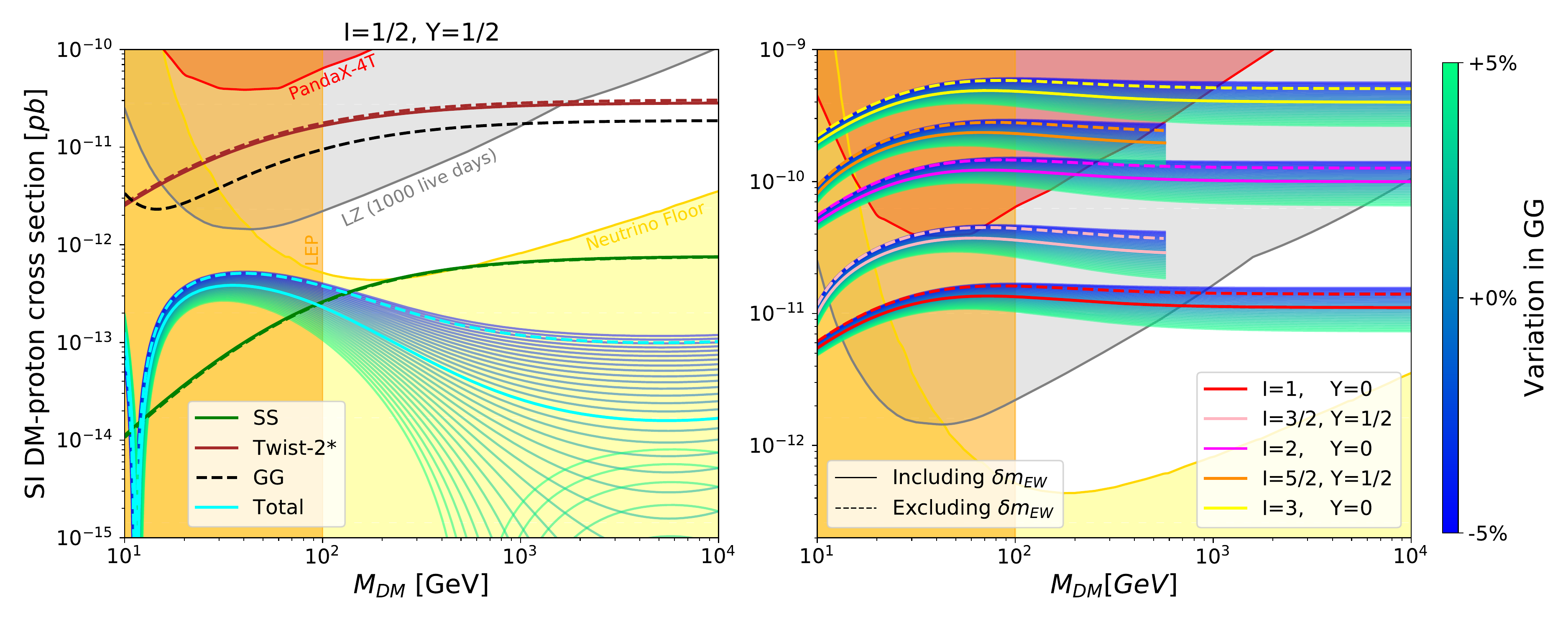}
\caption{Impact of mass splits on the one-loop cross sections for pseudo-Dirac multiplet. On the left, we show the contribution of various operators with and without mass splits, and the impact on the total. The variation is due to a 5\% variation in the GG contribution. On the right we show the total contribution with variation for other multiplets with larger isospin, $I\geq 1$. \label{fig:dmdd-loops-dm}}
\end{figure}

Our calculation takes into account the mass splits in the one-loop results for both SS and twist-2 operators, however this effect is not included yet in the two-loop computation for the gluon couplings. To supply to this, we include a 5\% variation to the latter.
In the left panel of Fig.~\ref{fig:dmdd-loops-dm}, we show the impact of the radiative mass split ($\delta m_{EW}$) for the doublet case. We show separately the contribution of SS and twist-2 operators, comparing the result with (dashed) and without (solid) mass splits taken into account. While the effect on each amplitude is small, the total cross section sees a substantial reduction once the mass splits are taken into account. As the two-loop GG contribution does not include the mass splits, we consider an additional 5\% variation: as shown by the band, this small effect can cause cancellations to occur for $M_{DM} \gtrsim 400$~GeV. Numerically, this effect is similar to the impact of the uncertainties, hence showing that the two are comparable.

In the right panel of Fig.~\ref{fig:dmdd-loops-dm}, we show the total contributions for models with $I \geq 1$. An enhanced cancellation is also observed in this case when comparing the solid lines (with the mass splits) to the dashed ones (without). The 5\% variation in the GG contribution, instead, generates an uncertainty that derives in an uncertainty of several hundred GeV on the DM mass.
The results presented here show that both mass splits and nuclear uncertainties produce similar effects on the SI cross sections, hence motivating further studies in this direction.

%% file: 04-DM+onemore_multiplet.tex

\section{Fermionic Dark Matter with one additional multiplet} \label{sec:mediator}

In this section we present the classification of models that contain one additional multiplet (mediator multiplet), in addition to the DM one. The mediator multiplet can be either odd or even under the symmetry protecting the stability of the DM candidate, and its quantum numbers are limited (and defined) by the requirement of the renormalisability and gauge invariance of its interaction with the DM multiplet (and SM fields).
We recall that we use different labels $F$/$\tilde{F}$ and $M$/$\tilde M$ for Dirac and Majorana fermion multiplets, respectively, since the two choices often lead to rather different models when a mediator is present, as we discuss below.

\subsection{Even scalar mediator ($\tilde{F}^I_Y S^{I'}_{Y'}$ and $\tilde{M}^I_0 S^{I'}_0$).}

\label{subsec:even-scalar-med}

The case of a scalar mediator that couples to the SM has been one of the first scenarios considered in simplified models (see e.g.~\cite{DiFranzo:2013vra,Abdallah:2014hon,Buckley:2014fba,Baek:2015lna,Abdallah:2015ter,Abercrombie:2015wmb,Boveia:2016mrp}), however it has been by now established that it is not simple nor minimal to achieve phenomenologically relevant models once the simplified case is included in a fully gauge-invariant model~\cite{Bell:2016ekl,Bauer:2017ota}. In particular, couplings of a single scalar to SM fermions are hard to obtain without breaking the EW symmetry, while couplings to gauge bosons only arise at dim--5 operator level unless the scalar is allowed to develop a non-zero vacuum expectation value. In the following, we will limit ourselves to the most minimal scenarios and not consider higher dimensional operators.

The models we consider feature a Yukawa coupling connecting two DM multiplets and a (peudo)scalar mediator multiplet.
They can be classified as follows (for the sake of minimality we consider  multiplets with zero hypercharge, $S^I_0$, to be  real):
\begin{itemize}
\item[D1 -] $\Delta \mathcal{L}_{D1} = - y_{\psi 1}\, \Phi\ \bar{\Psi} \Psi$, where the scalar multiplet has $I' = 0, 1, \dots 2I$ and $Y'=0$. For Dirac multiplets, this coupling preserves a U(1)$_{\rm DM}$ global symmetry acting on the DM multiplet. The scalar multiplet $S^I_0$ is real and CP-even.

\item[D2 -] $\Delta \mathcal{L}_{D2} = - i y_{\psi 2}\, \Phi\ \bar{\Psi} \gamma^5 \Psi$ is similar to the previous case, except for the presence of the $\gamma^5$ implying, simply, that $\Phi$ is a real CP-odd multiplet.

\item[D3 -] {$\Delta \mathcal{L}_{D3} = - y_{\psi 3}\, \Phi\ \bar{\Psi} \Psi^C + \mbox{h.c.}$}, where the scalar multiplet has {$Y'=2Y$} and {$I' = 0, 2, \dots 2I$
for $I$ integer, and 
$I' = 1, 3, \dots 2I$
  for $I$ semi-integer.} If $Y \neq 0$, the global U(1)$_{\rm DM}$ can be extended by {giving an internal charge $Q_{\rm DM}$ to $\Psi$ and respective charge {$2 Q_{\rm DM}$} to $\Phi$}
{(a linear coupling of $\Phi$ to SM would violate U(1)$_{DM}$, thus only a $\mathbb{Z}_2$ acting on the DM multiplet would survive).} For $Y=0$, we can still define a $\mathbb{Z}_4$ under which $\Phi \to - \Phi$ and $\Psi \to i \Psi$ (this can be broken to a dark $\mathbb{Z}_2$ in presence of linear couplings of the scalar, or a concomitant presence of D1/D2 couplings). For a Majorana multiplet this coupling is equivalent to D1.
 
\item[D4 -] {$\Delta \mathcal{L}_{D4} = - i y_{\psi 4}\, \Phi\ \bar{\Psi} \gamma^5 \Psi^C + \mbox{h.c.}$} is similar to the previous one, except that the CP properties of the scalar are altered. For a Majorana multiplet this coupling is equivalent to D2.

\end{itemize}

The properties of all the possible models are summarised in Table~\ref{tab:FtildeS}, where we identified five template scenarios with distinct properties. {Note that only integer isospin and hypercharges of the scalar mediator are allowed.} In the 4th and 5th columns (``DM sym." and ``Etx. sym.", respectively), we list the largest Dark symmetry allowed by the above Yukawa couplings, which could be broken by the couplings of the scalar mediator multiplet $\Phi$ to the SM. The last column contains the scalar mediators that can have linear (renormalisable) couplings to the SM.

\begin{table}[htb] \begin{center}
\begin{tabular}{|c|c|c|c|c|c|c|}
\hline
Model & D1/D2 & D3/D4 & DM sym. & Ext. sym. & Ext. Charges & Linear to SM \\
\hline \hline
 &  &  &  &  &  & 
\\
$\tilde{F}^I_{Y} S^{I'}_0$ & $\surd$ & - & U(1) & - & - & $S^0_0$, $S^1_0$  \\
 &  &  &  &  &  & 
\\
\hline
\multirow{2}{*}{$\tilde{F}^{I=\mbox{int.}}_0 S^{I' = \mbox{even}}_0$}  
& \multirow{2}{*}{$\surd$} 
& \multirow{2}{*}{$\surd$} 
& \multirow{2}{*}{$\mathbb{Z}_2$} 
& \multirow{2}{*}{-} 
& \multirow{2}{*}{-}
& \multirow{2}{*}{$S^0_0$} \\
&  &  &  &  &  & 
\\
\hline
\multirow{2}{*}{$\tilde{F}^{I=\mbox{int.}}_0 S^{I' = \mbox{even}}_0$}  
& \multirow{2}{*}{-} 
& \multirow{2}{*}{$\surd$} 
& \multirow{2}{*}{$\mathbb{Z}_2$} 
& \multirow{2}{*}{$\mathbb{Z}_4$} 
&  $\Phi \to - \Phi$
& \multirow{2}{*}{$S^0_0$} \\
&  &  &  &   & $\Psi \to i \Psi$ &  \\
\hline
\multirow{2}{*}{$\tilde{F}^I_Y S^{I'}_{2Y}$} & \multirow{2}{*}{-} & \multirow{2}{*}{$\surd$} & \multirow{2}{*}{$\mathbb{Z}_2$} & \multirow{2}{*}{U(1)} & $\Phi \to e^{i 2 Q_{\rm DM}} \Phi$ & \multirow{2}{*}{$S^1_1$, $S^0_2$} \\
 &  &  &  & & $\Psi \to e^{i Q_{\rm DM}} \Psi$ &  \\
\hline
 &  &  &  &  &  & 
\\
$\tilde{M}^I_0 S^{I'=\mbox{even}}_{0}$ & $\surd$ & $\surd$ & $\mathbb{Z}_2$ & -  & - & $S^0_0$
\\
 &  &  &  &  &  & 
\\
\hline
\end{tabular} \end{center}
\caption{Classification of models with a scalar even mediator multiplet. The extended symmetry in the fifth column refers to charges assigned to the scalar multiplet, as shown in the sixth column. In the last column we highlight scalar multiplets that allow for linear couplings to the SM that break the extended symmetry. For Majorana multiplets (last row), D3$\equiv$D1 and D4$\equiv$D2.} \label{tab:FtildeS}
\end{table}

The most general Lagrangians, for the real and complex scalar multiplets {(with integer hypercharges)}, read:
\beq
\Delta \mathcal{L}_{S^{I'}_0} &=& \frac{1}{2} (D_\mu \Phi)^2  - V(\Phi) - \frac{1}{2} \lambda\, (\Phi^2) (\phi_H^\dagger \phi_H) + V_{\rm linear}\,,  \label{eq:LagSreal}\\
\Delta \mathcal{L}_{S^{I'}_{Y'}} &=&  | D_\mu \Phi |^2  - V(\Phi, \Phi^\dagger) - \lambda\, (\Phi^\dagger \Phi) (\phi_H^\dagger \phi_H)  - \lambda'\, (\Phi^\dagger T^a_{I'} \Phi) (\phi_H^\dagger \tau^a\phi_H) + V_{\rm linear}\,; \label{eq:LagScomplex} 
\eeq
where $V$ is a generic potential for the scalar.
Note that, for the real case, only one Higgs portal coupling is allowed due to the fact that $\Phi$ has integer isospin. The term $V_{\rm linear}$ contains eventual linear couplings of $\Phi$ to a SM operator, which can be made of the Higgs field or leptons. Only 5 such cases occur:
\beq
S^0_0 & \Rightarrow & V_{\rm linear} = - \mu_{00}\ \Phi\ \phi_H^\dagger \phi_H\,, \;\; \mbox{(CP-even)}\,;\\
S^1_0 & \Rightarrow & V_{\rm linear} = - \mu_{10}\ \Phi^a\ \phi_H^\dagger \tau^a \phi_H\,,\;\; \mbox{(CP-even)}\,; \\
S^1_1 & \Rightarrow & V_{\rm linear} = - \mu_{11}\ \Phi^a\ \phi_H^\dagger \tau^a \phi_H^\dagger + \mbox{h.c.}\,; \label{eq:muS11}\\
S^0_2 & \Rightarrow & V_{\rm flinear} = - \xi_{02}^{ij}\ \Phi\ \bar{l}_R^{i\ C} l_R^j\,; \label{eq:xiS02}\\
S^1_1 & \Rightarrow & V_{\rm linear} = - \xi_{11}^{ij}\ \Phi^a\ \bar{l}_L^{i\ C}\tau^a l_L^j\,, \label{eq:xiS11}
\eeq
where $i,j$ are flavour indices.

The linear Higgs portal coupling, allowed only for CP--even $S^0_0$ and $S^1_0$ and for the charged iso-triplet $S^1_1$, necessarily implies that the scalar mediator acquires a vacuum expectation value {$\langle \Phi \rangle \neq 0$} via the Higgs one, thus a universal coupling to SM fermions is generated via the mixing with the physical Higgs boson~\cite{MarchRussell:2008yu,LopezHonorez:2012kv}. 
However, this mixing is strongly suppressed in the triplet cases because of three--level contributions to the $\rho$ parameter, while in the singlet case milder (but still important) bounds derive from the measurement of the $125$~GeV Higgs couplings~\cite{Arcadi:2019lka}. 
This shows that the couplings of the scalar mediator to SM fermions and gauge bosons are deemed to be small.
If the scalar mediator multiplet is much heavier than the EW scale and the DM mass, the bound on the coupling can be weakened. Integrating out the scalar multiplet generates the dim--5 operators between the DM multiplet and the Higgs boson we introduced in Section~\ref{sec:DMmult_Higgs}. Depending on the quantum numbers of the scalar multiplet, the following possibilities are realised for Dirac multiplets:
\begin{eqnarray}
S_0^0 & \Rightarrow & \frac{\kappa'}{\Lambda} = \frac{y_{\psi_1} \mu_{00}}{m_\Phi^2}\,, \qquad \kappa = \kappa_M = 0\,; \\
S_0^1 & \Rightarrow & \frac{\kappa}{\Lambda} = \frac{y_{\psi_1} \mu_{10}}{m_\Phi^2}\,, \qquad \kappa' = \kappa_M = 0\,; \\
S_1^1 & \Rightarrow & \frac{\kappa_M}{\Lambda} = \frac{2 y_{\psi_3} \mu_{11}}{m_\Phi^2}\,, \qquad \kappa = \kappa' = 0\,.
\end{eqnarray}
The singlet $S_0^0$ also generates the operator for the Majorana DM multiplet case via the coupling D1:
\begin{equation}
S_0^0 \Rightarrow  \frac{\kappa'}{\Lambda} = \frac{2 y_{\psi_1} \mu_{00}}{m_\Phi^2}\,. 
\end{equation}
{The couplings $\kappa$, $\kappa'$ and $\kappa_M$ are defined in Eqs~\eqref{eq:dim5} and \eqref{eq:dim5Maj}.}


{The other two cases only allow for couplings to leptons. The triplet $S^1_1$ in Eq.~\eqref{eq:xiS11}} corresponds to type-II see-saw models~\cite{Magg:1980ut,Cheng:1980qt,Lazarides:1980nt} for neutrino mass generation, and it has been studied in connection to DM in Refs~\cite{Chen:2014lla,Biswas:2017dxt,Gu:2019ogb,Lineros:2020eit}. The {doubly-charged scalar $S^0_2$ in Eq.~\eqref{eq:xiS02}} also contributes to neutrino masses, as it breaks lepton number by two units, and it has been studied in Ref.~\cite{Gustafsson:2012vj} paired with a scalar DM multiplet.

We finally note that a vacuum expectation value for the scalar mediator can be induced in all cases, in particular via the quartic coupling to the Higgs field {yielding $\langle \Phi^\dagger \Phi \rangle \neq 0$}, and with all the limitations and bound described above. The phenomenology of such cases follow the analyses done in the simplified models~\cite{Buchmueller:2015eea,Bell:2016ekl}.

If a vacuum expectation value is not generated, then the presence of the coupling to the scalar mediator multiplet does not affect the mass spectrum of the DM multiplet: this implies that models with $Y\neq 0$ are excluded by $Z$-mediated direct detection, while $Y=0$ models are probed at one-loop level, as discussed in Section~\ref{sec:DDminimal}, with additional contributions of the scalar mediator if it couples to quarks.

There is, however, a new class of mediators that arise from our classification: scalar mediators that only have bilinear couplings to the SM Higgs field. Such models have new interesting features that we will study in detail in the next section. For now, we content ourselves to classify the relevant models:

\begin{itemize}

\item[(a)] {\it Accidental stability:} the scalar mediator multiplet can be accidentally stable if all linear couplings to the SM are forbidden, and it is lighter than {twice the mass of} the fermionic DM multiplet. {One should note, however, that if the DM multiplet has $I\neq 0$,
then  due to the couplings D1--D4, 
triangle loops generate couplings of the scalar mediator multiplet to two EW gauge bosons, hence making it unstable. The only model with accidental stability is, therefore, $\tilde{F}_0^0 S_0^0$ with coupling D2 (i.e. CP--odd mediator).~\footnote{For $\tilde{M}_0^0 S_0^0$, the scalar mediator is CP--even, thus linear coupling to the Higgs cannot be forbidden.}}


\item[(b)] {\it Protected stability by $\mathbb{Z}_4$:} in models with an extended $\mathbb{Z}_4$ symmetry, i.e. $\tilde{F}_0^I S_0^{I'}$ with couplings D3/D4, the stability of the mediator is guaranteed by a discrete charge. Direct detection bounds apply via loop induced couplings for the two DM components, like the ones discussed in Section~\ref{sec:DDminimal}. The most minimal surviving model involves gauge singlets, $\tilde{F}_0^0 S_0^0$ with D4.  

\item[(c)]  {\it Protected stability by $\mbox{U(1)}$:} similarly, stability can be guaranteed by a U(1) symmetry in $\tilde{F}_Y^I S_{2Y}^{I'}$ models. In such cases, however, the fact that $Y\neq 0$ requires that a Majorana mass split is generated in the neutral DM fermionic candidate. This can only be achieved in models $\tilde{F}_{1/2}^I S_1^1$ with the couplings in Eq.~\eqref{eq:muS11} included: this however explicitly breaks the U(1) symmetry and allows decays of the mediator.
By integrating out $S_1^1$, or by a small vacuum expectation value, the same mass split induced by the Higgs operator discussed in Section~\ref{sec:DwithMhiggs} will arise. The only difference would be the presence of a coupling to a scalar mediator, which can affect the relic density computation.

\end{itemize}
We note that in all cases, direct detection from the fermionic DM candidate may be avoided if the dominant contribution to the relic density is coming from the stable scalar, however this case will best fit under a scalar DM multiplet study~\cite{dm-consistent-scalar}.

One should pay a special attention to constraints from EW precision data, in particular from the $\rho_0$ parameter~\cite{Ross:1975fq}, which is very close to one in the SM and measured with per-mille precision. In the general case of an arbitrary number of SU(2) scalar multiplets,  $\rho_0$  takes the form (see Eq.(10.58) in Ref.\cite{Beringer:1900zz}):
	\beq
	\rho_0=
	\frac{\sum_{n} [I_n (I_n+1)-I_{3n}^2] |v_n|^2}{2\sum_n I_{3n}^2 |v_n|^2} \ \ ,
	\label{eq:rho}
	\eeq	
where $I_n$, $I_{3n}$ and $v_n$ are the isospin, the third component  isospin of the vacuum state and the the vacuum expectation value for the $n^{th}$ scalar  multiplet, respectively.
Hence, besides the known cases with doublet and singlet,  strong bounds from $\rho$ can be avoided in several other non-trivial cases: for example, in the model with a septet scalar ($S^3_2$)~\cite{Hisano:2013sn} 
that can couple to a DM  quintuplet,
or the model with custodial combinations like triplets in the Georgi-Machacek model ($S^1_0 + S_1^1$)~\cite{Georgi:1985nv}. The latter would rather be the part of less minimal models, but still possibly quite interesting.
From Eq.~\eqref{eq:rho} one can see that 
the case $\tilde{F}_{1/2}^I S_1^1$ described in point (c) can better fit in a Georgi-Machacek scenario, where the triplet VEV is not too constrained. However, for the scenario with  a custodial violating triplet only, the coupling may be enough to generate a large enough mass split to avoid constraints from $\rho_0$.

To summarise this section, we found a new class of relevant minimal models with a scalar mediator that is (accidentally) stable: this includes a model with two singlets, $\tilde{F}_0^0 S_0^0$ with D2 or D4, which we study in more details in Section~\ref{sec:new-model-pheno}.

\subsection{Odd scalar mediator ($\tilde{F}^I_Y \tilde{S}^{I'}_{Y'}$ and $\tilde{M}^I_0 \tilde{S}^{I'}_{Y'}$)} \label{sec:oddscalar}

In this class of models, the DM fermion multiplet $\Psi$ couples to the odd scalar $\tilde \varphi$ and  to a SM fermion
via a Yukawa coupling: the quantum numbers of the scalar multiplet are, therefore, fixed by the properties of the chosen SM fermion.
As the SM fermions are chiral, one can classify two cases, distinguished by their chirality (a SU(2)$_L$  doublet $f_L$ or a singlet $f_R$): 
\begin{itemize}
\item[-] for left-handed SM fermions, the respective interactions read:
\beq
\Delta \mathcal{L} = - h_{f_L}^i \; \tilde \varphi_{f_L} \bar{\Psi}_R f^i_L+ \mbox{h.c.}
\eeq
hence, $\tilde\varphi_{f_L} = \{ I \pm 1/2, Y - Y_f\}$ (and an anti-triplet of QCD colour if $f_L$ is a quark);
\item[-] for right-handed SM fermions:
\beq
\Delta \mathcal{L} = - h_{f_R}^i\; \tilde\varphi_{f_R} \bar{\Psi}_L f^i_R + \mbox{h.c.}
\eeq
hence, $\tilde\varphi_f = \{ I, Y - Y_f\}$ (and an anti-triplet of QCD colour if $f_R$ is a quark).
\end{itemize}
Note that $i = 1,2,3$ is a SM family index, and the two types of couplings cannot co-exist with the same multiplet in minimal models. In other words, the couplings of the mediator necessarily involve one chirality and only one type of SM fermions.
The scalar multiplet will also have couplings to the Higgs~\cite{Hambye:2009pw}, in a form analogous to that of Eqs~\eqref{eq:LagSreal} or~\eqref{eq:LagScomplex}, but in the absence of any linear coupling forbidden by the DM parity.
As the cases of quarks and leptons lead to rather different physics, we will discuss them in detail separately.

\subsubsection{Quark-type mediators}

Firstly, as quark partners $\tilde{\varphi}_{q_{L/R}}$ carry QCD charges, they cannot constitute part of the DM relic density and  are always required to be heavier than the DM fermion candidate.
Besides the effect of EW interactions discussed in the previous section, the scalar mediator will contribute a new tree-level process to direct detection, $D q \to \tilde{\varphi} \to D q$,
whose rate is determined  by the value of the $h_{qL/R}$ coupling and the mass of $\tilde{\varphi}$
mediator for any given DM mass.
The most minimal, and safe, cases involve $\tilde{F}_0^0$ and $\tilde{M}_0^0$, for which the scalar mediator has the same quantum numbers as the corresponding SM fermion. This case is a template of supersymmetry ($\tilde{\varphi}_{q_{L/R}}$ being one of the squarks), and has been studied in detail in simplified models with $\tilde\varphi_{q_{L}}$ mediator and Majorana DM~\cite{Garny:2014waa,Ko:2016zxg,Arina:2020tuw}.

\subsubsection{Lepton-type mediators}

In this case, the scalar multiplet may contain a neutral state and therefore {also} play the role of DM (this case will be covered in a future work~\cite{dm-consistent-scalar}). In the case where the DM arises form the fermionic multiplet,
direct detection (for the  only surviving ``safe" cases of $\tilde{F}_0^0$ and $\tilde{M}_0^0$) occurs only at one-loop level contrary to the  case of the coloured scalar mediators discussed above. DM direct detection rates, in this case, are defined by the respective $h_{lL/R}$ Yukawa coupling and the mass of the scalar multiplet, occurring in the loop. This case also corresponds to the supersymmetry template with sleptons, and has been covered in Refs~\cite{Fukushima:2014yia,Baker:2018uox}.

\subsection{Even fermion mediator ($\tilde{F}^I_Y F^{I'}_{Y'}$)}

This case does not allow for renormalisable couplings between the mediator and the DM multiplet, however we list it here for completeness and because it leads to interesting new models of leptophilic DM. The only allowed coupling involves one mediator multiplet, $\Sigma$, and three DM multiplets $\Psi$.
In turn, the even multiplet $\Sigma$ needs to couple to the SM via a Yukawa-type coupling to leptons (quarks are excluded to avoid QCD charged DM).

The DM mediator coupling comes from a dim--6 operator:
\beq
\mathcal{L}_{\text{dim--6}} \supset \frac{1}{\Lambda^2}  (\bar{\Psi}^C \Psi) (\bar{\Psi}^C \Sigma) + \mbox{h.c.}
\eeq
which preserves a $\mathbb{Z}_3$ DM parity~\cite{DEramo:2010keq,Cai:2015zza} for a complex Dirac multiplet $\tilde{F}_Y^I$.~\footnote{For Majorana DM multiplets, the $\mathbb{Z}_3$ would be broken by the mass term. Furthermore, a coupling in the form $(\bar{\Psi} \Psi) (\bar{\Psi} \Sigma)$ does not preserve any DM parity nor U(1) charge.} Moreover, the hypercharges are related by:
\beq
Y' = - 3 Y\,.
\eeq
The last relation imposes a significant constraint on the mediator multiplet, as the hypercharge of the DM one needs to be semi-integer for semi-integer isospin and integer for integer isospin in order to have a neutral component.

As a consequence, the only allowed cases (with Yukawa couplings to leptons) are:
\beq
\mbox{Class A:} \phantom{xxx}& \Delta L = - \xi_L\ \bar{l}_L \phi_H^\dagger \Sigma + \mbox{h.c.} \,;& \phantom{xxx} \tilde{F}_0^{I=\mbox{int.}} F_0^{0,1}\,; \\
\mbox{Class B:} \phantom{xxx} & \Delta L = - \xi_R\ \bar{l}_R \phi_H \Sigma + \mbox{h.c.}\,; & \phantom{xxx} \tilde{F}_{1/2}^{I=\mbox{semi--int.}} F_{-3/2}^{1/2}\,.
\eeq
Due to direct detection constraints, and the role played by gauge interactions in the thermal relic abundance (which would make the mediator irrelevant), the only interesting case appears for a singlet DM, $\tilde{F}_0^0 F_0^0$, which belongs to class A. Note that $\Sigma$ is effectively a heavy right-handed neutrino. The relic density will thus be determined by the processes:
\beq
\Psi \Psi \leftrightarrow \bar{\Psi} \nu\,, \quad \Psi \Psi \to \bar{\Psi} \nu H\,.
\eeq
If the coupling to the SM is very small, being related to neutrino mass generation, then this could be an effective FIMP model.

\subsection{Odd fermion mediator ($\tilde{F}^I_Y \tilde{F}^{I'}_{Y'}$, $\tilde{M}^I_0 \tilde{F}^{I'}_{1/2}$ and $\tilde{F}^{I}_{1/2} \tilde{M}^{I'}_0$)}

In the case of the odd fermionic mediators, the only renormalisable coupling is a Yukawa with the Higgs boson. In general, therefore, the DM state will be the lightest mass eigenstate from the neutral components of the two multiplets. Notable examples of this class of models come from supersymmetry, where the lightest neutralino can be a mixture of bino-Higgsino ($\tilde{M}_0^0 \tilde{F}^{1/2}_{1/2}$ or $\tilde{F}^{1/2}_{1/2} \tilde{M}_0^0$) or wino-Higgsino ($\tilde{M}^1_0 \tilde{F}^{1/2}_{1/2}$ or $\tilde{F}^{1/2}_{1/2} \tilde{M}^1_0$). Note that in our notation the first multiplet is the one that has the largest component in the DM physical state.

The possible models can be classified based on the form of the Yukawa coupling:
\beq \label{eq:FtildeFtilde}
\Delta \mathcal{L} = - \lambda\, \bar{\Psi}' \phi \Psi + \mbox{h.c.}\,, \quad \mbox{with}\;\; I'=I\pm 1/2\;\; \mbox{and}\;\; \left\{
\begin{array}{l}
Y' = Y + 1/2\;\; \mbox{if}\;\; \phi = \phi_H\,, \\
Y' = Y - 1/2\;\; \mbox{if}\;\; \phi = \tilde{\phi}_H \equiv (i \sigma^2) \phi^\ast_H\,, 
\end{array} \right.
\eeq
where in our convention $\Psi'$ indicates the mediator multiplet. Note that the Higgs field may appear as is, or in the form of the complex conjugate $\tilde{\phi}_H$. Also, either the mediator or the DM multiplet can be of Majorana nature if either $Y'=0$ or $Y=0$.
In general, this class of mediator models have similar features as the simple DM multiplet cases, with an additional coupling to the Higgs boson that could make direct detection more critical.

One point of interest, though, is the fact that in the case of large mediator mass, i.e. $M' \gg m$, by integrating out the mediator multiplet one can generate the dim--5 couplings to the Higgs discussed in Sec.~\ref{sec:DMmult_Higgs}. In the case of Dirac multiplets, the coefficient of Eq.~\eqref{eq:dim5} are matched to the Yukawa coupling and mediator mass $M'$ as
\beq
\frac{\kappa}{\Lambda} = \pm \epsilon \frac{\lambda^2}{M'} \frac{2}{2I+1}\,, \quad \frac{\kappa'}{\Lambda} =  \frac{\lambda^2}{M'}  \frac{1}{2} \left( 1\pm \frac{1}{2I+1} \right)\,, \quad \mbox{for}\;\; I' = I \pm \frac{1}{2}\,;
\eeq
where $\epsilon = -1$ if $\tilde{\phi}_H$ appears in the Yukawa in Eq.~\eqref{eq:FtildeFtilde} (and $\epsilon = 1$ otherwise).
If the mediator is a Majorana multiplet, then only the coupling in Eq.~\eqref{eq:dim5Maj} is generated, with
\beq
\frac{\kappa_M}{\Lambda} = \pm \epsilon \frac{\lambda^2}{M'} \frac{2}{2I+1}\,, \quad \mbox{for}\;\; I' = I \pm \frac{1}{2}\,.
\eeq

\subsection{Even vector mediators ($\tilde{F}^I_Y V^{I'}_0$ and $\tilde{M}^I_0 V^{I'}_0$)}

Vector mediators are very popular in the simplified model approach to DM phenomenology (see e.g.~\cite{Petriello:2008pu,Khalil:2008ps,Mizukoshi:2010ky,An:2012va,An:2012ue,Frandsen:2012rk,Barger:2012ey,Basso:2012gz,Arcadi:2013qia,Alves:2013tqa,Alves:2015pea,Okada:2016gsh,Fairbairn:2016iuf,Arcadi:2017hfi,Okada:2016tci,Belyaev:2017vsx,Okada:2018ktp,Han:2018zcn,Cosme:2021baj}) 
mainly because they allow for ``gauge invariant'' couplings to vector currents of SM fermions. Nevertheless, it is not a simple task to find a consistent, truly gauge invariant, renormalisable model containing vector mediator multiplets. As the vector multiplet couples to a current containing the DM multiplet,
the  Lagrangian takes the form
\beq
\Delta \mathcal{L}_V = V_\mu\, \bar{\Psi} \gamma^\mu (g_{VL} P_L + g_{VR} P_R) \Psi\,, \qquad \mbox{with}\;\; I'=0, \dots 2I\,, \;\; Y'=0\,; 
\eeq
where $P_{L/R}$ are chirality projectors. As the hypercharge always vanishes (and the isospin is integer), we can always consider real multiplets.

For a generic vector field $V_\mu$, the most general Lagrangian up to renormalisable couplings reads~\cite{Belyaev:2005ew}:
\beq
\mathcal{L}_{V^{I'}_{0}} &=& \frac{1}{2} (D_\mu V_\nu - D_\nu V_\mu)^2 - \frac{1}{2} M_V^2 V^\mu V_\mu+ \xi_2\, W^a_{\mu \nu} (V_\mu T^a_{I'} V_\nu) +  \mbox{self int.} \nonumber \\
& & + \sum_{f \in {\rm SM}} \ V_\mu\, \bar{f} \gamma^\mu (g^f_{VL} P_L + g^f_{VR} P_R) f + g_{VH} V_\mu\ \left( \phi_H^\dag (D^\mu \phi_H) - (D^\mu \phi_H^\dag) \phi_H\right) \nonumber \\
&& + \lambda_0\, (V_\mu V^\mu) \phi_H^\dagger \phi_H + \lambda_1\, (V_\mu T_{I'}^a V^\mu)  \phi_H^\dagger \tau^a \phi_H\,, \label{eq:lagrV}
\eeq
where $W_{\mu \nu}^a$ is the energy-stress tensor of  SU(2)$_L$. The second line contains couplings to currents of SM fermions and the Higgs field, compatible with the quantum numbers of the vector multiplet: they are allowed only for the singlet $V_0^0$ and a triplet $V_0^1$. 

The Lagrangian in Eq.~\eqref{eq:lagrV},
which we require to be renormalisable  and consistent,  needs an additional scalar sector which breaks the gauge symmetry,  for which these vector mediators are being  gauge bosons (see e.g. \cite{Bell:2016fqf}). These gauge bosons can come from different theory space,  including supersymmetric, extra-dimensional or composite/technicolor origin.

 In Ref.~\cite{Zerwekh:2012bf} it has been shown that the self-interactions of the multiplet can be fixed in order to preserve perturbative unitarity in the scattering amplitude of vector multiplets, however Ref.~\cite{Belyaev:2018xpf} later showed that violation of perturbative unitarity occurs once the vector multiplet couples to massive gauge bosons (i.e. it is charged under a broken gauge group, like SU(2)$_L$) and/or to the Higgs: thus new states need to be included in order to restore the consistency of the model. They might affect the low energy properties of the theory by introducing phenomenologically relevant operators. In theories of this kind, the vector mediator may arise as a composite spin-1 meson of a confining strong dynamics, like in models of composite Goldstone Higgs.

One way to avoid these issues is to introduce the vector multiplet as a gauge field: in general, though, a vector carrying isospin needs to come from a model where the gauge symmetry SU(2)$_L$ is extended and broken at higher scales. Now, generating the couplings to the SM fermions becomes the challenge, as new fermions are likely to be needed in order to complete multiplets of the extended EW gauge symmetry. Note that here the chiral nature of the SM fermions is the main obstacle, as it may imply the presence of other chiral fermions.

One case that does not suffer from such problem is the singlet, $V_0^0$, as it could arise from a broken gauged U(1) symmetry under which some SM fermions are charged. Once again, though, a consistent theory would require an anomaly-free U(1), thus either additional charged heavy states are added, or one has very limited choices, as discussed in various DM $Z'$-portal  studies
cited above, see e.g.~\cite{Okada:2016gsh} and references therein.

\subsection{Odd vector mediators ($\tilde{F}^I_Y \tilde{V}^{I'}_{Y'}$)}

In the case of odd vector mediators, the only allowed couplings must involve the DM multiplet and a SM fermion. The classification of mediators, therefore, follows the same as the scalar odd mediators in Sec.~\ref{sec:oddscalar}:
\begin{itemize}
\item[-] for left-handed SM fermions, the coupling reads:
\beq
\Delta \mathcal{L} =  g_{VfL}^i \; V_{fL}^\mu \bar{\Psi}_L \gamma_\mu f^i_L+ \mbox{h.c.}
\eeq
thus $V_{fL} = \{ I \pm 1/2, Y - Y_f\}$ (and an anti-triplet of QCD colour if $f$ is a quark);
\item[-] for right-handed SM fermions:
\beq
\Delta \mathcal{L} = g_{fR}^i\; V_{fR}^\mu \bar{\Psi}_R \gamma_\mu f^i_R + \mbox{h.c.}
\eeq
thus $V_{fR} = \{ I, Y - Y_f\}$ (and an anti-triplet of QCD colour if $f$ is a quark).
\end{itemize}
As the mediator typically has non-zero hypercharge, the Lagrangian~\eqref{eq:lagrV} needs to be extended:
\beq
\mathcal{L}_{\tilde{V}^{I'}_{Y'}} &=& |D_\mu V_\nu - D_\nu V_\mu|^2 - M_V^2 V^\dag_\mu V^\mu + \xi_1\, B_{\mu \nu} (V^\dag_\mu V_\nu)  + \xi_2\, W^a_{\mu \nu} (V^\dag_\mu T^a_{I'} V_\nu)+ \xi_3\, G^c_{\mu \nu} (V^\dag_\mu \lambda^c V_\nu) \nonumber \\
&&  +  \mbox{self int.}  + \lambda_0\, (V^\dag_\mu V^\mu) \phi_H^\dagger \phi_H + \lambda_1\, (V^\dag_\mu T_{I'}^a V^\mu)  \phi_H^\dagger \tau^a \phi_H\,. \label{eq:lagrV2}
\eeq
Similarly to the case of even mediators, the above Lagrangian cannot be complete because of perturbative unitarity violation or the need to extend the gauge symmetries of the SM to generate $\tilde{V}$ as a gauge boson.

In such a scenario  $\tilde{V}$ can play a role of  a DM candidate if it is lighter than the  fermionic DM candidate. An example of complete model for vector DM involved in the weak interactions has been suggested in~\cite{Abe:2020mph}, where the authors introduce  two additional SU(2) triplets -- one odd and another even -- to make the model consistent.

%% file: 05_pheno.tex

\section{Phenomenology of a new representative model:  $\tilde F^0_0S^0_0$(CP-odd)}
\label{sec:new-model-pheno}

In this section we study the model $\tilde F^0_0S^0_0$(CP-odd)  with a
Dirac  fermion singlet ($\Psi \equiv \psi$) and a pseudo-scalar (CP-odd) singlet 
($\Phi\equiv a$) -- 
probably the simplest two component DM model  introduced in section 
\ref{subsec:even-scalar-med}. 
We have reported a preliminary study on this model in~\cite{dan-pascos2019,belyaev-moriond2021}. During completion of this work, an alternative, partly overlapping, analysis of the same model (without the study of the loop effects) appeared in~\cite{DiazSaez:2021pmg}.

The Lagrangian of the dark sector, to be added to the SM one, reads:
\begin{equation}
\Delta \mathcal{L}_{\tilde F^0_0S^0_0} = i \bar{\psi} \partial_\mu \gamma^\mu  \psi - m_\psi \bar{\psi}\psi + \frac{1}{2} (\partial_\mu a)^2 -  \frac{m_\Phi^2}{2} a^2 + i Y_\psi a \bar{\psi} \gamma^5 \psi  - \frac{\lambda_{aH}}{4} a^2 \phi_H^\dagger \phi_H - \frac{\lambda_a}{4}a^4\,,
\end{equation}
where $\phi_H$ is the SM Higgs doublet field.
A similar model has been investigated in Ref.~\cite{Baek_2017} where, however, a linear coupling of the pseudo-scalar with the Higgs was also allowed (hence breaking CP),  which leads to a very different phenomenology,  as this coupling implies that the pseudo-scalar develops a vacuum expectation value.

The model contains three new couplings: the Yukawa coupling $Y_\psi$ connecting the 
scalar mediator $a$ to the fermion DM $\psi$, the  self-interaction $\lambda_a$ of the pseudo-scalar $a$
and the quartic coupling to the Higgs ${\lambda_{aH}}$. The latter is the only coupling
connecting the new sector to the SM via a Higgs portal.  We recall that a linear coupling of $a$ to the Higgs
field is forbidden by different CP properties of the Higgs and $a$.
The invariance under CP is preserved as long as $a$ does not develop a vacuum expectation value, i.e. if
\begin{equation}
m_a^2 = m_\Phi^2 + \frac{\lambda_{aH} v^2}{8} \geq 0\,, \quad \lambda_a > 0\,,
\label{eq:cp}
\end{equation} 
where $m_a$ is the physical mass of the scalar particle,
which together with $m_\psi$ and three couplings comprises the set of five  parameters defining the model:
\begin{equation}
	m_a, \ \ \ m_\psi, \ \ \ Y_\psi, \ \ \ \lambda_{aH} \ \ \ \mbox{and}\ \ \ \lambda_a.
\end{equation}
The first four parameters only are relevant to the phenomenology we discuss here.
 We will be working in  the region of the parameter space
 defined by Eq.~\eqref{eq:cp}, where the phenomenology is very different from the model in Ref.~\cite{Baek_2017} as we have mentioned earlier. As $\psi$ couples exclusively and bi-linearly to $a$, it is a stable fermionic DM candidate protected by a dark U(1) global symmetry. The pseudo-scalar mediator $a$ can only decay into a pair of DM fermions. Hence, if $m_a < 2 m_\psi$, $a$ is said to be ``accidentally" stable and can contribute to the relic density as a second DM component. 
In this case the stability of $a$ is protected from its decays
to SM particles at all loops because the CP symmetry is conserved
in the dark sector.
Indeed, $a$ only couples bilinearly to the SM via the Higgs portal and only CP violation can allow for a linear coupling of $a$ to a SM operator. In this sense, it is the CP symmetry itself that prevents $a$ from decaying into SM states.

The interesting dynamics of this model, where $a$ 
is in touch with the SM via the Higgs portal coupling $\lambda_{aH}$ while $\psi$ only interacts with $a$,
leads to four distinct regimes of relevance for DM phenomenology,  as summarised in Table~\ref{tab:PhenoRegimes}:

\begin{itemize}
\item  In scenario A, both fermion and pseudo-scalar can thermalise with the SM states. If $m_a\leq 2 m_\psi$, then $a$ is stable and contributes to the relic abundance. Conversely, if $m_a>2m_\psi$, then it is unstable and merely acts as a mediator for the interactions of the fermionic DM to the SM. 
\item In scenario B, the relic abundance of $\psi$ is determined by the freeze-in mechanism,  driven by the very small value of $Y_\psi$, while $a$ contributes as a thermal DM component for $m_a<2m_\psi$. In the parameter space, where $m_a>2m_\psi$, the smallness of $Y_\psi$ can lead to $a$ being long-lived, decaying into a pair of $\psi$. 
\item In scenario C, both new particles can freeze-in via their small couplings to the SM sector (including the loop-induced coupling of $\psi$, as we discuss below), before thermalisation between the two species.
\item In scenario D, both particles have very small couplings. While $a$ can freeze-in via its coupling to the Higgs portal, the coupling of the fermion is too small and would lead to a negligible contribution to the total amount of relic density.
Depending on its mass, $a$ can be the only significant DM candidate, or decay promptly to the fermion $\psi$ after being produced in the early universe. 
\end{itemize}

\begin{table}[htbp]
	\begin{tabular}{l|l|l|l}
		Scenario & $Y_\psi$                 & $\lambda_{aH}$               & DM thermal properties                                             \\ \hline
		A & $\mathcal{O}(10^{-3}-1)$ & $\mathcal{O}(10^{-3}-1)$ & $\psi$ and $a$ thermal with SM                            \\
		B & $<\mathcal{O}(10^{-8})$  & $\mathcal{O}(10^{-3}-1)$ & $\psi$ non-thermal, $a$ thermal with SM                   \\
		C & $\mathcal{O}(10^{-3}-1)$ & $<\mathcal{O}(10^{-8})$  & $\psi$ and $a$ thermal with each other, non-thermal to SM \\
		D & $<\mathcal{O}(10^{-8})$  & $<\mathcal{O}(10^{-8})$  & $\psi$ and $a$ non-thermal with each other and SM        
	\end{tabular}
	\caption{\label{tab:PhenoRegimes}
		Table of distinct phenomenological DM scenarios possible in this model.}
\end{table}
Note that any other range of the couplings is excluded by DM over-production or out of control because of perturbativity loss.
Furthermore, in scenarios C and D, direct and indirect detection experiments, as well as colliders, would be unable to observe either of these new particles due to the feeble couplings to the SM. In contrast, in scenarios A and B, $a$ may be observable due to the sizeable Higgs portal coupling. In scenario A, the fermion may also be directly observables due to a loop-induced coupling to the Higgs, as we will discuss below. 

Implementation of this model along with the LANHEP \cite{Semenov:2008jy} source and libraries required for one-loop calculations have been made publicly available at HEPMDB \cite{hepmdb:FDM+a}.
\vspace{0.3cm}

\begin{figure}[htb]\label{fig:SigmaLoopPsi.pdf}
\centering
\includegraphics[width=0.8\textwidth]{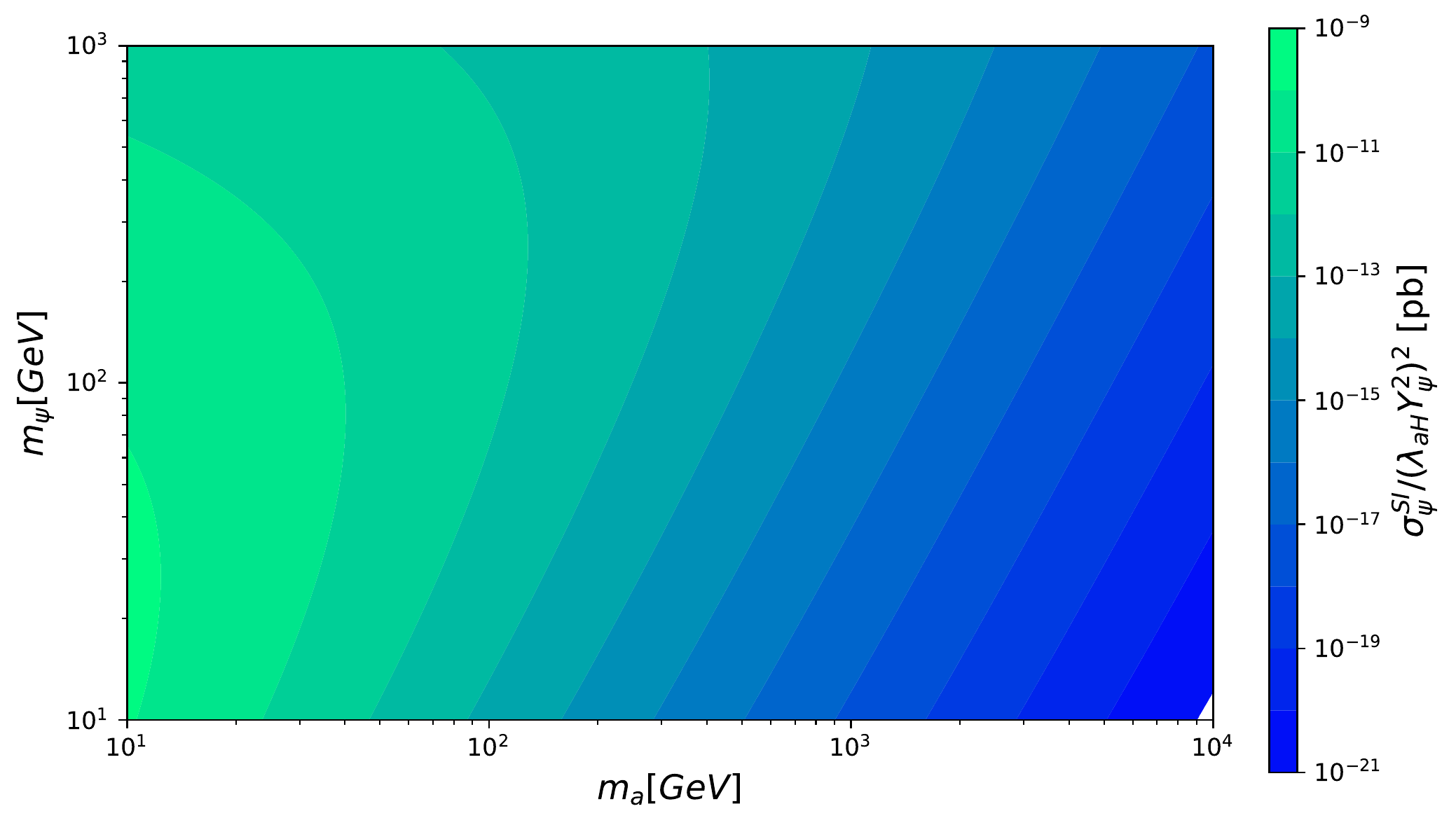}
\caption{Loop-induced direct detection cross section for $\psi$ scattering on nucleons $\sigma^{SI}_\psi$, scaled by the tree-level couplings $(\lambda_{aH} Y_\psi^2)^2$, as a function of the masses in GeV. \label{fig:SigmaLoopsPsi} }
\end{figure}

Let us start the discussion of the model's phenomenology by presenting some generic features of the new states, $\psi$ and $a$.
If $a$ is stable, its DM fraction can be revealed via direct detection thanks to the following SI elastic cross section on nuclei:
\begin{equation}
\sigma^{SI}_a (a N \to a N) = \frac{\lambda_{aH}^2 v^2 \lambda_N^2}{8 \pi m_H^4} \frac{m_N^2}{(m_a + m_N)^2}\,.
\end{equation}
where the nucleon effective coupling $\lambda_N$ ($N$ labels the nucleon type)
can be written in terms of the nucleon form factors presented in section \ref{sec:DDminimal}
as:
\begin{align}\label{eqn:lambdaN}
\lambda_N &= \frac{m_N}{2 v} \Bigg[ \sum_{q \in \{u,d,s\}} \frac{f_{Tq}^{(N)}m_q(\mu)}{m_q(\mu_{LHC})} + \frac{2}{27}f_{TG}^{(N)} \sum_{q \in \{c,b,t\}} \frac{m_q(\mu)}{m_q(\mu_{LHC})} \Bigg]\,.
\end{align}
The fermion DM, $\psi$, which is always stable, couples to the SM only via the mediator $a$. The  coupling of $\psi$ to the Higgs boson is, however, generated at one loop level.  The complete expression for this coupling is given in Appendix \ref{App:FDM+a_DDloop}, where $\delta Y_{DD}$ refers to $\delta Y$ (Eq.~\ref{eqn:deltaY}) evaluated at the direct detection scale, $t=0$. In the limit of small $m_a$, the effective $H\psi\psi$ Yukawa coupling, is given by
\begin{equation} \label{eq:1-loopPsiH}
\mathcal{L}_{1-loop} \supset \delta Y_{DD}\, H\ \bar{\psi} \psi\,, \left. \quad \delta Y_{DD}  \right|_{m_a \to 0} \approx - \frac{Y_\psi^2 \lambda_{aH} v}{32 \pi^2 m_\psi} \left( \ln \frac{m_\psi}{m_a}-1\right)\,.
\end{equation}
For larger $m_a$, the loop-induced coupling decreases monotonically, with $\delta Y \propto m_a^{-2}$ asymptotic for large $a$ masses.
This coupling is only relevant when both $Y_\psi$ and $\lambda_{aH}$ are sizeable, and it contributes to direct detection via the following SI cross section of $\psi$ on nucleons:
\begin{equation}
\sigma^{SI}_\psi (\psi N \to \psi N) = \frac{4 \delta Y_{DD}^2 \lambda_N^2}{\pi m_H^4} \left(\frac{m_\psi m_N}{m_\psi + m_N}\right)^2\,.
\end{equation}
As an illustration, we show in Fig.~\ref{fig:SigmaLoopsPsi} the SI cross section as a function of the masses, rescaled by the tree-level couplings. 
Taking into account that the current  direct detection limit is in the  $10^{-10}-10^{-9}$~pb  range,
 we can infer that this process provides relevant limits only for relatively small $\psi$ masses and couplings of order unity.

Another important constraint arises in the region, where the pseudo-scalar and/or the fermion are lighter than half the Higgs mass, i.e. $m_a,\ m_\psi < m_H/2$, thanks to the LHC limits on Higgs invisible decays. For the pseudo-scalar, the partial decay width is generated at tree-level: 
\begin{equation}
\Gamma_{H\to aa} = \frac{\lambda_{aH}^2 v^2}{128 \pi m_H} \sqrt{1-\frac{4 m_a^2}{m_H^2}}\,.
\end{equation}
For the fermion, the decay is induced via the one-loop induced coupling in Eq.~\eqref{eq:1-loopPsiH}.
Hence, the loop-induced $H \to \psi\psi$ partial decay width is given by
\begin{equation}
\Gamma_{H \to \psi\psi} = \frac{\delta Y_{H \to \psi \psi}^2 m_H}{8 \pi} \left(1-\frac{4 m_\psi^2}{m_H^2}\right)^\frac{3}{2}\,,
\end{equation}
where the effective coupling $\delta Y_{H \to \psi \psi}$ depends on a loop function $\Upsilon_{H\to\psi\psi}\equiv \Upsilon(s=m_H^2)$ (see Appendix \ref{App:FDM+a_DDloop}) 
\begin{equation}
\delta Y_{H\to\psi\psi} = -\frac{Y_\psi^2 \lambda_{aH} v}{32\pi^2} \Upsilon_{H\to\psi\psi}\, .
\end{equation}
We recall that $a$ always leads to missing energy, even when it decays promptly.
One should also note  that  the loop-induced coupling $\delta Y_{Relic} \equiv \delta Y(s \approx 4m_\psi^2(1+1/(2x)))$ (where x is mass to temperature ratio) coupling also plays a role for the relic density computation (see Appendix \ref{App:FDM+a_DDloop}), and is fully taken into account in our numerical results.

\subsection{Scenario A: 2-component thermal Dark Matter regime}\label{sec:scenario_A}

In this scenario (see Table~\ref{tab:PhenoRegimes}), the 
$\lambda_{aH}$ and $Y_\psi$ couplings are large enough
to thermalise both DM components in the early universe.

\begin{figure}[htb]
	\includegraphics[width=\textwidth]{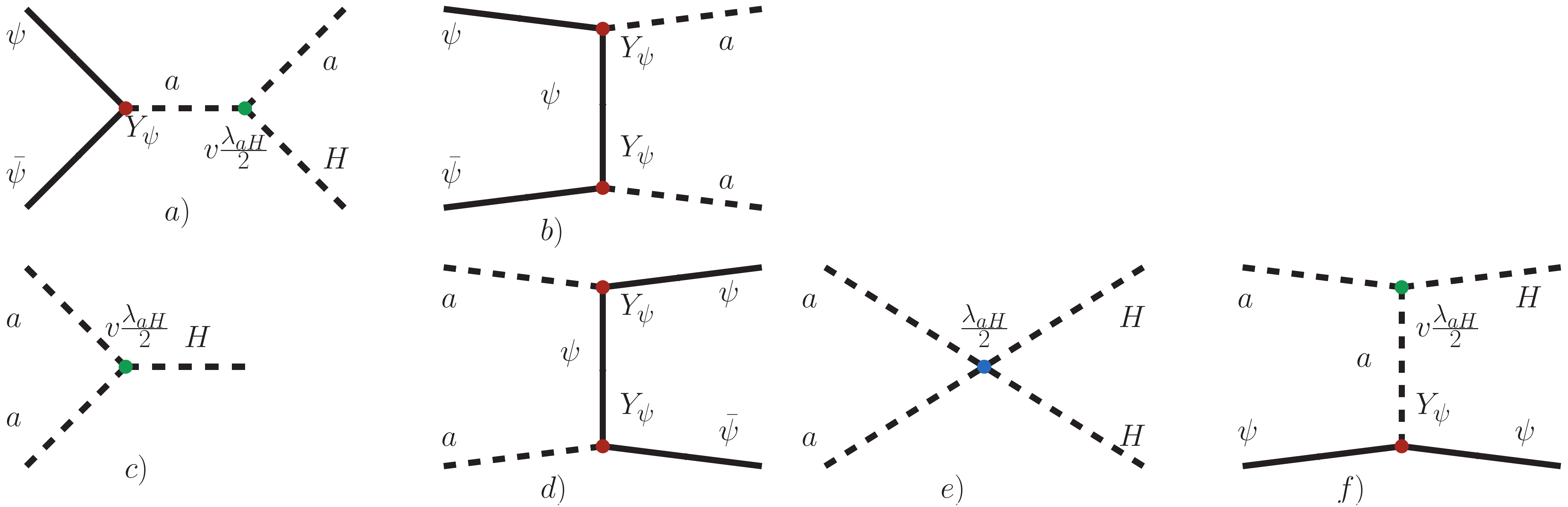}
	\caption{Tree-level Feynman diagrams for DM (co)annihilation:
		a)-b)  for $\bar{\psi}-\psi$ annihilation; c)-e)  for $aa$ annihilation and 
		f) for $\psi-a$ co-annihilation.  \label{fig:FD}}
\end{figure}

The relic density in this regime can be evaluated using two coupled Boltzmann equations (see Eq.(5) of Ref.~\cite{Belanger:2014vza}), which are defined by  the 
annihilation and co-annihilation processes, Feynman diagrams of which are shown in Fig.~\ref{fig:FD}. The equations for the two relic densities $n_a$ and $n_\psi$ read:
\begin{eqnarray}
\frac{dn_\psi}{dt} &=&  
-\sigma_v^{\psi\psi  \to a H} (n_\psi^2-n_a   \frac{\bar{n}_\psi^2}{\bar{n}_a}) 
-\sigma_v^{\psi\psi  \to a a} (n_\psi^2-n_a^2 
\frac{\bar{n}_\psi^2}{\bar{n}_a^2}) 
 -3Hn_\psi\,,
 \\
	\frac{dn_a}{dt} &=& 
- (\sigma_v^{a a \to H}+\sigma_v^{a a \to H H})(n_a^2 - \bar{n}_a^2 )  
- \sigma_v^{a a \to \psi \psi} (n_a^2-n_\psi^2 \frac{\bar{n}_a^2}{\bar{n}_\psi^2}) 
	\nonumber\\
    &&
- \frac{1}{2}  \sigma_v^{a \psi \to \psi H} ( n_a n_\psi - n_\psi\bar{n}_a) 
+ \frac{1}{2}  \sigma_v^{\psi \psi \to a H} ( n_\psi^2 - n_a 
\frac{\bar{n}_\psi^2}{\bar{n}_a}) -3H n_a 
 \,,
\end{eqnarray}
where $\bar{n}_a$ and $\bar{n}_\psi$ denote the equilibrium number densities for 
the two components, and $\sigma_v\equiv \braket{\sigma v}$.

We have performed a random scan of the 4-dimensional parameter space of the model
and have used MicrOMEGAs  \cite{Belanger:2001fz,Belanger:2004yn} 
to evaluate the DM  relic density and direct detection
rates
in the following range of the parameter space:
\begin{align}
10~\mbox{GeV} <  m_\psi < 10~\mbox{TeV}\,, &\qquad
10^{-1} < Y_\psi < 10\,, \nonumber \\
10~\mbox{GeV} <  m_a < 1~\mbox{TeV}\,,   & \qquad
10^{-4} <  \lambda_{aH} < 10\,.
\end{align}
The upper limit  on the couplings is defined  by the loss of perturbativity criteria.
We determine the allowed regions surviving  after imposing the following constraints:
\begin{itemize}
	\item
	We use the relic density fit from 
	PLANCK~\cite{Adam:2015rua} 
		$\Omega_{\rm PLANCK}h^{2}=0.1186\pm0.0020$
    and require
	\begin{equation}
	\Omega_h^2 < 0.12\,,
	\end{equation}
     which allows the  under-abundant model points.
    \item
    We impose the DM direct detection 
	constraints from PandaX-4T~\cite{PandaX-4T:2021bab}, which are 
	dominant over the DM indirect detection constraints, as we have explicitly checked.
	\item
	We use the invisible 
	Higgs decay constraints at the 
	LHC from ATLAS~\cite{ATLAS:2020kdi}, requiring
	\begin{equation}
	 Br(H\to\mbox{invis})<0.11\,.
	\end{equation}
\end{itemize}

\begin{figure}[htb]
\includegraphics[width=0.5\textwidth]{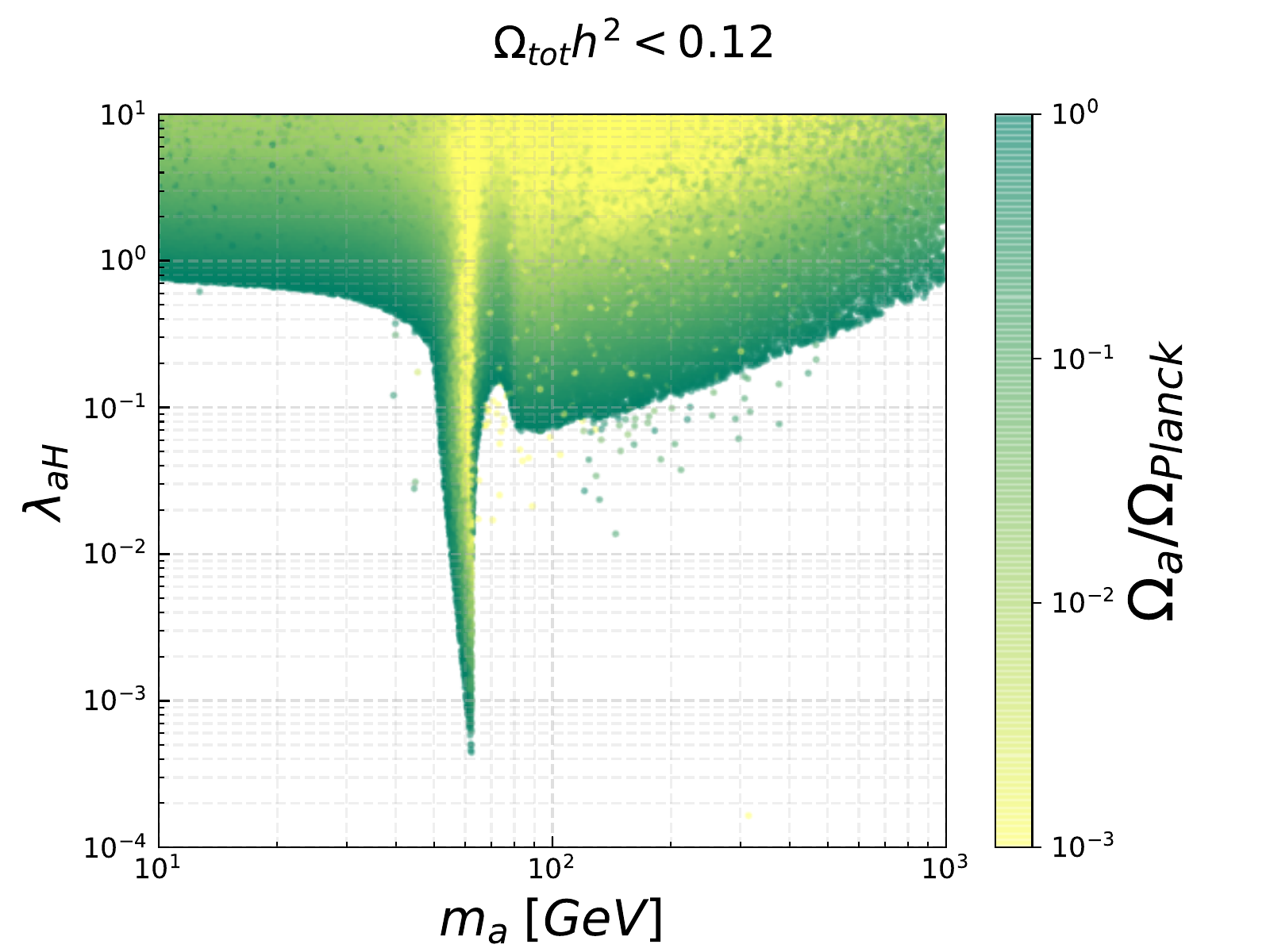}%
\includegraphics[width=0.5\textwidth]{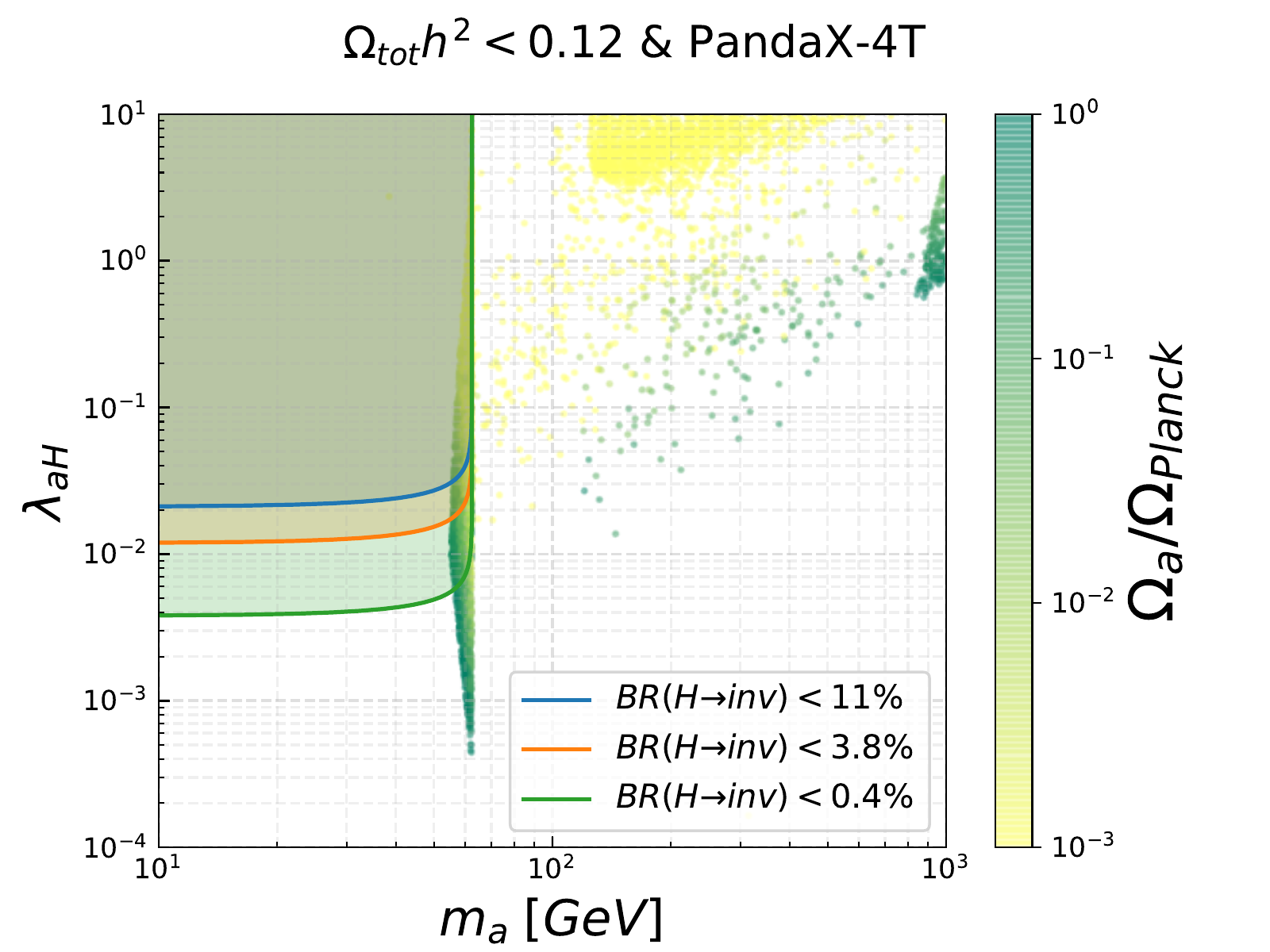}\\
\vskip -6cm(a)\hspace*{0.5\textwidth}\hspace{-0.5cm}(b)\vskip 5cm
\includegraphics[width=0.5\textwidth]{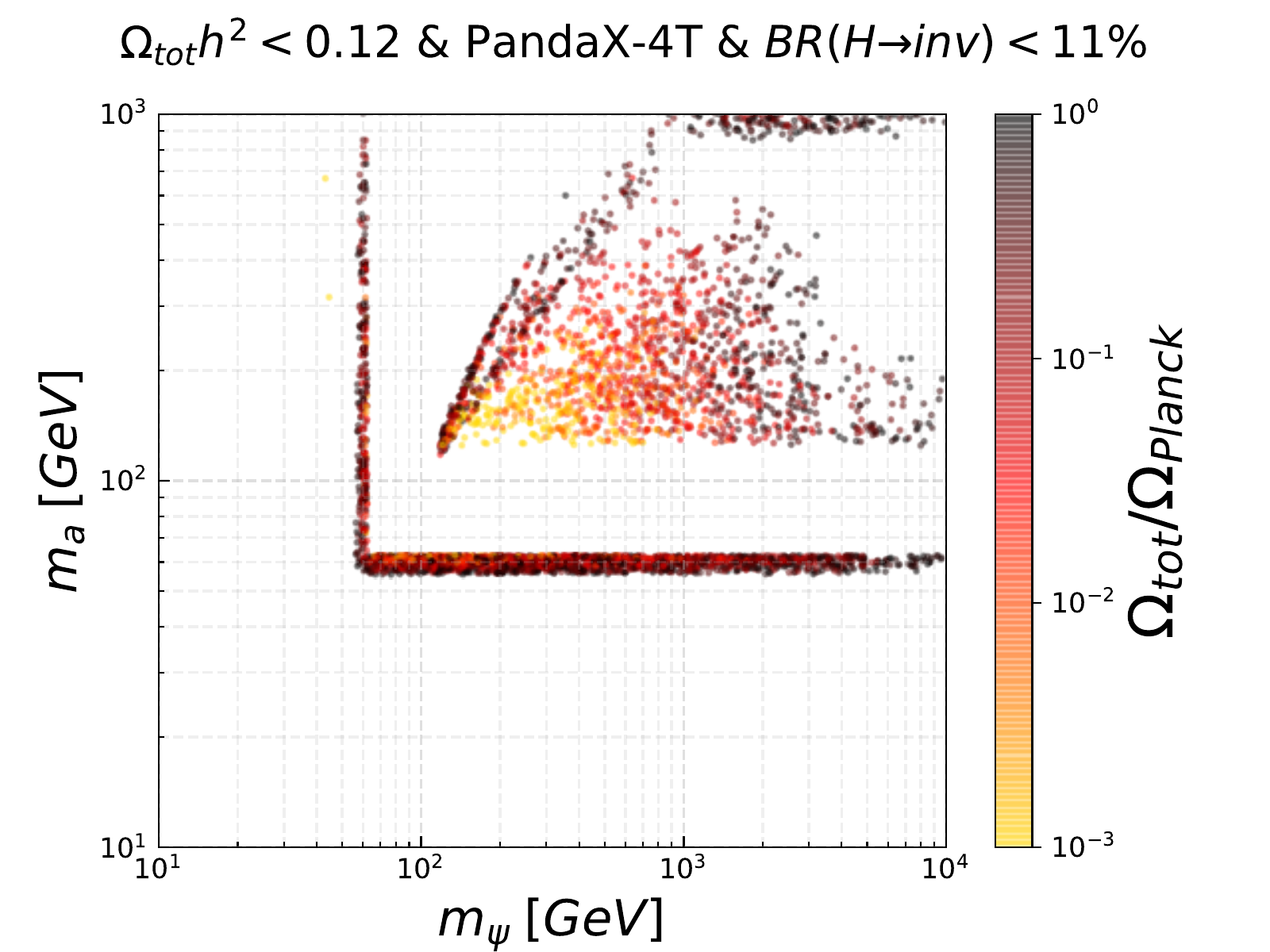}%
\includegraphics[width=0.5\textwidth]{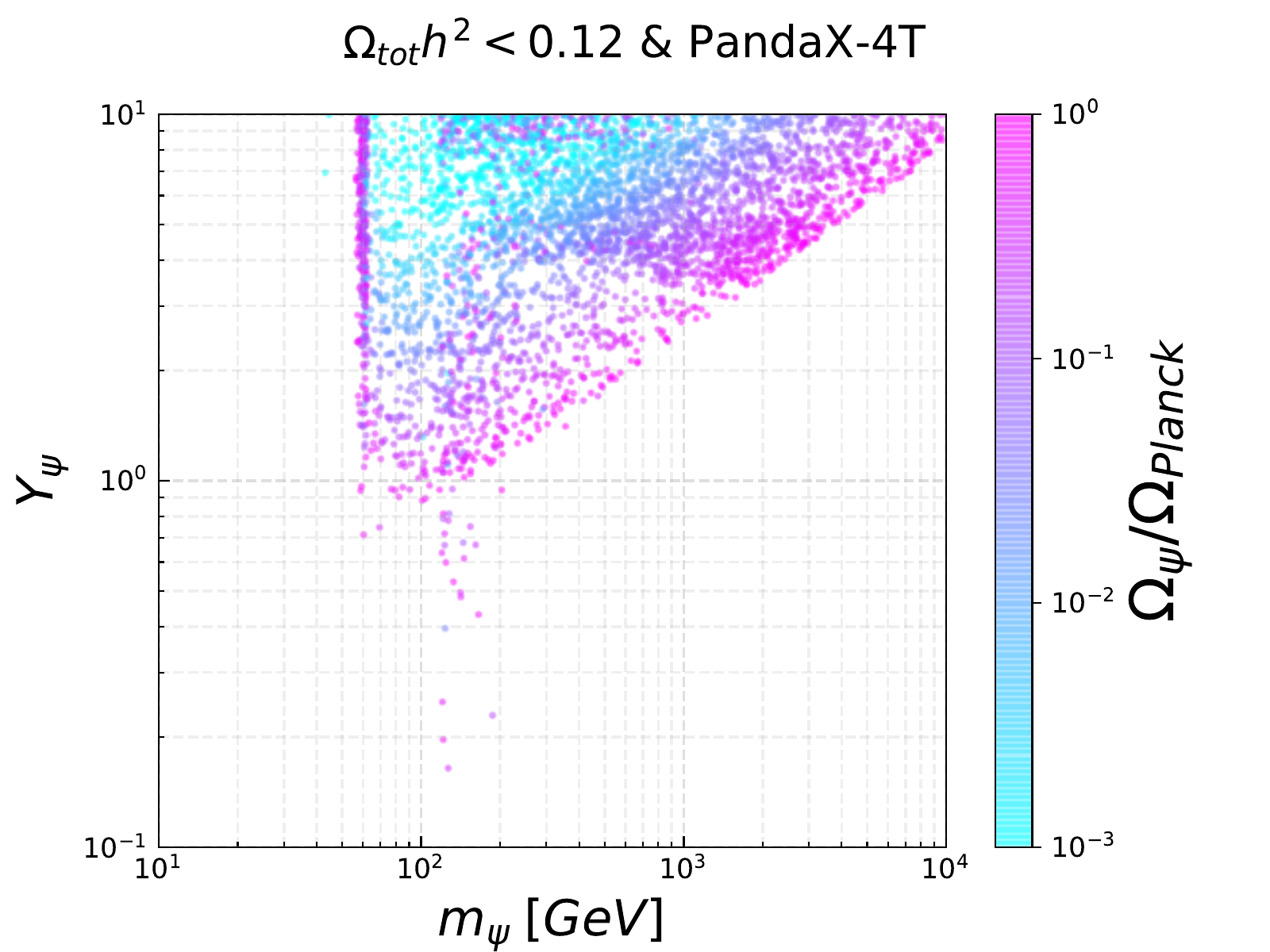}
\\
\vskip -6cm(c)\hspace*{0.5\textwidth}\hspace{-0.5cm}(d)\vskip 5cm

\caption{\label{fig:FDM-PS}
2D projections of the allowed parameter space for $\tilde F^0_0S^0_0$ (CP-odd)  model (after constraints given at the top of each frame)
with the colour map indicating the individual relative relic density of two DM components  $a$ ($\Omega_a/\Omega_{\text{\rm PLANCK}}$), 
$\psi$ ($\Omega_\psi/\Omega_{\text{\rm PLANCK}}$) or their sum ($\Omega_{\text{tot}}/\Omega_{\text{\rm PLANCK}}$). The points with relic density below $10^{-3}\, \Omega_{\rm PLANCK}$ are shown with colour corresponding to the smallest value.
}
\end{figure}

The results of the scan are presented in Fig.~\ref{fig:FDM-PS}, where we show 2D projections of the 
allowed parameter space for the $\tilde F^0_0S^0_0$(CP-odd)  model after imposing the constraints listed in the top of each frame.
The colour map indicates the relic density normalised to the PLANCK value ($\Omega_{\rm PLANCK} h^2=0.12$) for the 
two DM components  $a$ ($\Omega_a/\Omega_{\rm PLANCK}$, shown in green fading to yellow) and
$\psi$ ($\Omega_\psi/\Omega_{\rm PLANCK}$, shown in magenta fading to cyan), or their sum ($\Omega_{\text{tot}}/\Omega_{\rm PLANCK}$, shown in black fading to red).

In the top row of Fig.~\ref{fig:FDM-PS} we show the projection in the ($m_a$, $\lambda_{aH}$) plane, where
the colour map corresponds to values of $\Omega_a/\Omega_{\rm PLANCK}$ with dark green marking model points that saturate the 
relic density with $a$ alone. Recall that we keep all points with $\Omega_{\rm tot} h^2< 0.12$.
In Fig.~\ref{fig:FDM-PS}(a), no other constraint except the relic density is added: it clearly demonstrates the correlation between $\Omega_a$  
and the value of $\lambda_{aH}$, driven by the Feynman diagrams c)/e) and a)/f) of Fig.~\ref{fig:FD}. One can 
also see the region of the resonant annihilation through the Higgs boson, 
$aa\to H$, which takes place for $m_a \simeq m_H/2$. Due to its efficiency, it 
allows the value of $\lambda_{aH}$ to go as low as $\simeq 4\times 10^{-4}$ 
while being consistent with the $\Omega_{\rm PLANCK}$ constraint. Outside of the
resonant region, values of $\lambda_{aH}$ 
below $10^{-2}$  are excluded by
overclosure of the universe.
Furthermore, in Fig.~\ref{fig:FDM-PS}(b) we present the same 2D projection with points 
satisfying, in addition, the DM direct detection constraints from PandaX-4T experiment (both on $a$ and on $\psi$).
The plot illustrates how PandaX-4T excludes all points for $m_a \lesssim m_H$, except for a sliver
close to the Higgs resonance, which has small couplings or small relic density for the $a$ component, {and a few points with very low $a$ relic density (in yellow)}. 
One can see that all points with  $m_a \lesssim m_H/2$ below the $aa\to H$ resonant annihilation region are excluded  by  PandaX-4T experiment. This happens since in this region the $\Omega h^2\le  0.12$ constraint
requires the value of the $\lambda_{aH}$ coupling to be above $0.1$ that, in turn, 
leads to the SI DM direct detection rates to be above the  PandaX-4T
limits.

One should also note that, due to the specific set  of DM annihilation and co-annihilation diagrams shown in Fig.~\ref{fig:FD} and their interplay with each other, the relic density constraint requires $m_a < m_\psi$ in the whole parameter space, except the loop-induced $\psi\psi \to H$ annihilation region (we comment on this region below in more details). This region appears as a vertical strip in Fig.~\ref{fig:FDM-PS}(c)
for $m_\psi \simeq m_H/2$. Remarkably, this implies that $a$ is a stable DM component in the whole allowed parameter space, except for the Higgs funnel region for $\psi$, where $a$ can decay in the fermion DM component.

In Fig.~\ref{fig:FDM-PS}(b) we also superimpose the LHC bound on the Higgs invisible decays into $a$, 
$Br(H\to\mbox{invis})<0.11$, which excludes the Higgs resonant sliver for  $\lambda_{aH}\gtrsim 3\times 10^{-2}$,
as shown by the shaded region above the blue line. One can see that this bound is very complementary to the PandaX-4T constraint.
 Future collider projections are considered as well, showing that
the exclusion on $\lambda_{aH}$ will improve by a factor of about 3 at the High Luminosity LHC run (HL-LHC) (projected bound of $Br(H\to\mbox{invis}) <  3.8\%$ \cite{Atlas:2019qfx}), as shown by the orange line. The International Linear Collider (ILC) running at $\sqrt{s}=250$~GeV and with an integrated luminosity of $1.15~\mbox{ab}^{-1}$ will be able to exclude $\lambda_{aH}\gtrsim 4\times 10^{-3}$, as indicated by the green line, corresponding to a projected bound $Br(H\to\mbox{invis})<0.4\%$ \cite{Asner:2013psa}. One should also note that even the ILC will not be able to fully exclude  the Higgs resonant region, since $\lambda_{aH}$ goes below the ILC sensitivity by one order of magnitude.

Besides the Higgs sliver, a second viable region in the parameter space emerges for $m_a \gtrsim m_H$, 
as shown in plot~\ref{fig:FDM-PS}(b). It is defined by the interplay of the co-annihilation processes $\psi\psi \to aH$ and $a \psi \to H \psi$,
involving both new states of the dark sector. This is clearly illustrated by Figs~\ref{fig:FDM-PS}(c--d), in the plane defined by the masses
and the $\psi$ mass and coupling, respectively. Fig.~\ref{fig:FDM-PS}(c), showing a colour map  corresponding to the total
relic density $\Omega_{\text{tot}}$, offers the best view of this region. Besides the Higgs sliver for $a$, appearing as a horizontal band,
the allowed points highlight a vertical strip corresponding to the Higgs resonant region for $\psi$ via the one-loop induced
coupling, with
\begin{equation}
m_a \gtrsim m_\psi \quad 	
\label{eq:ma-mpsi-loop}
\end{equation}
and a wedge defined by
\begin{equation}
 m_\psi \gtrsim m_a\, .
	\label{eq:ma-mpsi}
\end{equation}

An interesting feature is the fact that masses below $m_H/2$ are 
excluded for both DM candidates: while for $a$ this is due to direct detection and (more marginally) by the Higgs invisible width, 
for $\psi$ this comes from the fact that for low masses the only efficient annihilation channel is $\psi\bar{\psi} \to a a $. This is efficient enough only for
$m_\psi \gtrsim m_a$, thus, $m_\psi < m_H/2$ would result in too much relic density due to the limit on $m_a$. 
In Fig.~\ref{fig:FDM-PS}(d)  we show the allowed
points projected on the $m_\psi$--$Y_\psi$ space, with colour map corresponding to the individual
relic density of $\psi$. We can see a clearly defined triangular shape, which emerges from the $\psi\psi \to a a$ annihilation process
and which requires the coupling $Y_\psi \gtrsim \mathcal{O}(1)$ to be fairly large to avoid overclosure
of the universe. On top of this, there is a ``leakage'' of points for $m_\psi \gtrsim m_H$, which 
emerge from the interplay with the process $\psi\psi \to a H$, which becomes relevant for 
$m_\psi \gtrsim m_a \sim m_H$. This means that for each value of $Y_\psi$, one can find a value
for $\lambda_{aH}$ that fixes the relic density below the limit. 
We also observe points with small $Y_\psi$ for masses below $m_H$: this is due to an interplay between the two processes
 $aa\to H$ and $\psi\psi \to a H$ above the threshold $m_\psi \gtrsim \frac{3}{4} m_H$. This value
 comes from the fact that the first process, $aa\to H$, dominates for $m_a\simeq m_H/2$ in the Higgs
 resonant region, while the second, $\psi\psi \to a H$, opens up for $m_\psi \simeq (m_a+m_H)/2$.

\begin{figure}[htb]
\includegraphics[width=0.5\textwidth]{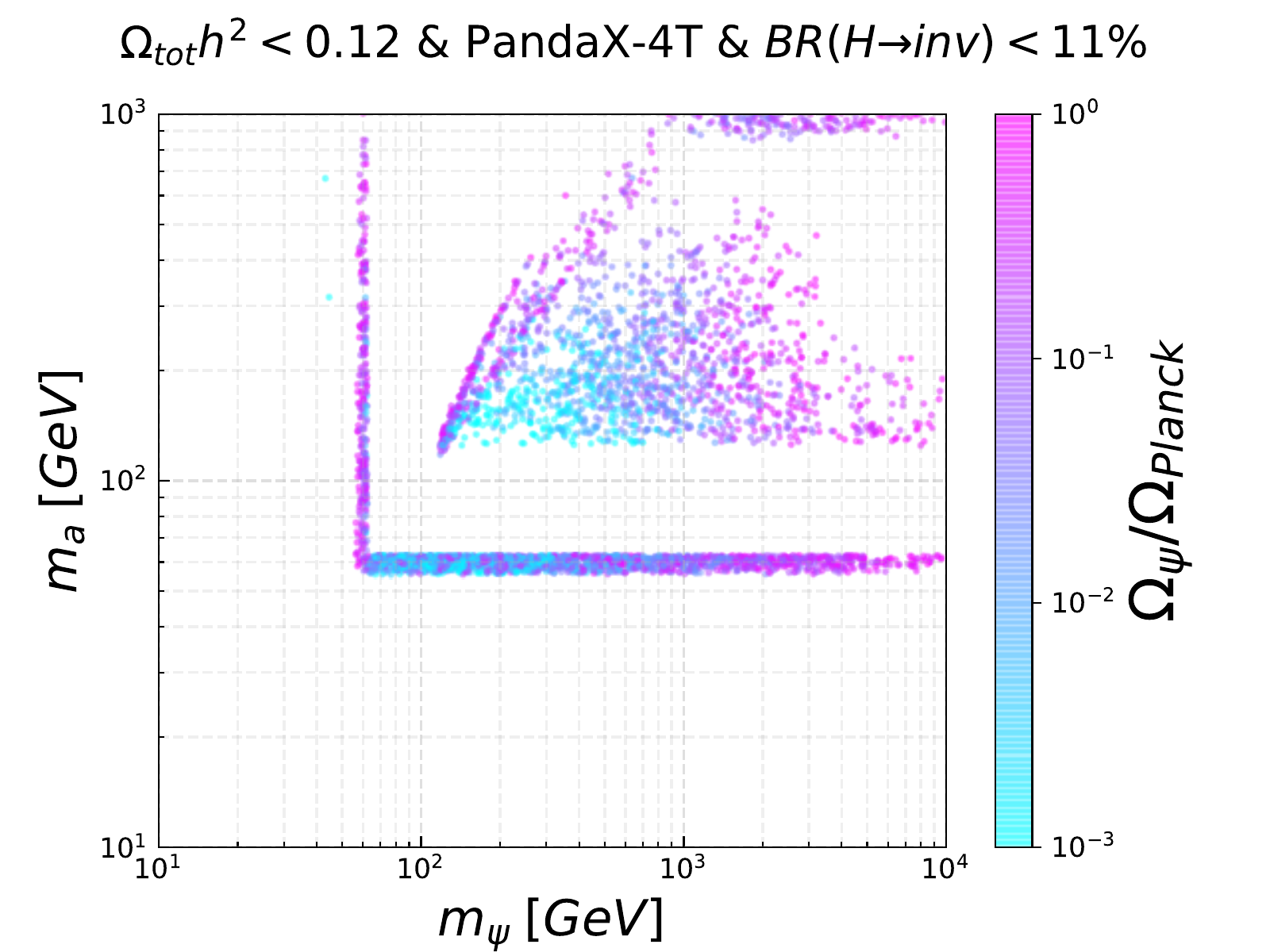}%
\includegraphics[width=0.5\textwidth]{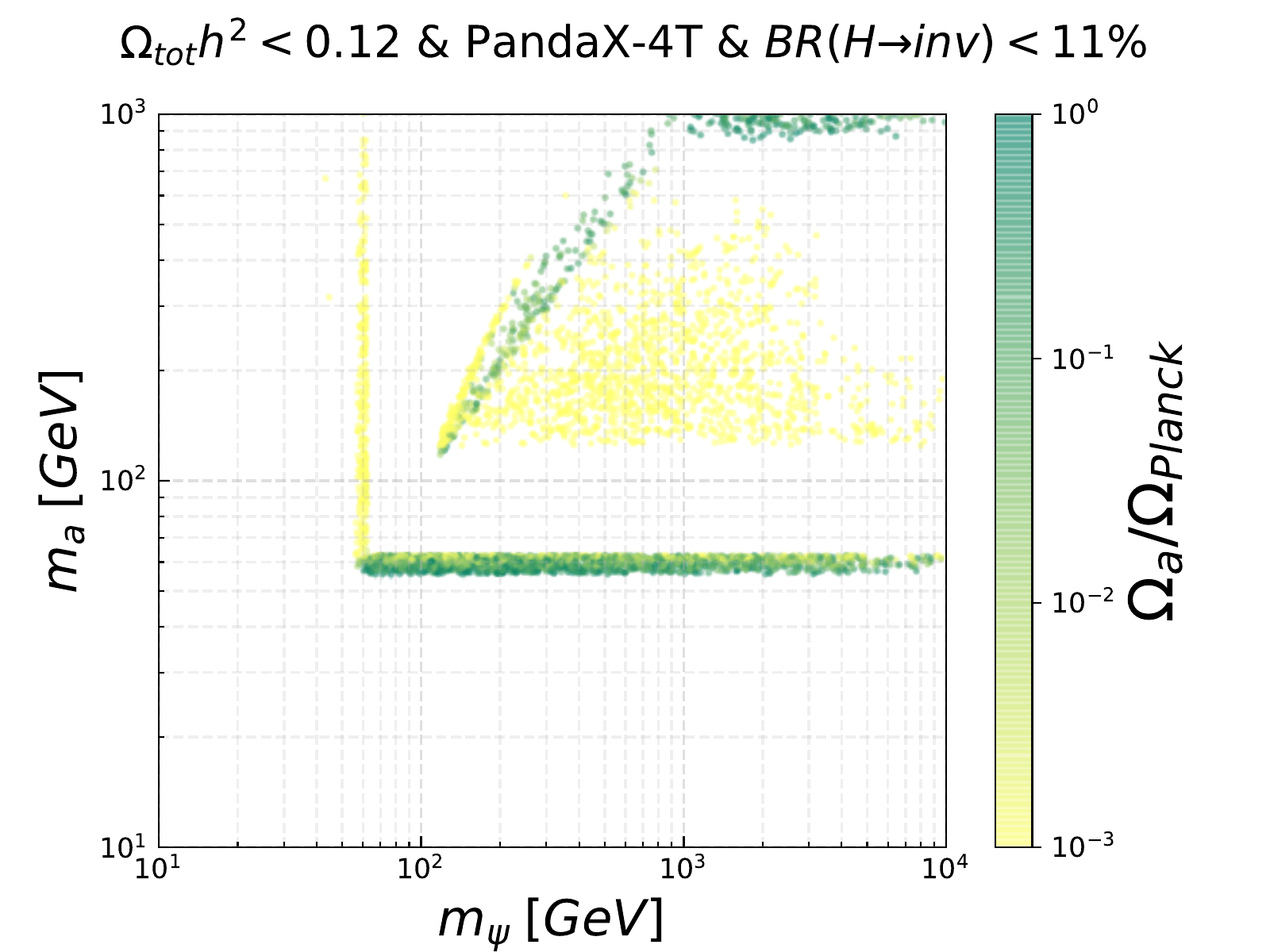}

\caption{\label{fig:FDM-PS_massPlane}
The distribution of relic density among species in the mass plane. Note that points with relic density below $10^{-3}\, \Omega_{\rm PLANCK}$ are shown with colour corresponding to the smallest value.}
\end{figure}

\begin{figure}[htb]
\includegraphics[width=0.5\textwidth]{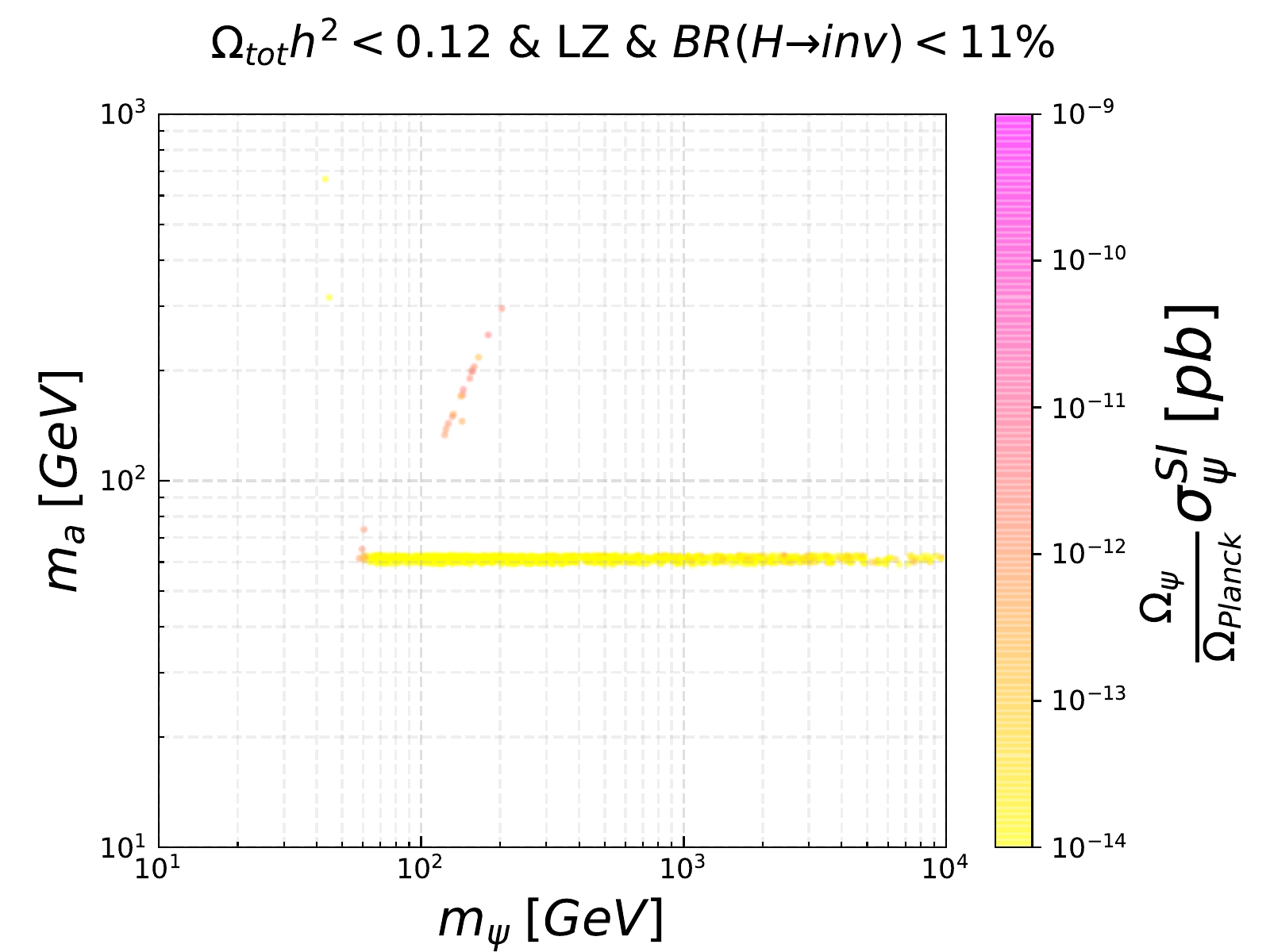}%
\includegraphics[width=0.5\textwidth]{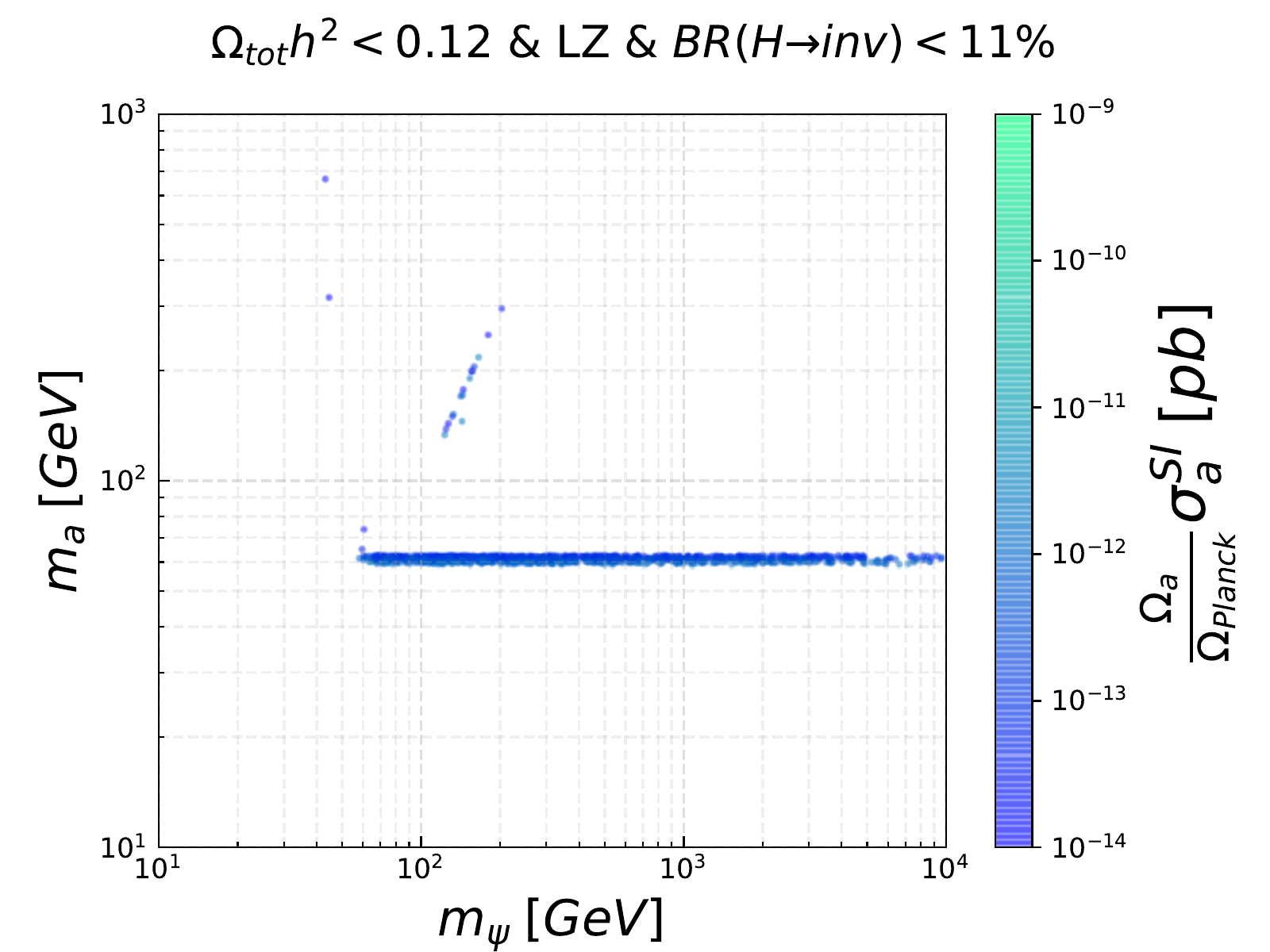}

\caption{\label{fig:LZ_massPlane}
Direct detection cross-sections (scaled by relevant relic abundance fraction) for the two DM species plotted in the mass plane, with constraints applied from future experiment LZ \cite{Akerib:2018lyp}. Note that small values below the range are shown with colour corresponding to the smallest value.}
\end{figure}

We remark from Fig.~\ref{fig:FDM-PS}(c) that points saturating
the measured relic density exist in almost the whole allowed parameter region, thanks to the interplay between 
the two components $a$ and $\psi$. In Fig.~\ref{fig:FDM-PS_massPlane} we show 
the contribution of each specie to the total relic (left for $\psi$ and right for $a$)
in the $(m_\psi,m_a)$ plane. Interestingly, the region with $m_a \sim m_H/2$ contains
points with sizeable and dominant relic from $\psi$, while $m_\psi \sim m_H/2$ 
	is {always} dominated by relic from $\psi$.
 The remaining parameter space contains a region with $m_a \sim m_\psi$ where both 
species can receive competitive relic densities, and regions dominated by $a$ for $m_a \gtrsim 300$~GeV and by $\psi$ for $m_\psi \gtrsim 1$~TeV.
Future direct detection experiments will be able to probe most of the remaining points, as demonstrated in Fig.~\ref{fig:LZ_massPlane}, where we impose the projected exclusion by the LZ next generation experiment \cite{Akerib:2018lyp}. The surviving points consist of  the Higgs sliver for $a$, with points dominated by the pseudo-scalar relic, and points with $m_a \sim m_\psi$. The latter ones still have sizeable SI cross-sections, discernible from the neutrino floor at future direct detection experiments. 

One should also note that the LZ
experiment will be able to almost exclude
the whole Higgs resonance region, $\psi\psi\to H$, which can also be probed, independently,  
at future colliders via invisible Higgs decays.
\begin{figure}[htb]
	\centering
	\includegraphics[width=0.5\textwidth]{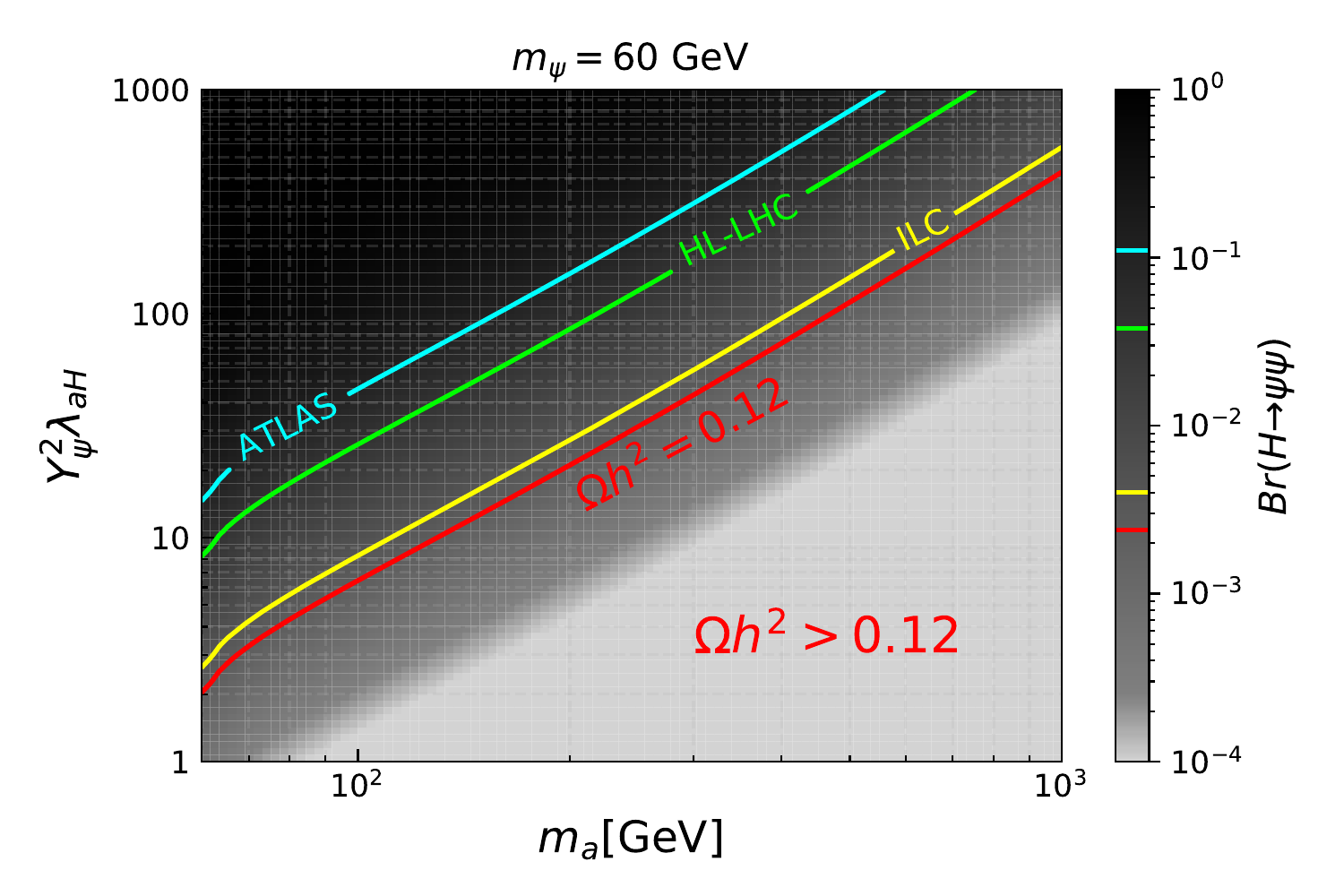}%
	\includegraphics[width=0.5\textwidth]{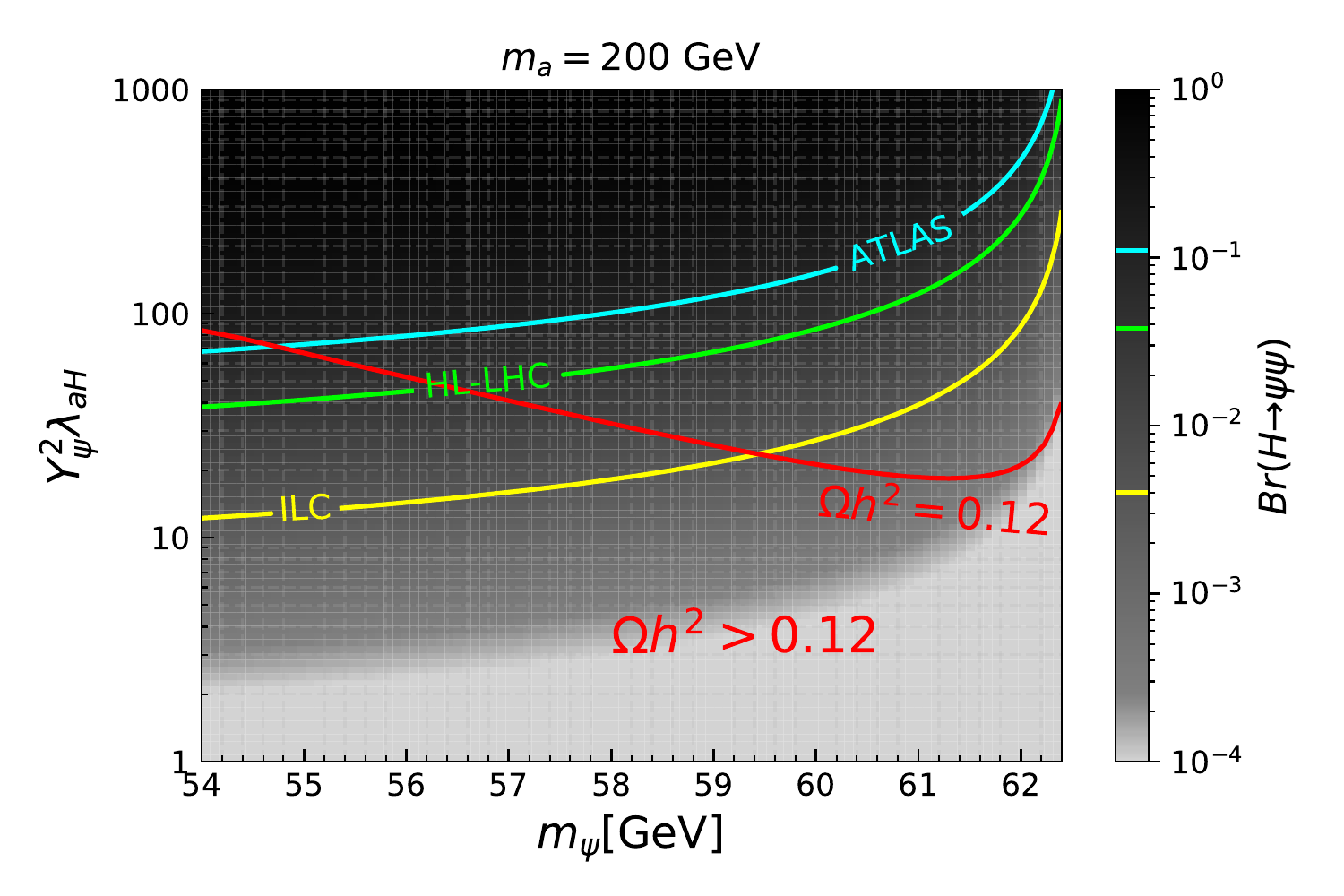}%
	\caption{\label{fig:Hinv_psipsi}
		Potential of the current LHC, HL--LHC and the ILC colliders
			to probe the loop induced branching ratio of Higgs to $\psi\psi$ for $m_\psi = 60$~GeV (left) and $m_a = 200$~GeV (right).}
\end{figure}
In this region $a$ is
heavier than $m_H/2$, thus contributing a very small fraction to the relic density
as shown in Fig.~\ref{fig:FDM-PS_massPlane}.
This region of the parameter space can also be efficiently probed by searches for invisible Higgs decays, especially at the ILC that will have the strongest sensitivity.
In Fig.~\ref{fig:Hinv_psipsi}
we present the comparison of the potential of the current LHC, HL-LHC and the ILC colliders
to probe this loop-induced $\psi\psi\to H$ region. In the left panel we fix $m_\psi = 60$~GeV and show the limits as a function of $m_a$. It is remarkable that, for this mass point, the ILC will be able to probe the Higgs invisible decay close to the value corresponding to the $\psi$ relic density saturating the PLANCK limit. The latter corresponds to $Br(H\to\psi\psi)\simeq 0.24\%$ to be compared to the projected ILC reach of $Br(H\to\psi\psi)\leq 0.4\%$.
In the right panel, instead, we fix $m_a = 200$~GeV and show the limits as a function of $m_\psi$. We can see that the ILC will be able to completely exclude $m_\psi \gtrsim 59.5$~GeV, while a region with the correct relic density will still be allowed for larger masses. Remarkably, the current ATLAS reach excludes $m_\psi \gtrsim 55$~GeV, while the HL-LHC will be able to push the limit to $m_\psi \gtrsim 56.5$~GeV.
\vskip 0.5cm
To summarise, the viable regions of the parameter space for Scenario A are:
\begin{itemize}
	\item 
	The $aa\to H$ annihilation region 
     with $m_a \simeq m_H/2$
	and $\lambda_{aH}\gtrsim 10^{-4}$,
	where the right amount of  relic density is provided by the diagram in Fig.~\ref{fig:FD}(c).
	This region can be probed by DM direct detection experiments and collider experiments looking for invisible Higgs decay.
	The main contribution to DM comes from $a$.
	\item The wedge region defined by $m_a, m_\psi> m_H$ and $m_a \lesssim m_\psi$,
	 where both components can be sizeable. This region can be probed by DM direct detection experiments.
	 The effective annihilation and co-annihilation are provided by the diagrams in Figs.~\ref{fig:FD}(a),(b),(d),(e),(f).
	\item	
	The $\psi\psi\to H$ annihilation region 
	with $m_\psi \simeq m_H/2$
	and $Y_\psi^2\lambda_{aH}>1$ and coupling generated at one-loop level.
	The dominant contribution to the relic density comes from $\psi$.
	This is the \emph{only} region where $a$ can be unstable, provided
	that $m_a> m_\psi/2$. This region can be effectively probed and even potentially closed by future ILC searches for invisible Higgs decay channels.
\end{itemize}
%

\subsection{Scenario B: $\psi$ FIMP regime with thermal $a$}

As we have seen, small values of $Y_\psi \lesssim \mathcal{O}(10^{-1})$ are excluded due to 
an excessive relic density of the fermionic component $\psi$. However, for extremely small values, $Y_\psi \lesssim \mathcal{O}(10^{-8})$,
$\psi$ will not be in thermal equilibrium at early times and it will freeze-in by means of the scattering of $a$ with the Higgs, $a H \to \bar{\psi} \psi$.
On the other hand, sizeable values of $\lambda_{aH}$ would guarantee that $a$ remains thermalised and contributes
with a thermal relic component (as a second specie when $m_a < 2 m_\psi$ or by decaying into the fermionic DM). 

\begin{figure}[htb]
\includegraphics[width=0.5\textwidth]{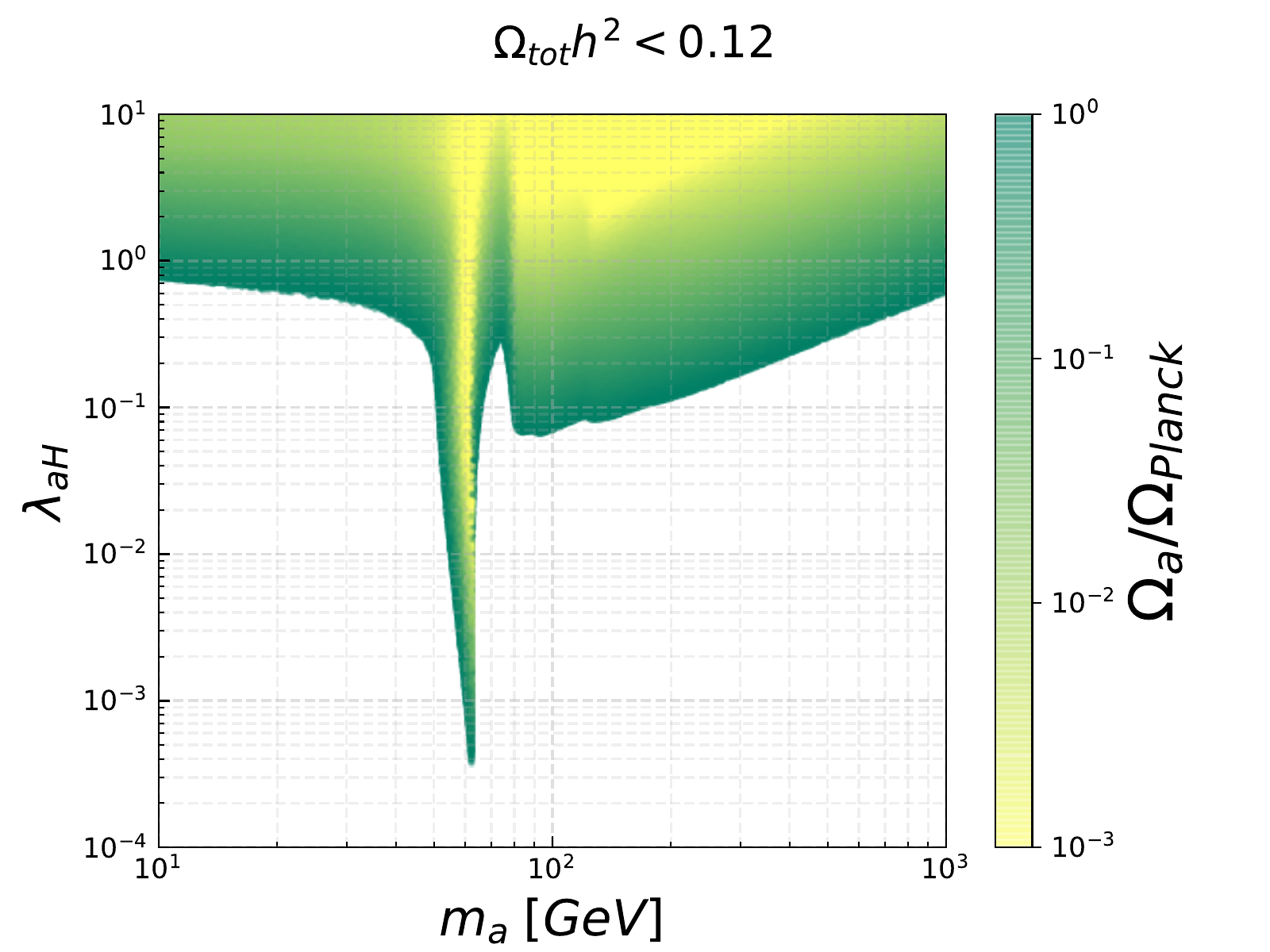}%
\includegraphics[width=0.5\textwidth]{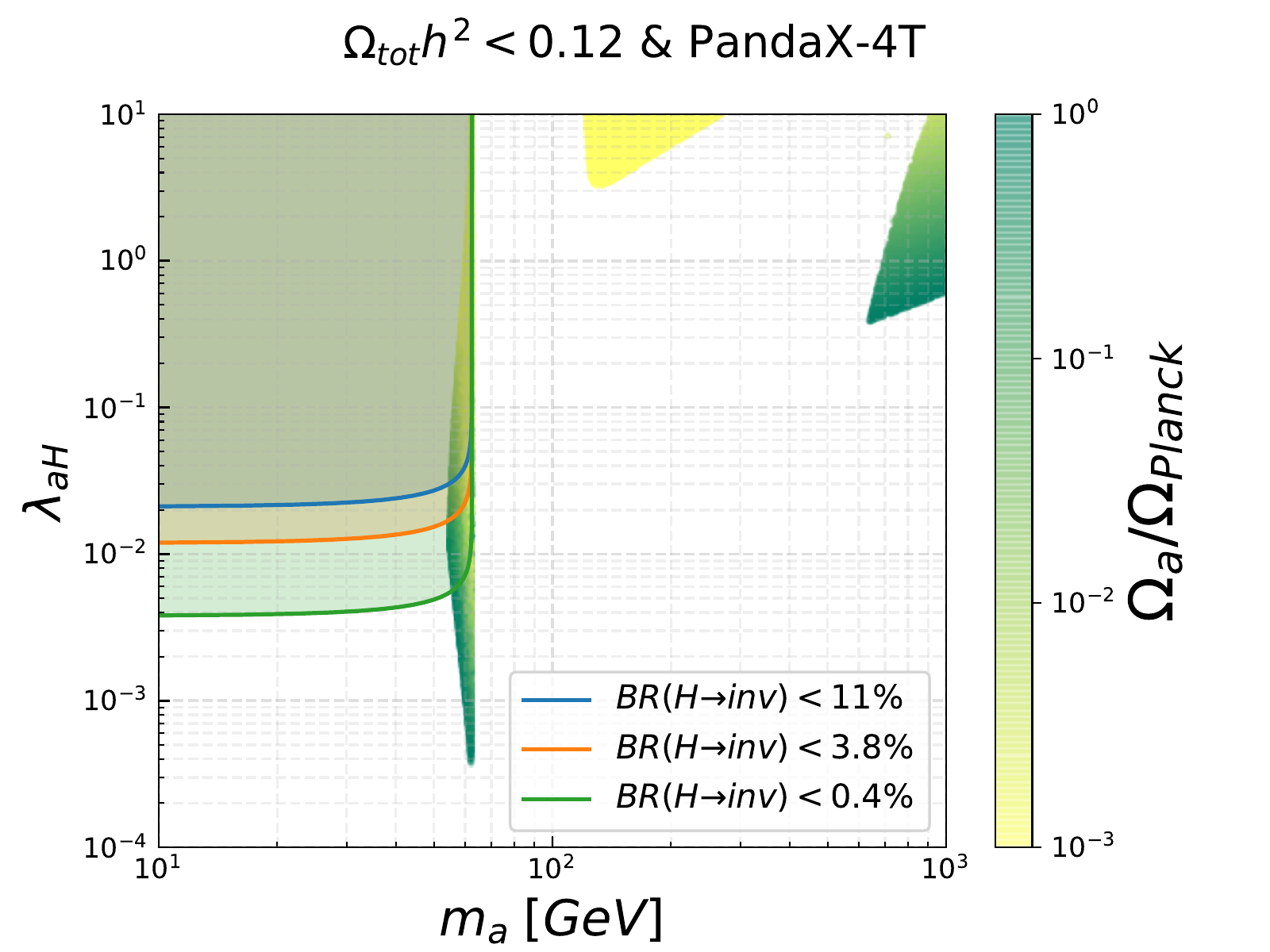}\\
\vskip -6cm(a)\hspace*{0.5\textwidth}\hspace{-0.5cm}(b)\vskip 5cm
\includegraphics[width=0.5\textwidth]{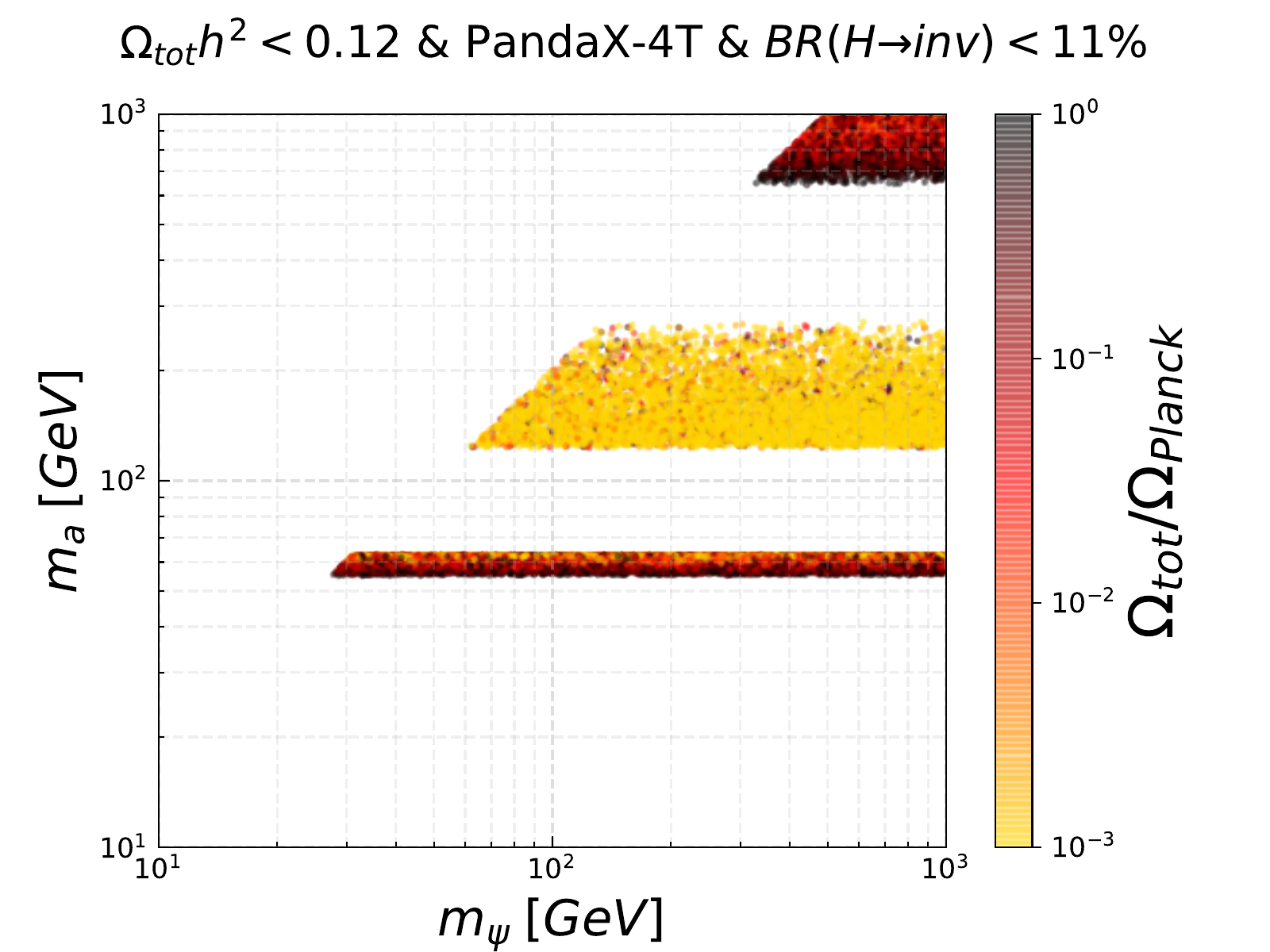}%
\includegraphics[width=0.5\textwidth]{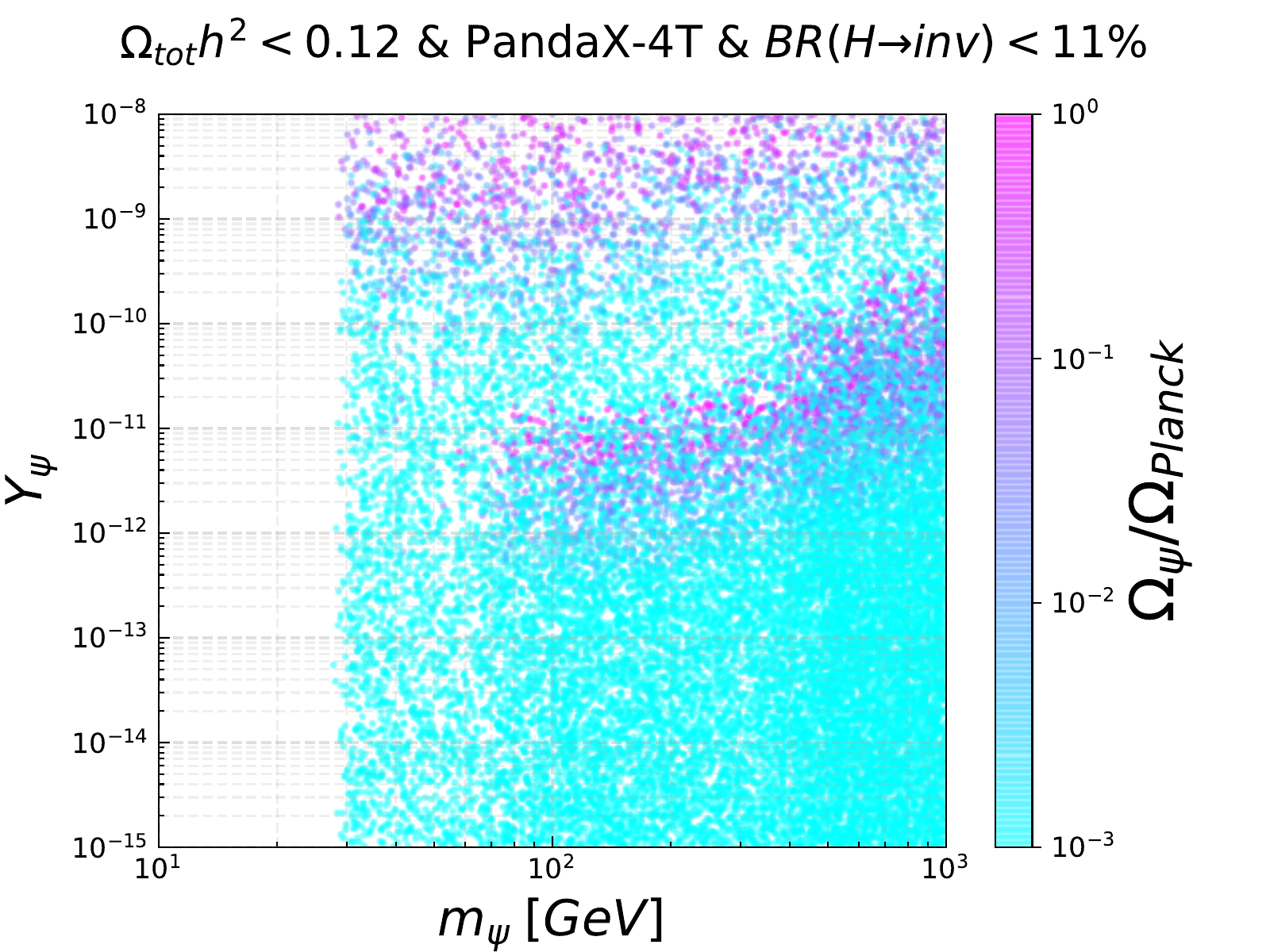}
\\
\vskip -6cm(c)\hspace*{0.5\textwidth}\hspace{-0.5cm}(d)\vskip 5cm
\caption{\label{fig:FDM-PS-fimp}
2D projections of the allowed parameter space for $\tilde F^0_0S^0_0$(CP-odd)  model with $m_a \leq 2m_\psi$ in the FIMP scenario B, after constraints given at the top of each frame.
The colour maps indicate the individual relative relic density of the two DM components  $a$ ($\Omega_a/\Omega_{Planck}$), 
$\psi$ ($\Omega_\psi/\Omega_{Planck}$) or their sum ($\Omega_{tot}/\Omega_{Planck}$).}
\end{figure}

In Fig.~\ref{fig:FDM-PS-fimp}
we present the results of this regime
for the $m_a < 2 m_\psi$ case, corresponding to two-component DM.
The first two plots in the top row -- Figs~\ref{fig:FDM-PS-fimp}(a) and (b) -- show the $\Omega_a h^2$ in the ($m_a$, $\lambda_{aH}$) plane, bearing similarity with  Figs~\ref{fig:FDM-PS}(a) and (b)
and demonstrating that the allowed regions are dominated by the thermal production of $a$. The only remarkable difference is the absence of ``leaking'' points, which were due to the co-annihilation
processes involving $\psi$ (so the smaller values of $\lambda_{aH}$ were allowed), which are now suppressed by the small value of $Y_\psi$. 
The contribution of $\psi$ via freeze-in is shown in the bottom frames of the figure.
In Fig.~\ref{fig:FDM-PS-fimp}(d), in particular, we show the relic 
density of $\psi$ in the ($m_\psi$, $Y_\psi$) plane. In this plot we can identify two distinct regions where sizeable values of $\Omega_\psi h^2$ can be attained (including saturating the whole
DM relic density): one for  $Y_\psi \gtrsim 10^{-9}$ starting from masses $m_\psi \gtrsim 30$~GeV (region BI), and another one for lower couplings,
$10^{-12} \lesssim Y_\psi \lesssim 10^{-9}$, starting at $m_\psi \gtrsim m_H/2$ (region BII). 
These two regions can be better understood by looking at the complementary plane, ($m_a$, $m_\psi$), shown in Fig.~\ref{fig:FDM-PS-fimp}(c): the region BI corresponds to points where $a$ is in the Higgs resonant sliver represented by the horizontal band; the
region BII  corresponds to triangle region at large $a$ mass,  where $m_a > m_H$. 
The scenario B can be probed only via the $a$ component of the DM relic and at colliders:
BI region is accessible via the Higgs invisible decay searches at colliders,
while DM direct detection experiments would be mainly sensitive to the region BII,
 as one can observe from Fig.~\ref{fig:FDM-PS-fimp}(b) demonstrating the effect of these searches.

\begin{figure}[htb]
	\includegraphics[width=0.5\textwidth]{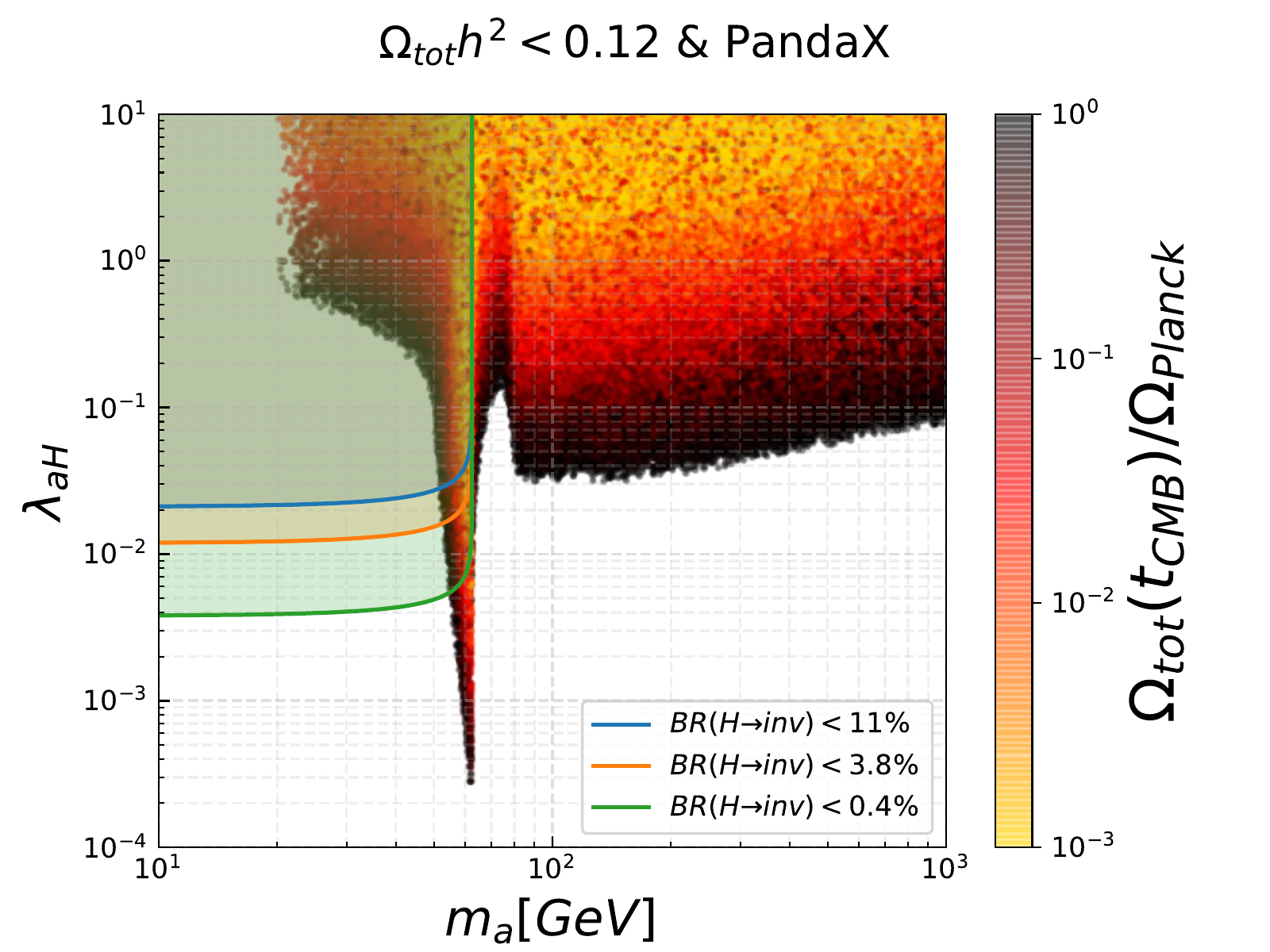}%
	\includegraphics[width=0.5\textwidth]{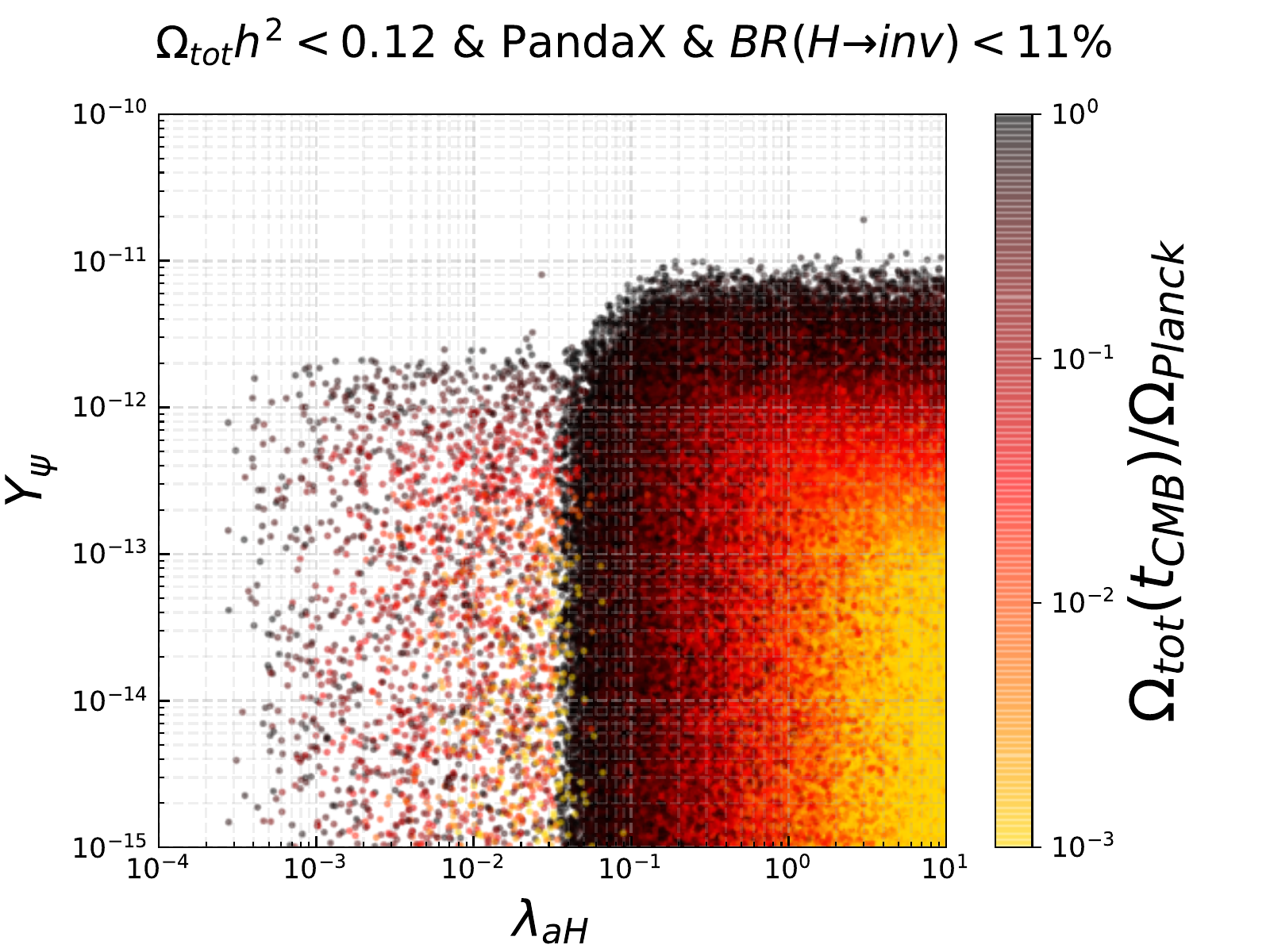}\\
	\vskip -6cm(a)\hspace*{0.5\textwidth}\hspace{-0.5cm}(b)\vskip 5cm
	\includegraphics[width=0.5\textwidth]{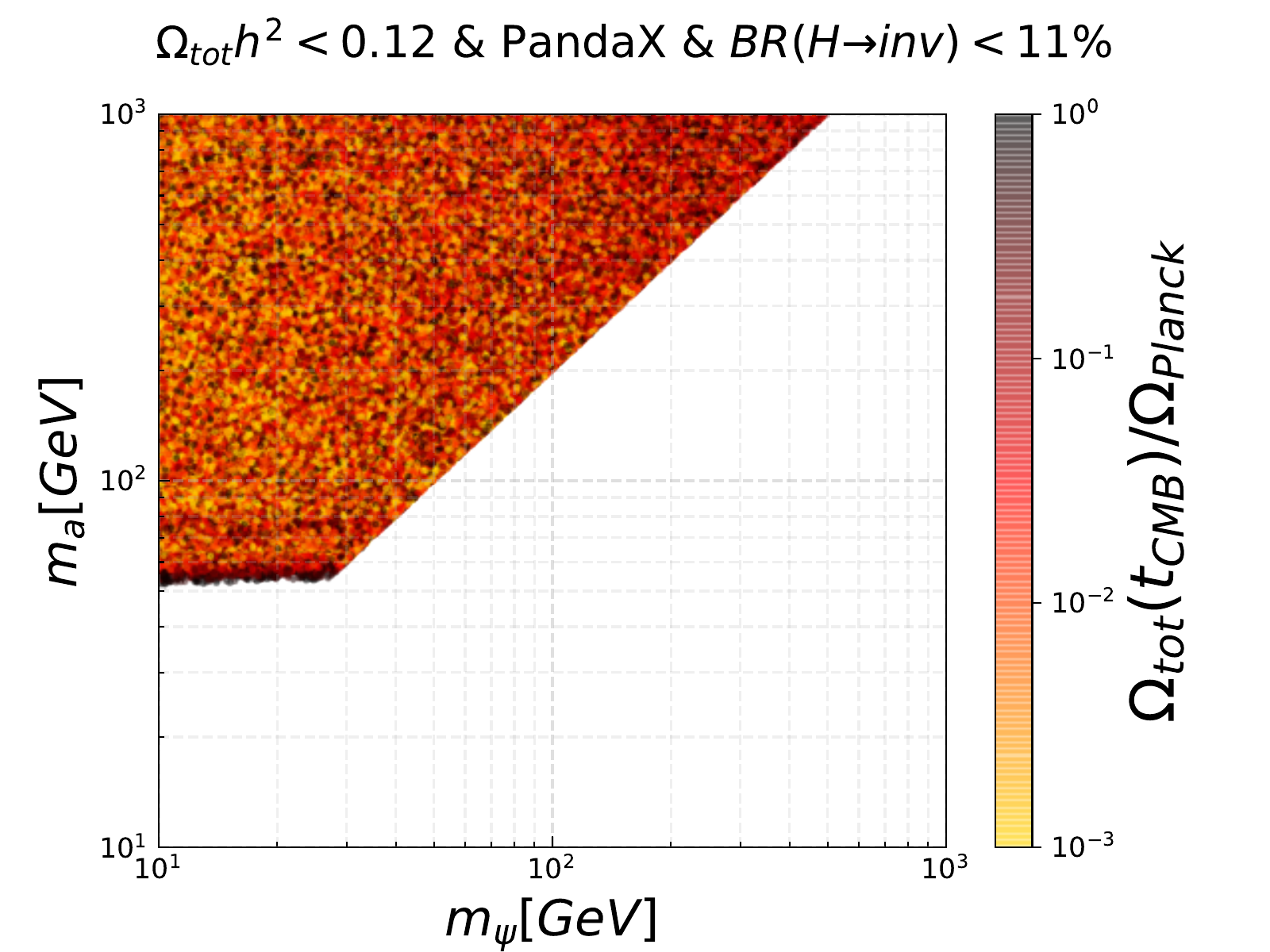}%
	\includegraphics[width=0.5\textwidth]{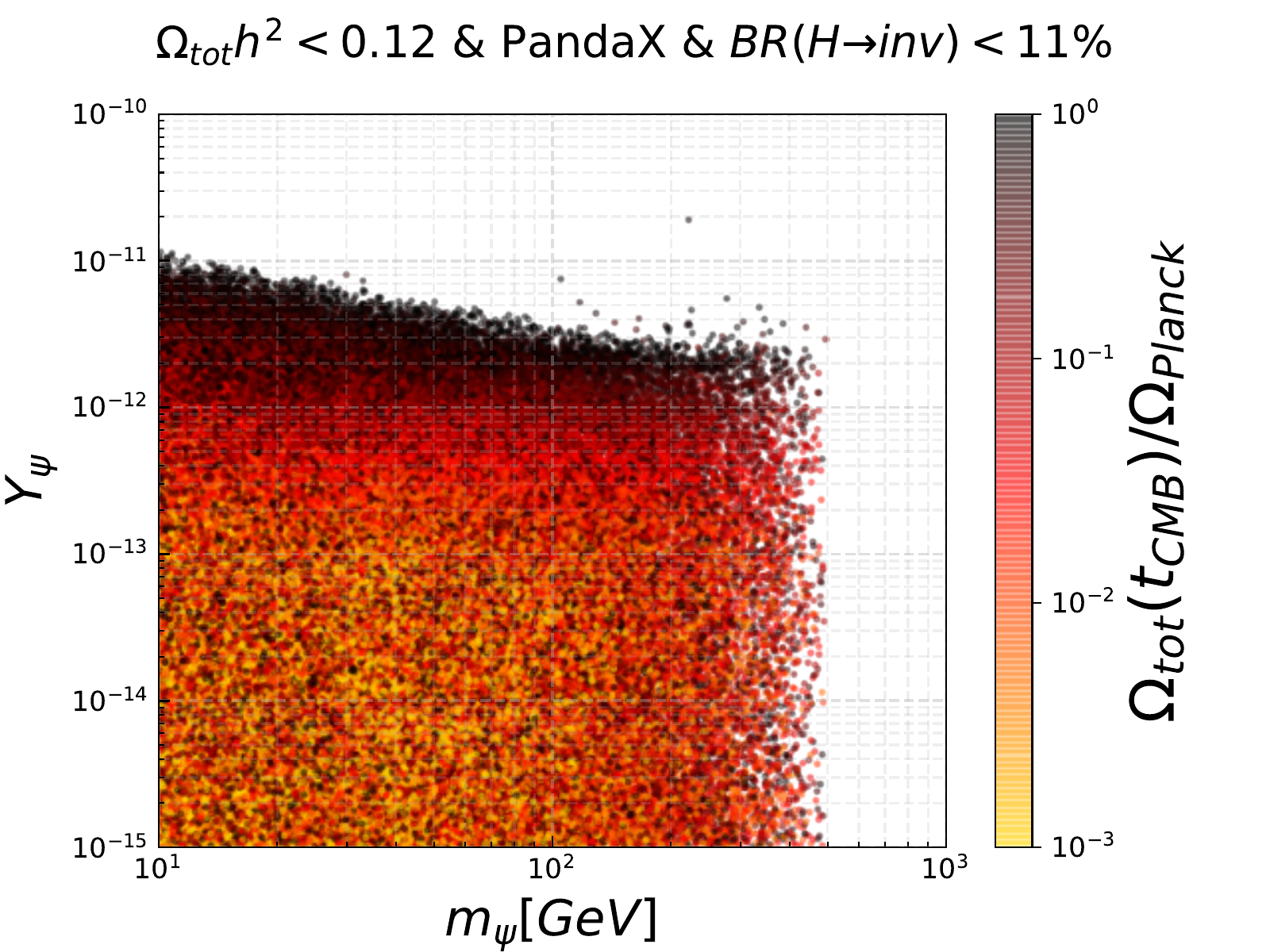}
	\\
	\vskip -6cm(c)\hspace*{0.5\textwidth}\hspace{-0.5cm}(d)\vskip 5cm
	\caption{\label{fig:FDM-PS-fimp_adec}
		2D projections of the allowed parameter space for $\tilde F^0_0S^0_0$(CP-odd)  model with $m_a>2m_\psi$ in the FIMP scenario B, after constraints given at the top of each frame.
		The colour maps indicate the individual relative relic density of the two DM components  $a$ ($\Omega_a/\Omega_{\rm PLANCK}$), 
		$\psi$ ($\Omega_\psi/\Omega_{\rm PLANCK}$) or their sum ($\Omega_{tot}/\Omega_{\rm PLANCK}$).}
\end{figure}

Finally, in Fig.~\ref{fig:FDM-PS-fimp_adec}
	we present numerical results for the region of the parameter space where  $m_a > 2 m_\psi$, region BIII,
	corresponding to a one-component DM ($\psi$) that originates from 
	$\psi$ freeze-in as well as $a \to \psi \psi$ decay processes after $a$ freezes out.
	In general, the correct evaluation of $\Omega h^2$
	requires taking into account the fact that $a$ may be long-lived due to the small values of $Y_\psi$. 
	The final relic densities stored in the two species are given by
\begin{equation}
\Omega_{\psi}(t) =
\Omega_\psi(t_{FI}) + \Omega_a(t_{FO})\ \dfrac{2 m_\psi}{m_a} \
\left(1-e^{-\dfrac{t}{\tau}}\right) \  \ , 
\quad \Omega_{a}(t) =
\Omega_a(t_{FO})
\ e^{-\dfrac{t}{\tau}} \  \ , 
\label{eq:adec}
\end{equation} 
where $\Omega_\psi(t_{FI})$ is the $\psi$ relic density at its freeze-in time and $\Omega_a(t_{FO})$ is the $a$ relic density at its freeze-out time, which is typically much smaller
than its life-time $\tau$. 
In most of the parameter space, $\tau$,
is much smaller than the CMB time, $\tau \ll t_{CMB} \simeq 2\times 10^5$~years, 
hence the relic densities in Eq.~\eqref{eq:adec}
simplify to
\begin{equation}
\Omega_{tot}(t_{CMB}) = \Omega_{\psi} (t_{CMB}) =
\Omega_\psi(t_{FI}) + \Omega_a(t_{FO})
\dfrac{2 m_\psi}{m_a} \ \ ,
\label{eq:adec_CMB}
\end{equation} 
while the relic density of $a$ is negligibly small.

There are several important features of the BIII region.
One of them is that there is no sensitivity
from DM direct detection experiments neither through $a$, as its relic density is negligibly small,  nor through $\psi$ that has a very weak coupling to the SM.
This region can be only tested via the invisible Higgs decay search at colliders, as shown in Fig.~\ref{fig:FDM-PS-fimp_adec}(a).
Figure~\ref{fig:FDM-PS-fimp_adec}(b)
presents the interplay between 
$\lambda_{aH}$ ans $Y_\psi$ couplings: for $\lambda_{aH}\simeq 1$, the relic density 
saturating the PLANCK measurement is mainly provided by freezed-in $\psi$
with $10^{-12}<Y_\psi<10^{-11}$. This feature is also 
 clearly visible in
 Fig.~\ref{fig:FDM-PS-fimp_adec}(d)
 via the upper edge in the allowed $m_\psi$ values.
 When $Y_\psi<10^{-12}$, instead,
the main contribution to  the relic density 
comes from  $\lambda_{aH} \simeq 0.1$ via the $a$ relic at freeze-out, which then completely decays to $\psi\psi$.
One can also see one more pattern in Fig.~\ref{fig:FDM-PS-fimp_adec}(b) represented by ``scattered" points with
$0.001 < \lambda_{aH} < 0.1$, 
where $aa\to H$ annihilation takes place and provides the right amount of DM
via $a$ freeze-out. Figure~\ref{fig:FDM-PS-fimp_adec}(c) 
shows the range of $m_a$ and $m_\psi$ masses
viable in this scenario. In particular, it shows that the lower limit on $m_a$  is about 50 GeV. This limit comes from the current invisible decay search at the LHC,
which extends the $m_a \gtrsim m_H/4$ limit which comes from relic density constraints defined by $m_a$ and $m_\psi$ kinematics, as one can see from  Fig.~\ref{fig:FDM-PS-fimp_adec}(a).
One can also see from Fig.~\ref{fig:FDM-PS-fimp_adec}(a)
that the hierarchy between the masses $m_a$ and $m_\psi$ can be quite large. 
This means that a small value of the ratio $m_\psi/m_a$
 can provide viable parameter space
even if $\Omega_a (t_{FO})$ is too large, as one can see from the second term of Eq.~\eqref{eq:adec_CMB}.

\subsection{Scenarios C and D: 2-component FIMPs}

These two scenarios are characterised by a very small coupling of the Dark sector to the SM, i.e. a tiny $\lambda_{aH}$. As such, they are very difficult to test while they can provide the right amount of relic density. For this reason, we do not present any numerical scan, instead we will qualitatively discuss the main features of the two scenarios.

In case C, $\lambda_{aH}$ is very small while $Y_\psi$ is sizeable. Hence, both $a$ and $\psi$ can be produced via freeze-in via the couplings
to the Higgs (for $\psi$ loop induced). A large $Y_\psi$ would simply reshuffle the relic density of the two components at later times. When $m_a>2m_\psi$, then $a$ would promptly decay resulting in $\psi$ saturating the relic density. 

In case D, the smallness of both relevant couplings would lead to $\psi$ occupying an insignificant part of the relic as freeze-in for this species would be hampered doubly by the small couplings. This means that in this scenario  $\psi$  decouples
from $a$ and  the model is effectively reduces to the well-explored scalar  portal model 
with freeze-in scenario.

%% file: 06_conclusions.tex

\section{Conclusions and outlook} \label{sec:conclusions}

After the discovery of the Higgs boson, the search for a Dark Matter particle has become the new grail and hard-sought nirvana of the particle physics community. The diversity in the experimental techniques and the remarkable progress achieved in each one call for more sophisticated theoretical studies, especially when trying to combine and compare various experimental bounds. The main difficulty stands in the large array of energies probed by the experiments: from low energy interactions in direct and indirect detection, to high energies at colliders like the LHC and the future FCC-ee/hh, ILC and CEPC. Moreover, additional constraints come from Cosmology via the relic density, precisely determined via the cosmic microwave background measurements, and from precision measurements in the electroweak sector of the Standard Model. 

Complete models that contain a Dark Matter candidate, like supersymmetry or composite Higgs models, provide a consistent comparison at the price of specificities  that are hard to disentangle from the phenomenology {and generic features of the Dark Matter sector itself}. 
{Exploration of Dark Matter properties
independent of quite a few  details of the  complete model is, therefore, a challenge.}
In this work, we propose a systematic classification of minimal consistent Dark Matter models, which are required to respect the complete symmetries of the Standard Model, {as summarised in Table~\ref{tab:MCDM}}. They provide the missing link between effective field theory approaches and complete models, and offer a consistent and model-independent comparison between various experimental constraints. 
Moreover, because of their consistency, MCDM models can  serve as a complete theory by themselves or be used as a building blocks within
a bigger framework.
{This approach allows to create a convenient basis 
	for the DM  model space which can be used for a systematic DM exploration at various experiments}.

{In our framework} the Dark Matter particle is embedded in an electroweak multiplet, characterised by its weak Isospin and hypercharge. Similarly, a mediator multiplet is included with all renormalisable interactions. The only exception to the latter is given by dimension-5 couplings to the Higgs, which can split masses and, therefore, crucially influence direct detection bounds. We consider in this work fermionic Dark multiplets and discover that many models are still allowed by all constraints, beyond the simplified models currently considered in the literature. We also revisit one-loop contributions to direct detection, including for the first time the mass splits in the dark multiplet. Due to the presence of a fine cancellation among various contributions to the amplitudes, the presence of a small mass split affects significantly the total spin-independent cross section.  

Our main results can be summarised as follows:
\begin{itemize}
\item Dark multiplets with hypercharge equal of above 1 are excluded by the presence of a charged lightest component and $Z$ decay bounds. 
\item The loop-induced direct detection excludes multiplets with Isospin equal or above 3 (sextet), while other multiplets are probed by current or future experiments. The doublet escapes detection thanks to a cancellation among various contributions to the elastic scattering amplitude.
\item Dimension-5 couplings of the Higgs, potentially generated by a heavy scalar or fermion mediators
{(i.e. described by MCDMs with  a mediator multiplet)},  play a crucial role in splitting Dirac multiplets in Majorana mass eigenstates, hence removing the strong constraints from $Z$-mediated direct detection.
{On the other hand, the value of the mass split of the neutral states of the order of few GeV is being tested by DM direct detection experiments at present, while future DM direct detection experiments will be able to test it at sub-GeV level.
}
\item
{Besides the important role of the mass split effects for DM direct detection, 
we have also shown the role of the PDF and the QCD scale uncertainties, which can be similarly important 
to provide cancellations for  a loop-induced direct detection amplitudes. 
}
\end{itemize}

We also study in detail a new model with a Dirac singlet Dark multiplet and a CP-odd singlet scalar mediator. While the mediator is even under the Dark parity protecting the fermion, it can be accidentally stable if it is lighter that twice the fermion mass. Thanks to the interplay between the two components, the parameter space can be probed by the synergy between future direct detection and the measurement of the invisible Higgs decay width at colliders. Furthermore, in the small coupling regime, both fermion and scalar can be produced in the universe by freeze-in. This is one example of interesting models, neglected in the literature, which is highlighted by our complete {MCDM} classification.

In this paper, we provide a first complete and consistent classification of effective models for Dark Matter that allows for a consistent and systematic comparison between all constraints on the Dark Matter particle candidate. We focus here on fermionic spin-1/2 Dark multiplets, while the paradigm can be applied to any spin option. We leave for a future work to compile a classification for spin-0 and spin-1 Dark multiplets and their minimal one-mediator extensions. Many models are still allowed and viable, beyond the simplified cases analysed in the literature. A systematic study of all the cases can help us establish the feasibility of a Dark Matter candidate around the electroweak scale, which seems to be under siege by the non-discovery of the historical WIMP candidates. Furthermore, as our classification requires full invariance under the Standard Model symmetries, the models we present
{can be} easily embedded {into} more complete models and they can be further UV-completed in a consistent way. Henceforth, we believe that our classification can provide the required missing link between experimental searches and the model building required to obtain the new Standard Model that includes Dark Matter.

\section*{Acknowledgements}

Authors acknowledge the use of the IRIDIS High Performance
Computing Facility, and associated support services at the University of Southampton to complete
this work. 
AB and DL acknowledge  support from the STFC grant ST/L000296/1 and Soton-FAPESP
grant. 
GC is grateful to the LABEX Lyon Institute of Origins (ANR-10-LABX-0066) Lyon for its financial support within the program ``Investissements d'Avenir'' of the French government operated by the National Research Agency (ANR).
The work of AP  funded by the RFBR and CNRS project number20-52-15005.
The work of AP was
also supported  by the Interdisciplinary Scientific and Educational School of Moscow University
for Fundamental and Applied Space Research. 
We are grateful to the Mainz Institute for Theoretical Physics (MITP) of the DFG Cluster of Excellence PRISMA+ (Project ID 39083149) for its hospitality and support during the initial stages of this work.

%% file: Appendix.tex

\newpage
\appendix

\section{Appendix}

\subsection{Radiative mass corrections for single electroweak multiplet models} \label{app:radcorr}
For a vector-like fermion contained in an electroweak multiplet and in the absence of additional electroweak multiplets, the radiative mass split may be found from calculating the one-loop self-energy resulting from radiation and absorption of a single vector boson, $V$ (which may be a photon, Z or W boson). The amplitude for this diagram is given by

\begin{align}\label{singleLoop}
    i \Sigma(\slashed{p}) &= g^2 \int \frac{d^4k}{(2\pi)^4} \frac{\gamma^\mu (\slashed{p} + \slashed{k} + M_D) \gamma^\nu (-g_{\mu,\nu})}{[(p+k)^2 - M_{DM}^2][k^2-m_V^2]} \nonumber \\
    &= \frac{i}{(4\pi)^2} \left[ C_0 + C_D^A \bm{A}(M_D^2) + C_V^A\bm{A}(m_V^2) + C_B \bm{B}(M_D^2, m_V^2)  \right] \,,
\end{align}

\noindent
where $g$ is the coupling between the fermion and vector, $p$ is the external fermion momentum, $k$ the loop momentum, $M_{DM}$ is the DM mass and $m_V$ is the mass of the vector boson. Here, \textbf{A} and \textbf{B} are the 1 and 2 point Passarino-Veltman integrals, as defined in \cite{tHooft:1978jhc}. Their coefficients are found to be
\begin{align}
    C_0 &= (2g^2+\delta_M)M_{DM} + (-g^2+\delta_Z)\slashed{p} \,, \nonumber \\
    C_D^A &= -\frac{g^2}{p^2}\slashed{p}\,, \nonumber \\
    C_V^A &= \frac{g^2}{p^2}\slashed{p}\,, \nonumber \\
    C_B &= g^2 \left[ 2\slashed{p} - 4M_{DM} - \frac{\slashed{p}}{p^2}(p^2 + m_V^2 - M_{DM}^2) \right]\,, \nonumber \\
        &= \frac{g^2}{p^2}(p^2 + M_{DM}^2 - m_V^2)\slashed{p} - 4g^2M_{DM} \,.
\end{align}

\noindent
The divergent parts of these coefficients are absorbed using the counterterms in the $\overline{MS}$ scheme

\begin{equation}
    \delta_Z = -\frac{g^2}{\hat{\epsilon}} , \quad \delta_M = \frac{4g^2}{\hat{\epsilon}}, \quad \frac{1}{\hat{\epsilon}} \equiv \frac{2}{4-D} - \gamma_E + \log{4\pi},  
\end{equation}

\noindent
where $\delta_Z$ is the field renormalisation, $\delta_m$ is the mass renormalisation, D is the space-time dimension and $\gamma_E$ is the Euler-Mascheroni constant. The resulting amplitude may be expressed using the function  

 \begin{align}
     f(r) &= \frac{16\pi^2}{M_{DM} g^2} \Sigma(\slashed{p}=M_{DM}) \nonumber \\
     		&= \frac{r}{2} \left[2r^3 \log (r)-2r+\sqrt{r^2-4} \left(r^2+2\right) \log \left(A\right) \right] - 4 \,,
 \end{align}

\noindent
where $r\equiv\tfrac{m_V}{M_{DM}}$ and $A = \left( r^2 -2 - r\sqrt{r^2-4}\right)/2$. Combining contributions from all diagrams, we find the total self-energy as follows:

\begin{equation}
    \Sigma^{(tot)}(n,Q,Y) = \frac{M_{DM} g^2}{16\pi^2} \left[ Q^2f_\gamma + \frac{(Qc_w^2-Y)^2}{c_w^2}f_Z + \left[ C_{W^+}(n,Q,Y)^2+C_{W^-}(n,Q,Y)^2 \right]f_W \right]\,.
\end{equation}

\noindent
Here, $f_V \equiv f \left(\frac{m_V}{M_{DM}} \right)$, $n$ is the dimension of the multiplet, $g$ is the weak coupling, $c_w \equiv cos(\theta_w)$ where $\theta_w$ is the Weinberg angle, $Q$ is the electric charge of the fermion and $Y$ is the hypercharge (using convention of $Q=T_3+Y$). The coupling to $W^\pm$ may be expressed as $C_{W^\pm} = \tfrac{1}{2\sqrt{2}} \sqrt{n^2-(2Q-2Y \pm 1)^2}$. This leads to an expression for the difference between the pole masses of two members of a given multiplet with charges $Q$ and $Q'$ respectively,

\begin{align} \label{deltaM1}
    M_Q-M_{Q'} &= (-\Sigma_Q^{(tot)}(M_{DM})) - (-\Sigma_{Q'}^{(tot)}(M_{DM})) \nonumber \\ 
                &= \frac{M_{DM} g^2}{16\pi^2} (Q-Q') \left[ (Q+Q'-2Y)(f_W-f_Z) + (Q+Q')(f_Z-f_\gamma)s_w^2 \right] \,.
\end{align}



A more numerically stable expression also exists for f(r), given by

\begin{equation}
    f(r) = \frac{r}{2}  \left[2r^3 \log (r)-2r-\sqrt{r^2-4} \left(r^2+2\right) \log \left(B\right) \right] - 4 \,,
\end{equation}

\noindent
where $B = \left( r^2 -2 + r\sqrt{r^2-4}\right)/2$. The limits of this function for large and small $M_{DM}$ respectively are given by

\begin{align}
    \lim_{r\to 0} f(r) &= -4 +2 \pi  r -3 r^2+\frac{3 \pi  r^3}{4}+\mathcal{O}\left(r^4\right) \,, \nonumber \\ 
	\lim_{r\to \infty} f(r) &= 6\log{r} - \frac{5}{2} + \frac{1}{r^2} \left( 8 \log{r} - \frac{8}{3}\right) + \mathcal{O}\left( r^{-4}\right) \,.
\end{align}

%
%
%
\subsection{Loop induced direct detection calculation}\label{app:MDM_DD}
Both for psuedo-Dirac fermion DM candidates and for multiplets with hypercharge $Y=0$, the tree level scattering amplitude between DM and nucleons vanishes due to the absence of a $Z$ coupling. As such, loop-induced scattering is key to probe such models at DM direct detection experiments. The interaction Lagrangian relevant for the one-loop calculation for a Dirac multiplet (containing a Dirac DM candidate $D^0$) is given by

\begin{align}
\Delta \mathcal{L}_{\rm Dirac} &= \left[ \frac{g_2}{2\sqrt{2}} \sqrt{n^2-(2Y+1)^2} \bar{D^0} \gamma^\mu D^- W_\mu^+ + \frac{g_2}{2\sqrt{2}} \sqrt{n^2-(2Y-1)^2} \bar{D^0} \gamma^\mu D^+ W_\mu^- + \text{h.c.} \right] \nonumber \\
& \quad + \frac{g_2 (-Y)}{c_W} \bar{D^0}\gamma^\mu D^0 Z_\mu^0 \,.
\end{align}

For the psuedo-Dirac multiplet case, this Dirac fermion $D^0$ splits into two Majorana mass eigenstates, $D^0 \to ( \chi^0_1+i\chi^0_2)/\sqrt{2}$. Without loss of generality, we assume that $\chi^0_1$ is sufficiently lighter than $\chi^0_2$ to prevent tree-level inelastic scattering via Z boson and it is the only DM candidate. This leads to the relevant interaction Lagrangian

\begin{align}
\Delta \mathcal{L}_{\rm pseudo-Dirac} &= \left[ \frac{g_2}{4} \sqrt{n^2-(2Y+1)^2} \bar{\chi^0_1} \gamma^\mu D^- W_\mu^+ + \frac{g_2}{4} \sqrt{n^2-(2Y-1)^2} \bar{\chi^0_1} \gamma^\mu D^+ W_\mu^- + \text{h.c.} \right] \nonumber \\
& \quad + \frac{ig_2 (-Y)}{c_W} \bar{\chi^0_1}\gamma^\mu \chi^0_2 Z_\mu^0 \,.
\end{align}

Finally, for a Majorana candidate $\chi^0$ contained within a Majorana multiplet (where $Y=0$), the relevant Lagrangian reads

\begin{align}
\Delta \mathcal{L}_{\rm Majorana} &= \left[ \frac{g_2}{4\sqrt{2}} \sqrt{n^2-1} \bar{\chi^0} \gamma^\mu \chi^- W_\mu^+ + \frac{g_2}{4\sqrt{2}} \sqrt{n^2-1} \bar{\chi^0}^c \gamma^\mu (\chi^-)^c (W_\mu^+)^c + \text{h.c.} \right] \,.
\end{align}

First we discuss the box diagrams (A and B in Fig.~\ref{fig:triLoop}). For Dirac and pseudo-Dirac cases, $D^+$ and $D^-$ are distinct particles ($\bar{D}^\pm$ is used to refer to their respective antiparticles), which can have different masses. As such, a DM particle(antiparticle) will couple to up(down)-type quarks via diagram A containing a $D^+$($\bar{D}^+$) or diagram B containing a $D^-$($\bar{D}^-$) and down(up)-type quarks via diagram A containing a $D^-$($\bar{D}^-$) or diagram B containing a $D^+$($\bar{D}^+$), or to all quarks by diagrams A and B when two Z bosons are exchanged. In the Majorana case, the charged state in the two loops is the same. Hence, these diagrams can be combined when in the Majorana case, or when the masses are the same (as it is for $Y=0$ with mass splits generated by EW loops).

%
%

For simplicity, we perform our computation in the zero momentum transfer limit ($t \approx 0$) from the start. 
We recall that for the case of Dirac DM, couplings to quark types may be different in general, as such diagrams A and B (untwisted and twisted topologies) must be calculated independently. The amplitudes for diagram A and B are given by

\begin{align}\label{eqn:AB_amplitudes}
    iM_A = \xi_A \int \frac{d^4l}{(2\pi)^4} \frac{J_{D}^{\mu\nu}J_{q,A}^{\rho \sigma} g_{\mu\rho}g_{\nu\sigma}}{\mathcal{D}_A}\,, \quad\quad
    iM_B = \xi_B \int \frac{d^4l}{(2\pi)^4} \frac{J_{D}^{\mu\nu}J_{q,B}^{\rho \sigma} g_{\mu\sigma}g_{\nu\rho}}{\mathcal{D}_B} \,,
\end{align}

\noindent
respectively, where $\xi_A$,$\xi_B$ are the products of the four couplings (with vector, axial couplings removed - note $a_V,a_A = \tfrac{1}{2}$ for W exchange diagrams) for each diagram. 
The DM current is given by

\begin{equation}
    J_{D}^{\mu\nu} = \bar{\mathit{u}}(\text{p})\gamma ^{\mu }(\slashed{p}+\slashed{l}+M_{D^*})\gamma ^{\nu }\mathit{u}(\text{p}) \,,
\end{equation}

\noindent
where $p$ is the DM momentum and $l$ is the loop momentum. The quark currents contained in Eq.~\eqref{eqn:AB_amplitudes} are given by

\begin{align}
    J_{q,A}^{\rho \sigma} &= \bar{\mathit{u}}(q)\gamma ^{\rho }\left(a_V-a_A \gamma_5\right)(\slashed{q}-\slashed{l}+m_Q)\gamma ^{\sigma }\left(a_V-a_A
   \gamma_5\right)\mathit{u}(q) \,, \nonumber \\
    J_{q,B}^{\rho \sigma} &= \bar{\mathit{u}}(q)\gamma ^{\rho }\left(a_V-a_A \gamma_5\right)(\slashed{q}+\slashed{l}+m_Q)\gamma ^{\sigma }\left(a_V-a_A
   \gamma_5\right)\mathit{u}(q) \,,
\end{align}

\noindent
where q is the quark momentum. The denominators are given by
 
\begin{align}
    \mathcal{D}_A &= ((p+l)^2-M_{D^*}^2)(l^2-m_V^2)^2((q-l)^2-m_Q^2)\,, \nonumber \\
    \mathcal{D}_B &= ((p+l)^2-M_{D^*}^2)(l^2-m_V^2)^2((q+l)^2-m_Q^2)\,.
\end{align}

In these expressions, $m_q$($m_Q$) refers to the mass of the external(internal) quarks, $m_V$ is the mass of the vector running in the loop and $M_D*$ is the mass of the DM partner propagating inside the loop (i.e for W exchange diagram, this would be relevant charged DM partner mass). After removing the Lorentz structures not relevant to spin-independent scattering cross sections (terms involving $\gamma_5$ or $\sigma_{\mu \nu}$), these loop amplitudes may be expressed in terms of five Lorentz structures appearing in the numerators

\begin{align}
    \mathcal{N}_A &= J_{D}^{\mu\nu}J_{q,A}^{\rho \sigma} g_{\mu\rho}g_{\nu\sigma} \nonumber \\
    &= -4M_{D^*}m_Q(a_A^2-a_V^2)<\mathds{1}><\mathds{1}>\nonumber \\
    & +2(a_A^2+a_V^2) \left[(p+l).(q-l)<\gamma^\mu><\gamma_\mu> +<\slashed{q}-\slashed{l}> <\slashed{p}+\slashed{l}> \right]\nonumber \\
    & -2M_{D^*}(a_A^2+a_V^2)<\mathds{1}><\slashed{q}-\slashed{l}> +2m_Q(a_A^2-a_V^2)<\slashed{p}+\slashed{l}><\mathds{1}> \,,
\end{align}

\begin{align}
    \mathcal{N}_B &= J_{D}^{\mu\nu}J_{q,B}^{\rho \sigma} g_{\mu\sigma}g_{\nu\rho} \nonumber \\
    &= -4M_{D^*}m_Q(a_A^2-a_V^2)<\mathds{1}><\mathds{1}>\nonumber \\
    & +2(a_A^2+a_V^2) \left[(p+l).(q+l)<\gamma^\mu><\gamma_\mu> +<\slashed{q}+\slashed{l}> <\slashed{p}+\slashed{l}> \right]\nonumber \\
    & -2M_{D^*}(a_A^2+a_V^2)<\mathds{1}><\slashed{q}+\slashed{l}> +2m_Q(a_A^2-a_V^2)<\slashed{p}+\slashed{l}><\mathds{1}>\,,
\end{align}

\noindent
where we use a shorthand for spinors; the first(second) pair of angled brackets, $<\Gamma>$ designate the DM(quark) current $\bar{u}\Gamma u$ for Lorentz stucture $\Gamma$. Next we expand the combined integral around small quark momenta, analogously to Ref.~\cite{Hisano:2011cs}, under the assumption that $m_Q \approx m_q$. We may change basis to be in terms of Twist-2 operators using the identity

\begin{align}
        \bar{q} i\partial^\mu \gamma^\nu q &= \bar{q}\left[ \frac{i\partial^\mu \gamma^\nu + i\partial^\nu \gamma^\mu}{2} -\frac{1}{4}g^{\mu\nu}i\slashed{\partial}\right]q + \bar{q}\left[ \frac{i\partial^\mu \gamma^\nu - i\partial^\nu \gamma^\mu}{2}\right]q + \frac{1}{4}g^{\mu\nu}\bar{q}i\slashed{\partial}q \nonumber \\
        &=\mathcal{O}_{\mu\nu}^q + \frac{1}{4}g^{\mu\nu}m_q\bar{q}q \,,
\end{align}
 \noindent
where we used irreducible decomposition of the quark current and in last line the antisymmetric piece is dropped as it does not contribute to the nuclear matrix element. Combining amplitudes for diagrams A and B ($\xi \equiv \xi_A=\xi_B$), we arrive to the following expression

\begin{align} \label{eq:MAB}
\frac{i\mathcal{M}_{A+B}}{\xi} (4\pi)^2 & = \zeta_1 <\mathds{1}><\mathds{1}> + \zeta_2 <p^\mu p^\nu> <\gamma_\mu q_\nu> + \zeta_3 <\gamma^\mu p^\nu> <\gamma_\nu p_\mu> + \zeta_4 <\gamma^\mu p^\nu> <\gamma_\mu q_\nu > \nonumber \\
& = \left( \zeta_1 + \frac{\zeta_2 m_q M_{DM}^2}{4} + \frac{(\zeta_3+\zeta_4)m_q M_{DM}}{4} \right) <\mathds{1}><\mathds{1}> + \left(\zeta_3+\zeta_4\right) <\gamma^\mu p^\nu>[\mathcal{O}_{\mu \nu}] \nonumber \\
& \quad + \zeta_2 <p^\mu p^\nu>[\mathcal{O}_{\mu \nu}] \nonumber \\
& \equiv \frac{m_q\Delta_S(x,y,a_V,a_A)}{m_V^3} <\mathds{1}><\mathds{1}> + \frac{\Delta_{T1}(x,y,a_V,a_A)}{M_{DM} m_V^3} <\gamma^\mu p^\nu>[\mathcal{O}_{\mu \nu}] \nonumber \\
& \quad + \frac{\Delta_{T2}(x,y,a_V,a_A)}{M_{DM}^2 m_V^3} <p^\mu p^\nu>[\mathcal{O}_{\mu \nu}] \,,
\end{align}

\noindent
where the loop functions $\Delta_S$, $\Delta_{T1}$, $\Delta_{T2}$ are given below and depend on dimensionless quantities $x \equiv m_V^2/M_{DM}^2$ and $y \equiv (M_{D*}-M_{DM})/M_{DM}$.
The contribution from the triangle diagram (C in Fig.~\ref{fig:triLoop}), again with couplings extracted as $\xi$ is given by

\begin{equation} \label{eq:MC}
\frac{i\mathcal{M}_C}{\xi} (4\pi)^2 = \Delta_H(x,y)<\mathds{1}><\mathds{1}> \,,
\end{equation}
where $\Delta_H$ is given below. 

The full expressions for the loop functions appearing in Eqs.~\eqref{eq:MAB} and \eqref{eq:MC} are given by

\begin{align}\label{eqn:DeltaLoopFunctions}
\Delta_H(x,y) &= \sqrt{x} \Bigg[\frac{2 (b^2-2 (y+1) (-x+y^2+y+1)) \log \Big(\frac{b+x+c}{2 \sqrt{x} (y+1)}\Big)}{b}+(x-y^2) \log (\frac{(y+1)^2}{x})+2 \Bigg]\,, \nonumber \\
\Delta_S(x,y,a_V,a_A) &= \frac{1}{b c x^{3/2}} \Bigg[ 2 y (y+2) \log (\frac{b+c+x}{2 \sqrt{x} (y+1)}) (a_A^2 (b^4+b^2 x (c-2 x+5 y+7) \nonumber \\
& + x^2 (y+1) (5 c-5 x-2 y+8))+a_V^2 (b^4+3 b^2 x (y^2+y+1)+3 x^2 (y+1) (x-y^2))) \nonumber \\
& +2 b \log (\frac{1}{c}+1) (a_A^2 (c^4-2 c^3 x+c^2 (5 x y+x)+3 x^2 (2 y+1))+a_V^2 (c^4-3 c^2 x (y+1) \nonumber \\
& -x^2 (2 y+1)))+b \log (\frac{x}{(y+1)^2}) (a_A^2 (c^4-2 c^3 x+c^2 (5 x y+x)+c x^3+6 x^2 (2 y+1)) \nonumber \\
& +a_V^2 (c^4-3 c^2 x (y+1)-c x^3-2 x^2 (2 y+1)))-2 b x^2 (2 y+1) (3 a_A^2-a_V^2) \log (\frac{x}{c}) \nonumber \\
& +2 b c x (a_A^2 (c-x)+a_V^2 (c+x)) \Bigg]\,, \nonumber \\
\Delta_{T1}(x,y,a_V,a_A) &=\frac{2 (a_A^2+a_V^2)}{3 \sqrt{x}} \Bigg[-\frac{2 b (b^2 (c+x-2)+6 (x-1) x) \log (\frac{b+c+x}{2 \sqrt{x} (y+1)})}{x} \nonumber \\
& -\frac{(b^4 (c+2 x-2)-2 b^2 x (-(c+6) x+x^2+5)+2 x^2 (c (x-2)-x^2+5 x+2)) \log (\frac{x}{(y+1)^2})}{c x} \nonumber \\
& -\frac{2 (c^5-2 c^4 (x+1)+6 c^2 x+6 x^2) \log (\frac{1}{c}+1)}{c x}+\frac{12 x \log (\frac{x}{c})}{c}-2 x^2+2 x y^2+4 x y+x \nonumber \\
& -2 y^4-8 y^3-4 y^2+8 y\Bigg]\,, \nonumber \\
\Delta_{T2}(x,y,a_V,a_A) &= \frac{2 (a_A^2+a_V^2)}{3 x^{3/2}} \Bigg[ (x (6 b^2+x (6 c+4 y+21)-8 c y) \nonumber \\
& +\frac{2 (b^4 (3 c+3 x-4 y)+6 b^2 x (x (y+2)-y^3+y)+6 x^2 (y+1) (x-y^2)) \log (\frac{b+c+x}{2 \sqrt{x} (y+1)})}{b} \nonumber \\
& -(c^3 (4 y-3 c)+2 x^3 (-3 c+y-3)+6 c (c+1) x (c-y)+3 x^4) \log (\frac{x}{(y+1)^2}) \nonumber \\
& +2 c \log (\frac{1}{c}+1) (3 c^3-2 c^2 (3 x+2 y)+6 c x (y-1)+6 x y) \Bigg]\,,
\end{align}
where we have made the convenient substitutions $b=\sqrt{x^2-2x(y^2+2y+2+y^2(y+2)^2}$ and $c=y(y+2)$.  Their derivation using Package-X \cite{Patel:2015tea} along with the implementation into a c library which computes the total cross-sections are given as supporting material at \cite{locke_daniel_2022_6308438}.

Taking the limit that the internal DM partner mass $M_{D*}$ equals the DM mass $M_{DM}$ (or $y \to 0$), we recover the result of \cite{Hisano:2011cs}

\begin{align}
\lim_{y\to 0} \Delta_H(x,y) &= g_H(x)/2\,, \nonumber \\
\lim_{y \to 0} \Delta_S(x,y,a_V,a_A) &=4(a_V^2-a_A^2) g_S(x)\,, \nonumber \\
\lim_{y \to 0} \Delta_{T1}(x,y,a_V,a_A) &= 8(a_A^2+a_V^2) g_{T1}(x)\,, \nonumber \\
\lim_{y \to 0} \Delta_{T2}(x,y,a_V,a_A) &= 8(a_A^2+a_V^2) g_{T2}(x)\,,
\end{align}

\noindent
where the loop functions without DM and partner mass splits ($g_i$) match those of \cite{Hisano:2011cs} and are given by

\begin{align}
	g_H(x) &= -\frac{2}{b_x} (2 + 2 x - x^2)\arctan{\frac{2b_x}{\sqrt{x}}} + 2 \sqrt{x} (2 - x \log{x})\,, \nonumber \\
    g_S(x) &= \frac{1}{4 b_x}(x^2-2x+4)\arctan{\frac{2b_x}{\sqrt{x}}} +\frac{1}{4}\sqrt{x}(2-x\log{x})\,,  \nonumber \\
    g_{T1}(x)&=\frac{1}{3}b_x(2+x^{2})\arctan{\frac{2b_x}{\sqrt{x}}}+\frac{1}{12}\sqrt{x}(1-2x-x(2-x)\log{x})]\,,  \nonumber \\
    g_{T2}(x) &= \frac{1}{4b_x}x(x^2-4x+2)\arctan{\frac{2b_x}{\sqrt{x}}} - \frac{1}{4}\sqrt{x}\left(1-2x-x(2-x)\log{x}  \right] \,,
\end{align}

\noindent
where $b_x \equiv \sqrt{1-x/4}$.



\subsection{Loop induced h-$\psi$-$\psi$ coupling in $\tilde F^0_0S^0_0$(CP-odd) representative model} \label{App:FDM+a_DDloop}
Here we present details of the calculation the loop which induces h-$\psi$-$\psi$ coupling, key to the phenomenology of the model studied in Section~\ref{sec:new-model-pheno}.

\begin{figure}[H]
	\centering
	\includegraphics[width=0.5\textwidth]{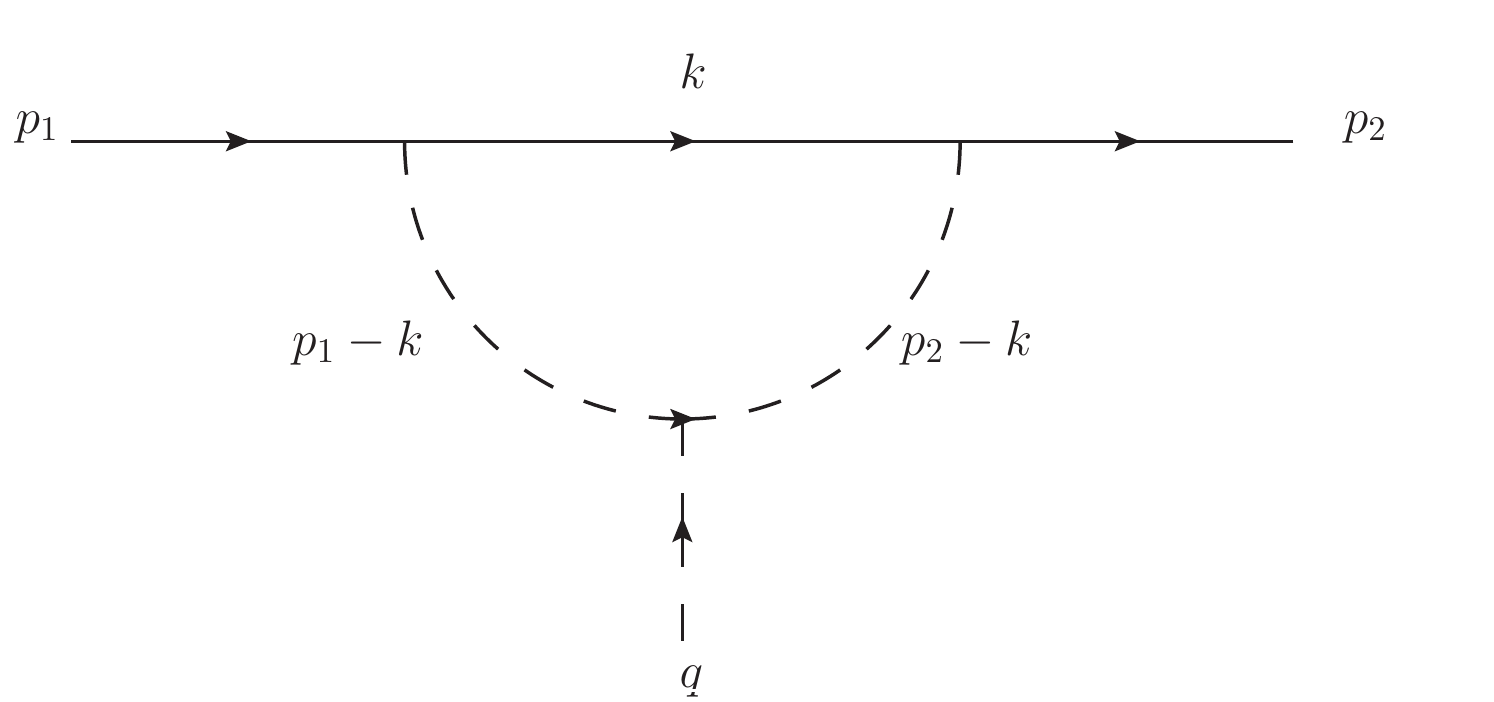}
	\caption{Feynman diagram for loop induced  h-$\psi$-$\psi$ coupling. Internal scalar lines are from propagating pseudo-scalar, $a$. \label{fig:FDM+a_loop}}
\end{figure}

The loop-induced h-$\psi$-$\psi$ coupling is generated by a loop containing the pseudo-scalar, as shown in Fig.~\ref{fig:FDM+a_loop}. This loop must be evaluated at three different scales: for DM direct detection the external Higgs momentum squared is $q^2=t\approx0$; for Higgs invisible decays to a pair of DM fermions, $q^2=s=m_H^2$; finally, for the relic abundance computation, the scale varies with the temperature, $q^2=s=4m_\psi^2 (1+\tfrac{1}{2x})$ where $x\approx 20$ around freeze-out temperature (in this region the loop factor is relatively insensitive to x). It is useful to define the quantities

\begin{align}
\Upsilon_{DD} &\equiv \Upsilon(s=0)\,, \nonumber \\
\Upsilon_{Relic} &\equiv \Upsilon(s=4m_\psi^2(1+1/(2x))\,, \nonumber \\
\Upsilon_{H\to\psi\psi} &\equiv \Upsilon(s=m_H^2)\,.
\end{align}

\noindent
The amplitude for this diagram is given by.

\begin{align}
\Upsilon(m_a, m_\psi, s) \equiv \left( \frac{ie^{-\gamma_E \epsilon}}{(4\pi)^{d/2}} \right)^{-1} \mu^{2\epsilon} & \int \frac{d^d k}{(2\pi)^d} \frac{(\slashed{k}+m_\psi)}{(k^2-m_\psi^2)((p_1-k)^2-m_a^2)((p_2-k)^2-m_a^2)} \nonumber \\
&= m_\psi \big[ C_0(m_\psi^2,s,m_\psi^2,m_\psi,m_a,m_a) + \\ \nonumber
& \quad C_1(m_\psi^2,s,m_\psi^2,m_\psi,m_a,m_a) + C_2(m_\psi^2,s,m_\psi^2,m_\psi,m_a,m_a)\big] \,,
\end{align}

\noindent
where $p_1$($p_2$) is the incoming(outgoing) momentum of fermion $\psi$ and q is the incoming momentum of the Higgs boson. The functions $C_i$ correspond to the 3-point Passarino-Veltman integrals \cite{Passarino:1978jh,Hahn:2000jm}. The derivation of this function using Package-X is provided at \cite{locke_daniel_2022_6308438} and a c library used by a LANHEP implementation of this model are given at \cite{hepmdb:FDM+a}. 

\begin{figure}[H]
	\centering
	\includegraphics[width=1.0\textwidth]{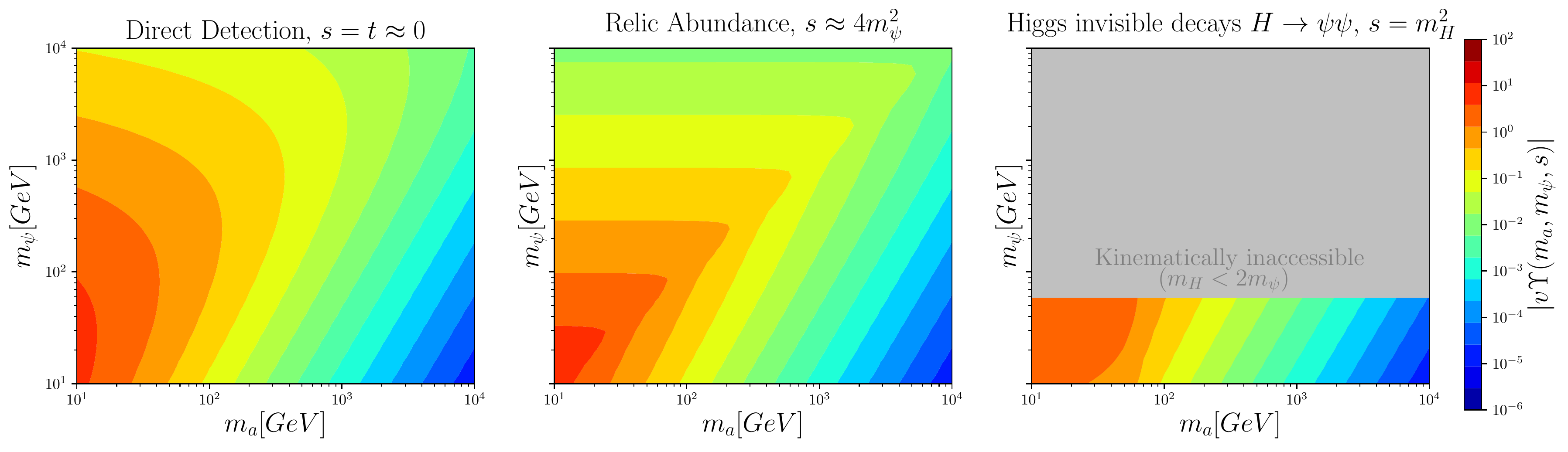}
	\caption{Dimensionless loop function $v\Upsilon(m_a, m_\psi, s)$ (where $v$ is the SM Higgs vacuum expectation value) for the three scales of interest to the phenomenology. \label{fig:FDM+a_LoopContours}}
\end{figure}

Fig.~\ref{fig:FDM+a_LoopContours} demonstrates the dimensionless function entering the one-loop result, $v\Upsilon (m_a, m_\psi, s)$ (where $v$ is SM Higgs vacuum expectation value) in the $(m_a, m_\psi)$ plane, evaluated at the three scales relevant for the phenomenology of this model; the direct detection scale (left), the scale around freeze-out of $\psi$ (centre) and the scale of the Higgs boson decay into $\bar{\psi} \psi$ (right).
This loop amplitude gives rise to an effective Yukawa couplings

\begin{align}\label{eqn:deltaY}
\delta Y = \frac{-\lambda_{aH}Y_\psi^2}{32\pi^2}  v \Upsilon(m_a, m_\psi, s) \,.
\end{align}

\begin{figure}[htb]
\centering
\includegraphics[width=\textwidth]{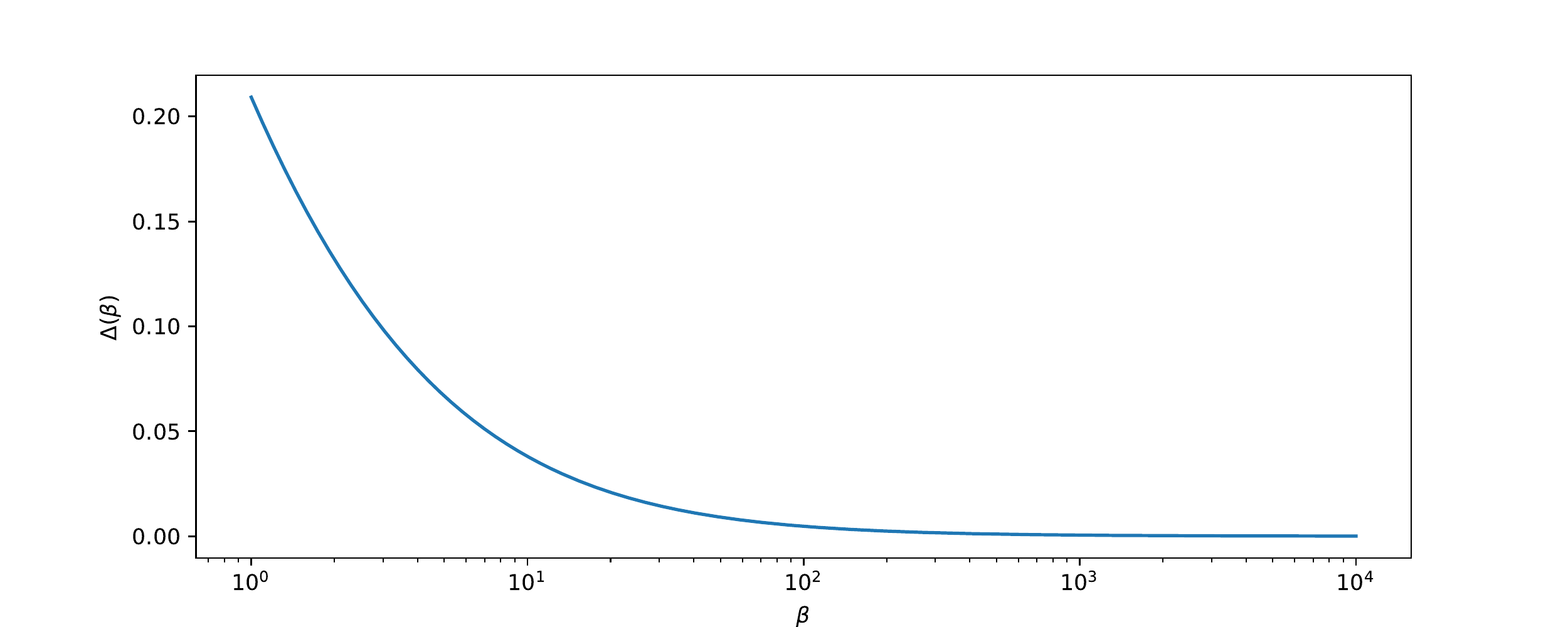}
\caption{Dimensionless loop function $\Delta$ as a function of $\beta$. \label{fig:Delta}}
\end{figure}

In the limit where $s \to 0$ as is relevant for DM direct detection, a compact expression may be found, 

\begin{align}
\Delta(\beta) & \equiv m_\psi \Upsilon(s=0) \nonumber\\
&= \frac{(\beta -4) (\beta -1) \log (\beta )-2 \left(\beta +(\beta -3) \sqrt{(\beta -4) \beta } \log \left(\frac{\beta +\sqrt{(\beta -4) \beta }}{2\sqrt{\beta }}\right)-4\right)}{2 (\beta -4)}\,,
\end{align}

\noindent
which is a function of a single variable, $\beta \equiv m_a^2 / m_\psi^2$. The value of this function over the range of $\beta$ is presented in Fig.~\ref{fig:Delta}.

%
%
%

%% file: fermion_consistent_DM.bbl
\providecommand{\href}[2]{#2}\begingroup\raggedright\begin{thebibliography}{100}

\bibitem{Bertone:2018krk}
G.~Bertone and T.~Tait, M.~P., {\it {A new era in the search for dark matter}},
   {\em Nature} {\bf 562} (2018), no.~7725 51--56
  [\href{http://arXiv.org/abs/1810.01668}{{\tt 1810.01668}}].

\bibitem{WMAP:2010qai}
{\bf WMAP} Collaboration, E.~Komatsu {\em et.~al.}, {\it {Seven-Year Wilkinson
  Microwave Anisotropy Probe (WMAP) Observations: Cosmological
  Interpretation}},  {\em Astrophys. J. Suppl.} {\bf 192} (2011) 18
  [\href{http://arXiv.org/abs/1001.4538}{{\tt 1001.4538}}].

\bibitem{Adam:2015rua}
{\bf Planck} Collaboration, R.~Adam {\em et.~al.}, {\it {Planck 2015 results.
  I. Overview of products and scientific results}},  {\em Astron. Astrophys.}
  {\bf 594} (2016) A1 [\href{http://arXiv.org/abs/1502.01582}{{\tt
  1502.01582}}].

\bibitem{Planck:2018vyg}
{\bf Planck} Collaboration, N.~Aghanim {\em et.~al.}, {\it {Planck 2018
  results. VI. Cosmological parameters}},  {\em Astron. Astrophys.} {\bf 641}
  (2020) A6 [\href{http://arXiv.org/abs/1807.06209}{{\tt 1807.06209}}].
  [Erratum: Astron.Astrophys. 652, C4 (2021)].

\bibitem{Aprile:2018dbl}
{\bf XENON} Collaboration, E.~Aprile {\em et.~al.}, {\it {Dark Matter Search
  Results from a One Tonne$\times$Year Exposure of XENON1T}},
  \href{http://arXiv.org/abs/1805.12562}{{\tt 1805.12562}}.

\bibitem{PandaX-4T:2021bab}
{\bf PandaX-4T} Collaboration, Y.~Meng {\em et.~al.}, {\it {Dark Matter Search
  Results from the PandaX-4T Commissioning Run}},  {\em Phys. Rev. Lett.} {\bf
  127} (2021), no.~26 261802 [\href{http://arXiv.org/abs/2107.13438}{{\tt
  2107.13438}}].

\bibitem{Akerib:2018lyp}
{\bf LUX-ZEPLIN} Collaboration, D.~Akerib {\em et.~al.}, {\it {Projected WIMP
  sensitivity of the LUX-ZEPLIN dark matter experiment}},  {\em Phys. Rev. D}
  {\bf 101} (2020), no.~5 052002 [\href{http://arXiv.org/abs/1802.06039}{{\tt
  1802.06039}}].

\bibitem{Goodman:2010ku}
J.~Goodman, M.~Ibe, A.~Rajaraman, W.~Shepherd, T.~M. Tait {\em et.~al.}, {\it
  {Constraints on Dark Matter from Colliders}},  {\em Phys.Rev.} {\bf D82}
  (2010) 116010 [\href{http://arXiv.org/abs/1008.1783}{{\tt 1008.1783}}].

\bibitem{Aad:2012fw}
{\bf ATLAS} Collaboration, G.~Aad {\em et.~al.}, {\it {Search for dark matter
  candidates and large extra dimensions in events with a photon and missing
  transverse momentum in $pp$ collision data at $\sqrt{s}=7$ TeV with the ATLAS
  detector}},  {\em Phys.Rev.Lett.} {\bf 110} (2013), no.~1 011802
  [\href{http://arXiv.org/abs/1209.4625}{{\tt 1209.4625}}].

\bibitem{Aad:2013oja}
{\bf ATLAS} Collaboration, G.~Aad {\em et.~al.}, {\it {Search for dark matter
  in events with a hadronically decaying W or Z boson and missing transverse
  momentum in $pp$ collisions at $\sqrt{s} =$ 8 TeV with the ATLAS detector}},
  {\em Phys.Rev.Lett.} {\bf 112} (2014), no.~4 041802
  [\href{http://arXiv.org/abs/1309.4017}{{\tt 1309.4017}}].

\bibitem{Aad:2014vka}
{\bf ATLAS} Collaboration, G.~Aad {\em et.~al.}, {\it {Search for dark matter
  in events with a Z boson and missing transverse momentum in pp collisions at
  $\sqrt{s}$=8 TeV with the ATLAS detector}},  {\em Phys.Rev.} {\bf D90}
  (2014), no.~1 012004 [\href{http://arXiv.org/abs/1404.0051}{{\tt
  1404.0051}}].

\bibitem{ATLAS:2014wra}
{\bf ATLAS} Collaboration, G.~Aad {\em et.~al.}, {\it {Search for new particles
  in events with one lepton and missing transverse momentum in $pp$ collisions
  at $\sqrt{s}$ = 8 TeV with the ATLAS detector}},  {\em JHEP} {\bf 1409}
  (2014) 037 [\href{http://arXiv.org/abs/1407.7494}{{\tt 1407.7494}}].

\bibitem{Khachatryan:2014tva}
{\bf CMS} Collaboration, V.~Khachatryan {\em et.~al.}, {\it {Search for physics
  beyond the standard model in final states with a lepton and missing
  transverse energy in proton-proton collisions at $\sqrt{s}$ = 8 TeV}},  {\em
  Phys.Rev.} {\bf D91} (2015), no.~9 092005
  [\href{http://arXiv.org/abs/1408.2745}{{\tt 1408.2745}}].

\bibitem{Khachatryan:2015nua}
{\bf CMS} Collaboration, V.~Khachatryan {\em et.~al.}, {\it {Search for the
  production of dark matter in association with top-quark pairs in the
  single-lepton final state in proton-proton collisions at sqrt(s) = 8 TeV}},
  \href{http://arXiv.org/abs/1504.03198}{{\tt 1504.03198}}.

\bibitem{Buchmueller:2013dya}
O.~Buchmueller, M.~J. Dolan and C.~McCabe, {\it {Beyond Effective Field Theory
  for Dark Matter Searches at the LHC}},  {\em JHEP} {\bf 1401} (2014) 025
  [\href{http://arXiv.org/abs/1308.6799}{{\tt 1308.6799}}].

\bibitem{Cheung:2013dua}
C.~Cheung and D.~Sanford, {\it {Simplified Models of Mixed Dark Matter}},  {\em
  JCAP} {\bf 1402} (2014) 011 [\href{http://arXiv.org/abs/1311.5896}{{\tt
  1311.5896}}].

\bibitem{Dutta:2014kia}
B.~Dutta, Y.~Gao and T.~Kamon, {\it {Probing Light Nonthermal Dark Matter at
  the LHC}},  {\em Phys.Rev.} {\bf D89} (2014), no.~9 096009
  [\href{http://arXiv.org/abs/1401.1825}{{\tt 1401.1825}}].

\bibitem{Busoni:2014sya}
G.~Busoni, A.~De~Simone, J.~Gramling, E.~Morgante and A.~Riotto, {\it {On the
  Validity of the Effective Field Theory for Dark Matter Searches at the LHC,
  Part II: Complete Analysis for the $s$-channel}},  {\em JCAP} {\bf 1406}
  (2014) 060 [\href{http://arXiv.org/abs/1402.1275}{{\tt 1402.1275}}].

\bibitem{Papucci:2014iwa}
M.~Papucci, A.~Vichi and K.~M. Zurek, {\it {Monojet versus the rest of the
  world I: t-channel models}},  {\em JHEP} {\bf 1411} (2014) 024
  [\href{http://arXiv.org/abs/1402.2285}{{\tt 1402.2285}}].

\bibitem{Bai:2014osa}
Y.~Bai and J.~Berger, {\it {Lepton Portal Dark Matter}},  {\em JHEP} {\bf 1408}
  (2014) 153 [\href{http://arXiv.org/abs/1402.6696}{{\tt 1402.6696}}].

\bibitem{Berlin:2014cfa}
A.~Berlin, T.~Lin and L.-T. Wang, {\it {Mono-Higgs Detection of Dark Matter at
  the LHC}},  {\em JHEP} {\bf 1406} (2014) 078
  [\href{http://arXiv.org/abs/1402.7074}{{\tt 1402.7074}}].

\bibitem{Hamaguchi:2014pja}
K.~Hamaguchi, S.~P. Liew, T.~Moroi and Y.~Yamamoto, {\it {Isospin-Violating
  Dark Matter with Colored Mediators}},  {\em JHEP} {\bf 1405} (2014) 086
  [\href{http://arXiv.org/abs/1403.0324}{{\tt 1403.0324}}].

\bibitem{Busoni:2014haa}
G.~Busoni, A.~De~Simone, T.~Jacques, E.~Morgante and A.~Riotto, {\it {On the
  Validity of the Effective Field Theory for Dark Matter Searches at the LHC
  Part III: Analysis for the $t$-channel}},  {\em JCAP} {\bf 1409} (2014) 022
  [\href{http://arXiv.org/abs/1405.3101}{{\tt 1405.3101}}].

\bibitem{Balazs:2014jla}
C.~Bal\'azs and T.~Li, {\it {Simplified Dark Matter Models Confront the Gamma
  Ray Excess}},  {\em Phys.Rev.} {\bf D90} (2014), no.~5 055026
  [\href{http://arXiv.org/abs/1407.0174}{{\tt 1407.0174}}].

\bibitem{Buchmueller:2014yoa}
O.~Buchmueller, M.~J. Dolan, S.~A. Malik and C.~McCabe, {\it {Characterising
  dark matter searches at colliders and direct detection experiments: Vector
  mediators}},  {\em JHEP} {\bf 1501} (2015) 037
  [\href{http://arXiv.org/abs/1407.8257}{{\tt 1407.8257}}].

\bibitem{Abdallah:2014hon}
J.~Abdallah, A.~Ashkenazi, A.~Boveia, G.~Busoni, A.~De~Simone {\em et.~al.},
  {\it {Simplified Models for Dark Matter and Missing Energy Searches at the
  LHC}},  \href{http://arXiv.org/abs/1409.2893}{{\tt 1409.2893}}.

\bibitem{Harris:2014hga}
P.~Harris, V.~V. Khoze, M.~Spannowsky and C.~Williams, {\it {Constraining Dark
  Sectors at Colliders: Beyond the Effective Theory Approach}},  {\em
  Phys.Rev.} {\bf D91} (2015), no.~5 055009
  [\href{http://arXiv.org/abs/1411.0535}{{\tt 1411.0535}}].

\bibitem{Racco:2015dxa}
D.~Racco, A.~Wulzer and F.~Zwirner, {\it {Robust collider limits on
  heavy-mediator Dark Matter}},  {\em JHEP} {\bf 1505} (2015) 009
  [\href{http://arXiv.org/abs/1502.04701}{{\tt 1502.04701}}].

\bibitem{Jacques:2015zha}
T.~Jacques and K.~Nordstrom, {\it {Mapping monojet constraints onto Simplified
  Dark Matter Models}},  \href{http://arXiv.org/abs/1502.05721}{{\tt
  1502.05721}}.

\bibitem{Khachatryan:2014rra}
{\bf CMS} Collaboration, V.~Khachatryan {\em et.~al.}, {\it {Search for dark
  matter, extra dimensions, and unparticles in monojet events in proton-proton
  collisions at $\sqrt{s}$ = 8 TeV}},
  \href{http://arXiv.org/abs/1408.3583}{{\tt 1408.3583}}.

\bibitem{Aad:2014vea}
{\bf ATLAS} Collaboration, G.~Aad {\em et.~al.}, {\it {Search for dark matter
  in events with heavy quarks and missing transverse momentum in $pp$
  collisions with the ATLAS detector}},  {\em Eur.Phys.J.} {\bf C75} (2015),
  no.~2 92 [\href{http://arXiv.org/abs/1410.4031}{{\tt 1410.4031}}].

\bibitem{Khachatryan:2014rwa}
{\bf CMS} Collaboration, V.~Khachatryan {\em et.~al.}, {\it {Search for new
  phenomena in monophoton final states in proton-proton collisions at
  $\sqrt{s}$ = 8 TeV}},  \href{http://arXiv.org/abs/1410.8812}{{\tt
  1410.8812}}.

\bibitem{Aad:2014tda}
{\bf ATLAS} Collaboration, G.~Aad {\em et.~al.}, {\it {Search for new phenomena
  in events with a photon and missing transverse momentum in $pp$ collisions at
  $\sqrt{s}=8$ TeV with the ATLAS detector}},  {\em Phys.Rev.} {\bf D91}
  (2015), no.~1 012008 [\href{http://arXiv.org/abs/1411.1559}{{\tt
  1411.1559}}].

\bibitem{Aad:2015zva}
{\bf ATLAS} Collaboration, G.~Aad {\em et.~al.}, {\it {Search for new phenomena
  in final states with an energetic jet and large missing transverse momentum
  in pp collisions at $\sqrt{s}=8$ TeV with the ATLAS detector}},
  \href{http://arXiv.org/abs/1502.01518}{{\tt 1502.01518}}.

\bibitem{ATLAS:2021fjm}
{\bf ATLAS} Collaboration, {\it {Combination and summary of ATLAS dark matter
  searches using 139 fb$^{-1}$ of $\sqrt {s}$ = 13 TeV $p p$ collision data and
  interpreted in a two-Higgs-doublet model with a pseudoscalar mediator}}, .

\bibitem{ATLAS:2020uiq}
{\bf ATLAS} Collaboration, G.~Aad {\em et.~al.}, {\it {Search for dark matter
  in association with an energetic photon in $pp$ collisions at $\sqrt{s}$ = 13
  TeV with the ATLAS detector}},  {\em JHEP} {\bf 02} (2021) 226
  [\href{http://arXiv.org/abs/2011.05259}{{\tt 2011.05259}}].

\bibitem{CMS:2020ulv}
{\bf CMS} Collaboration, A.~M. Sirunyan {\em et.~al.}, {\it {Search for dark
  matter produced in association with a leptonically decaying Z boson in
  proton-proton collisions at $\sqrt{s} =$ 13 TeV}},  {\em Eur. Phys. J. C}
  {\bf 81} (2021), no.~1 13 [\href{http://arXiv.org/abs/2008.04735}{{\tt
  2008.04735}}]. [Erratum: Eur.Phys.J.C 81, 333 (2021)].

\bibitem{CMS:2019ykj}
{\bf CMS} Collaboration, A.~M. Sirunyan {\em et.~al.}, {\it {Search for dark
  matter particles produced in association with a Higgs boson in proton-proton
  collisions at $ \sqrt{\mathrm{s}} $ = 13 TeV}},  {\em JHEP} {\bf 03} (2020)
  025 [\href{http://arXiv.org/abs/1908.01713}{{\tt 1908.01713}}].

\bibitem{CMS:2018ysw}
{\bf CMS} Collaboration, A.~M. Sirunyan {\em et.~al.}, {\it {Search for dark
  matter particles produced in association with a top quark pair at $\sqrt{s}
  =$ 13 TeV}},  {\em Phys. Rev. Lett.} {\bf 122} (2019), no.~1 011803
  [\href{http://arXiv.org/abs/1807.06522}{{\tt 1807.06522}}].

\bibitem{CMS:2018nlv}
{\bf CMS} Collaboration, A.~M. Sirunyan {\em et.~al.}, {\it {Search for dark
  matter produced in association with a Higgs boson decaying to $\gamma\gamma$
  or $\tau^+\tau^-$ at $\sqrt{s} =$ 13 TeV}},  {\em JHEP} {\bf 09} (2018) 046
  [\href{http://arXiv.org/abs/1806.04771}{{\tt 1806.04771}}].

\bibitem{Kahlhoefer:2015bea}
F.~Kahlhoefer, K.~Schmidt-Hoberg, T.~Schwetz and S.~Vogl, {\it {Implications of
  unitarity and gauge invariance for simplified dark matter models}},  {\em
  JHEP} {\bf 02} (2016) 016 [\href{http://arXiv.org/abs/1510.02110}{{\tt
  1510.02110}}].

\bibitem{Goncalves:2016iyg}
D.~Goncalves, P.~A.~N. Machado and J.~M. No, {\it {Simplified Models for Dark
  Matter Face their Consistent Completions}},  {\em Phys. Rev. D} {\bf 95}
  (2017), no.~5 055027 [\href{http://arXiv.org/abs/1611.04593}{{\tt
  1611.04593}}].

\bibitem{Duerr:2016tmh}
M.~Duerr, F.~Kahlhoefer, K.~Schmidt-Hoberg, T.~Schwetz and S.~Vogl, {\it {How
  to save the WIMP: global analysis of a dark matter model with two s-channel
  mediators}},  {\em JHEP} {\bf 09} (2016) 042
  [\href{http://arXiv.org/abs/1606.07609}{{\tt 1606.07609}}].

\bibitem{Deshpande:1977rw}
N.~G. Deshpande and E.~Ma, {\it {Pattern of Symmetry Breaking with Two Higgs
  Doublets}},  {\em Phys.Rev.} {\bf D18} (1978) 2574.

\bibitem{Cirelli:2005uq}
M.~Cirelli, N.~Fornengo and A.~Strumia, {\it {Minimal dark matter}},  {\em
  Nucl.Phys.} {\bf B753} (2006) 178--194
  [\href{http://arXiv.org/abs/hep-ph/0512090}{{\tt hep-ph/0512090}}].

\bibitem{Hambye:2009pw}
T.~Hambye, F.-S. Ling, L.~Lopez~Honorez and J.~Rocher, {\it {Scalar Multiplet
  Dark Matter}},  {\em JHEP} {\bf 0907} (2009) 090
  [\href{http://arXiv.org/abs/0903.4010}{{\tt 0903.4010}}].

\bibitem{Ding:2012sm}
R.~Ding and Y.~Liao, {\it {Spin 3/2 Particle as a Dark Matter Candidate: an
  Effective Field Theory Approach}},  {\em JHEP} {\bf 04} (2012) 054
  [\href{http://arXiv.org/abs/1201.0506}{{\tt 1201.0506}}].

\bibitem{Khojali:2016pvu}
M.~O. Khojali, A.~Goyal, M.~Kumar and A.~S. Cornell, {\it {Minimal Spin-3/2
  Dark Matter in a simple $s$-channel model}},  {\em Eur. Phys. J.} {\bf C77}
  (2017), no.~1 25 [\href{http://arXiv.org/abs/1608.08958}{{\tt 1608.08958}}].

\bibitem{Khojali:2017tuv}
M.~O. Khojali, A.~Goyal, M.~Kumar and A.~S. Cornell, {\it {Spin-3/2 Dark Matter
  in a simple $t$-channel model}},  {\em Eur. Phys. J.} {\bf C78} (2018),
  no.~11 920 [\href{http://arXiv.org/abs/1705.05149}{{\tt 1705.05149}}].

\bibitem{Asorey:2010zz}
M.~Asorey and D.~Garc{\'i}a-{\'A}lvarez, {\it {Higher spin dark matter}},  {\em
  AIP Conf. Proc.} {\bf 1241} (2010), no.~1 1192--1197.

\bibitem{Weinberg:1979sa}
S.~Weinberg, {\it {Baryon and Lepton Nonconserving Processes}},  {\em Phys.
  Rev. Lett.} {\bf 43} (1979) 1566--1570.

\bibitem{Essig:2007az}
R.~Essig, {\it {Direct Detection of Non-Chiral Dark Matter}},  {\em Phys. Rev.
  D} {\bf 78} (2008) 015004 [\href{http://arXiv.org/abs/0710.1668}{{\tt
  0710.1668}}].

\bibitem{Hisano:2011cs}
J.~Hisano, K.~Ishiwata, N.~Nagata and T.~Takesako, {\it {Direct Detection of
  Electroweak-Interacting Dark Matter}},  {\em JHEP} {\bf 07} (2011) 005
  [\href{http://arXiv.org/abs/1104.0228}{{\tt 1104.0228}}].

\bibitem{Billard:2013qya}
J.~Billard, L.~Strigari and E.~Figueroa-Feliciano, {\it {Implication of
  neutrino backgrounds on the reach of next generation dark matter direct
  detection experiments}},  {\em Phys. Rev. D} {\bf 89} (2014), no.~2 023524
  [\href{http://arXiv.org/abs/1307.5458}{{\tt 1307.5458}}].

\bibitem{PhenoData:PandaX-4T}
A.~Belyaev, G.~Cacciapaglia and D.~Locke, ``Phenodata, digitised pandax-4t
  wimp-nucleon cross section limit (figure. 4 from arxiv:2107.13438).''
  \url{hepmdb.soton.ac.uk/phenodata?p=2202.00132}, Feb, 2022.

\bibitem{PhenoData:LZ}
A.~Belyaev, G.~Cacciapaglia and D.~Locke, ``Phenodata, digitised lux-zeplin
  wimp-nucleon cross section limit (figure. 8 from arxiv:1802.06039).''
  \url{hepmdb.soton.ac.uk/phenodata?p=2202.00131}, Feb, 2022.

\bibitem{PhenoData:NeutrinoFloor}
A.~Belyaev, G.~Cacciapaglia and D.~Locke, ``Phenodata, digitised neutrino floor
  for wimp-nucleon cross section measurements, (figure. 5 (left) from
  arxiv:1307.5458).'' \url{hepmdb.soton.ac.uk/phenodata?p=2202.00133}, Feb,
  2022.

\bibitem{Ellis:2018dmb}
J.~Ellis, N.~Nagata and K.~A. Olive, {\it {Uncertainties in WIMP Dark Matter
  Scattering Revisited}},  {\em Eur. Phys. J. C} {\bf 78} (2018), no.~7 569
  [\href{http://arXiv.org/abs/1805.09795}{{\tt 1805.09795}}].

\bibitem{Hou:2019efy}
T.-J. Hou {\em et.~al.}, {\it {New CTEQ global analysis of quantum
  chromodynamics with high-precision data from the LHC}},  {\em Phys. Rev. D}
  {\bf 103} (2021), no.~1 014013 [\href{http://arXiv.org/abs/1912.10053}{{\tt
  1912.10053}}].

\bibitem{Buckley:2014ana}
A.~Buckley, J.~Ferrando, S.~Lloyd, K.~Nordstr\"om, B.~Page, M.~R\"ufenacht,
  M.~Sch\"onherr and G.~Watt, {\it {LHAPDF6: parton density access in the LHC
  precision era}},  {\em Eur. Phys. J. C} {\bf 75} (2015) 132
  [\href{http://arXiv.org/abs/1412.7420}{{\tt 1412.7420}}].

\bibitem{DiFranzo:2013vra}
A.~DiFranzo, K.~I. Nagao, A.~Rajaraman and T.~M.~P. Tait, {\it {Simplified
  Models for Dark Matter Interacting with Quarks}},  {\em JHEP} {\bf 11} (2013)
  014 [\href{http://arXiv.org/abs/1308.2679}{{\tt 1308.2679}}]. [Erratum:
  JHEP01,162(2014)].

\bibitem{Buckley:2014fba}
M.~R. Buckley, D.~Feld and D.~Goncalves, {\it {Scalar Simplified Models for
  Dark Matter}},  {\em Phys. Rev. D} {\bf 91} (2015) 015017
  [\href{http://arXiv.org/abs/1410.6497}{{\tt 1410.6497}}].

\bibitem{Baek:2015lna}
S.~Baek, P.~Ko, M.~Park, W.-I. Park and C.~Yu, {\it {Beyond the Dark matter
  effective field theory and a simplified model approach at colliders}},  {\em
  Phys. Lett.} {\bf B756} (2016) 289--294
  [\href{http://arXiv.org/abs/1506.06556}{{\tt 1506.06556}}].

\bibitem{Abdallah:2015ter}
J.~Abdallah {\em et.~al.}, {\it {Simplified Models for Dark Matter Searches at
  the LHC}},  {\em Phys. Dark Univ.} {\bf 9-10} (2015) 8--23
  [\href{http://arXiv.org/abs/1506.03116}{{\tt 1506.03116}}].

\bibitem{Abercrombie:2015wmb}
D.~Abercrombie {\em et.~al.}, {\it {Dark Matter benchmark models for early LHC
  Run-2 Searches: Report of the ATLAS/CMS Dark Matter Forum}},  {\em Phys. Dark
  Univ.} {\bf 27} (2020) 100371 [\href{http://arXiv.org/abs/1507.00966}{{\tt
  1507.00966}}].

\bibitem{Boveia:2016mrp}
A.~Boveia {\em et.~al.}, {\it {Recommendations on presenting LHC searches for
  missing transverse energy signals using simplified $s$-channel models of dark
  matter}},  {\em Phys. Dark Univ.} {\bf 27} (2020) 100365
  [\href{http://arXiv.org/abs/1603.04156}{{\tt 1603.04156}}].

\bibitem{Bell:2016ekl}
N.~F. Bell, G.~Busoni and I.~W. Sanderson, {\it {Self-consistent Dark Matter
  Simplified Models with an s-channel scalar mediator}},  {\em JCAP} {\bf 1703}
  (2017) 015 [\href{http://arXiv.org/abs/1612.03475}{{\tt 1612.03475}}].

\bibitem{Bauer:2017ota}
M.~Bauer, U.~Haisch and F.~Kahlhoefer, {\it {Simplified dark matter models with
  two Higgs doublets: I. Pseudoscalar mediators}},  {\em JHEP} {\bf 05} (2017)
  138 [\href{http://arXiv.org/abs/1701.07427}{{\tt 1701.07427}}].

\bibitem{MarchRussell:2008yu}
J.~March-Russell, S.~M. West, D.~Cumberbatch and D.~Hooper, {\it {Heavy Dark
  Matter Through the Higgs Portal}},  {\em JHEP} {\bf 07} (2008) 058
  [\href{http://arXiv.org/abs/0801.3440}{{\tt 0801.3440}}].

\bibitem{LopezHonorez:2012kv}
L.~Lopez-Honorez, T.~Schwetz and J.~Zupan, {\it {Higgs portal, fermionic dark
  matter, and a Standard Model like Higgs at 125 GeV}},  {\em Phys. Lett.} {\bf
  B716} (2012) 179--185 [\href{http://arXiv.org/abs/1203.2064}{{\tt
  1203.2064}}].

\bibitem{Arcadi:2019lka}
G.~Arcadi, A.~Djouadi and M.~Raidal, {\it {Dark Matter through the Higgs
  portal}},  {\em Phys. Rept.} {\bf 842} (2020) 1--180
  [\href{http://arXiv.org/abs/1903.03616}{{\tt 1903.03616}}].

\bibitem{Magg:1980ut}
M.~Magg and C.~Wetterich, {\it {Neutrino Mass Problem and Gauge Hierarchy}},
  {\em Phys. Lett.} {\bf 94B} (1980) 61--64.

\bibitem{Cheng:1980qt}
T.~P. Cheng and L.-F. Li, {\it {Neutrino Masses, Mixings and Oscillations in
  SU(2) x U(1) Models of Electroweak Interactions}},  {\em Phys. Rev.} {\bf
  D22} (1980) 2860.

\bibitem{Lazarides:1980nt}
G.~Lazarides, Q.~Shafi and C.~Wetterich, {\it {Proton Lifetime and Fermion
  Masses in an SO(10) Model}},  {\em Nucl. Phys.} {\bf B181} (1981) 287--300.

\bibitem{Chen:2014lla}
C.-H. Chen and T.~Nomura, {\it {Inert Dark Matter in Type-II Seesaw}},  {\em
  JHEP} {\bf 09} (2014) 120 [\href{http://arXiv.org/abs/1404.2996}{{\tt
  1404.2996}}].

\bibitem{Biswas:2017dxt}
A.~Biswas and A.~Shaw, {\it {Explaining Dark Matter and Neutrino Mass in the
  light of TYPE-II Seesaw Model}},  {\em JCAP} {\bf 1802} (2018), no.~02 029
  [\href{http://arXiv.org/abs/1709.01099}{{\tt 1709.01099}}]. [Erratum:
  JCAP1907,E01(2019)].

\bibitem{Gu:2019ogb}
P.-H. Gu, {\it {Double type II seesaw mechanism accompanied by Dirac fermionic
  dark matter}},  {\em Phys. Rev.} {\bf D101} (2020), no.~1 015006
  [\href{http://arXiv.org/abs/1907.10019}{{\tt 1907.10019}}].

\bibitem{Lineros:2020eit}
R.~A. Lineros and M.~Pierre, {\it {Dark Matter candidates in a Type-II
  radiative neutrino mass model}},  \href{http://arXiv.org/abs/2011.08195}{{\tt
  2011.08195}}.

\bibitem{Gustafsson:2012vj}
M.~Gustafsson, J.~M. No and M.~A. Rivera, {\it {Predictive Model for
  Radiatively Induced Neutrino Masses and Mixings with Dark Matter}},  {\em
  Phys. Rev. Lett.} {\bf 110} (2013), no.~21 211802
  [\href{http://arXiv.org/abs/1212.4806}{{\tt 1212.4806}}]. [Erratum: Phys.
  Rev. Lett.112,no.25,259902(2014)].

\bibitem{Buchmueller:2015eea}
O.~Buchmueller, S.~A. Malik, C.~McCabe and B.~Penning, {\it {Constraining Dark
  Matter Interactions with Pseudoscalar and Scalar Mediators Using Collider
  Searches for Multijets plus Missing Transverse Energy}},  {\em Phys. Rev.
  Lett.} {\bf 115} (2015), no.~18 181802
  [\href{http://arXiv.org/abs/1505.07826}{{\tt 1505.07826}}].

\bibitem{dm-consistent-scalar}
A.~Belyaev, G.~Cacciapaglia and D.~Locke, {\it Minimal consistent dark matter
  models for collider an direct detection characterisation: scalar dark
  matter},  {\em paper in progress} (2020).

\bibitem{Ross:1975fq}
D.~A. Ross and M.~J.~G. Veltman, {\it {Neutral Currents in Neutrino
  Experiments}},  {\em Nucl. Phys.} {\bf B95} (1975) 135--147.

\bibitem{Beringer:1900zz}
{\bf Particle Data Group} Collaboration, J.~Beringer {\em et.~al.}, {\it
  {Review of Particle Physics (RPP)}},  {\em Phys. Rev.} {\bf D86} (2012)
  010001.

\bibitem{Hisano:2013sn}
J.~Hisano and K.~Tsumura, {\it {Higgs boson mixes with an SU(2) septet
  representation}},  {\em Phys. Rev.} {\bf D87} (2013) 053004
  [\href{http://arXiv.org/abs/1301.6455}{{\tt 1301.6455}}].

\bibitem{Georgi:1985nv}
H.~Georgi and M.~Machacek, {\it {DOUBLY CHARGED HIGGS BOSONS}},  {\em Nucl.
  Phys.} {\bf B262} (1985) 463--477.

\bibitem{Garny:2014waa}
M.~Garny, A.~Ibarra, S.~Rydbeck and S.~Vogl, {\it {Majorana Dark Matter with a
  Coloured Mediator: Collider vs Direct and Indirect Searches}},  {\em JHEP}
  {\bf 06} (2014) 169 [\href{http://arXiv.org/abs/1403.4634}{{\tt 1403.4634}}].

\bibitem{Ko:2016zxg}
P.~Ko, A.~Natale, M.~Park and H.~Yokoya, {\it {Simplified DM models with the
  full SM gauge symmetry : the case of $t$-channel colored scalar mediators}},
  {\em JHEP} {\bf 01} (2017) 086 [\href{http://arXiv.org/abs/1605.07058}{{\tt
  1605.07058}}].

\bibitem{Arina:2020tuw}
C.~Arina, B.~Fuks, L.~Mantani, H.~Mies, L.~Panizzi and J.~Salko, {\it {Closing
  in on $t$-channel simplified dark matter models}},  {\em Phys. Lett. B} {\bf
  813} (2021) 136038 [\href{http://arXiv.org/abs/2010.07559}{{\tt
  2010.07559}}].

\bibitem{MCDM-sclar-dm}
A.~Belyaev, G.~Cacciapaglia and D.~Locke, {\it {Minimal Consistent Dark Matter
  models for Collider an Direct detection Characterisation: scalar dark
  matter}},  {\em to appear}.

\bibitem{Fukushima:2014yia}
K.~Fukushima, C.~Kelso, J.~Kumar, P.~Sandick and T.~Yamamoto, {\it {MSSM dark
  matter and a light slepton sector: The incredible bulk}},  {\em Phys. Rev. D}
  {\bf 90} (2014), no.~9 095007 [\href{http://arXiv.org/abs/1406.4903}{{\tt
  1406.4903}}].

\bibitem{Baker:2018uox}
M.~J. Baker and A.~Thamm, {\it {Leptonic WIMP Coannihilation and the Current
  Dark Matter Search Strategy}},  {\em JHEP} {\bf 10} (2018) 187
  [\href{http://arXiv.org/abs/1806.07896}{{\tt 1806.07896}}].

\bibitem{DEramo:2010keq}
F.~D'Eramo and J.~Thaler, {\it {Semi-annihilation of Dark Matter}},  {\em JHEP}
  {\bf 06} (2010) 109 [\href{http://arXiv.org/abs/1003.5912}{{\tt 1003.5912}}].

\bibitem{Cai:2015zza}
Y.~Cai and A.~P. Spray, {\it {Fermionic Semi-Annihilating Dark Matter}},  {\em
  JHEP} {\bf 01} (2016) 087 [\href{http://arXiv.org/abs/1509.08481}{{\tt
  1509.08481}}].

\bibitem{Petriello:2008pu}
F.~J. Petriello, S.~Quackenbush and K.~M. Zurek, {\it {The Invisible $Z^\prime$
  at the CERN LHC}},  {\em Phys. Rev. D} {\bf 77} (2008) 115020
  [\href{http://arXiv.org/abs/0803.4005}{{\tt 0803.4005}}].

\bibitem{Khalil:2008ps}
S.~Khalil and H.~Okada, {\it {Dark Matter in B-L Extended MSSM Models}},  {\em
  Phys. Rev. D} {\bf 79} (2009) 083510
  [\href{http://arXiv.org/abs/0810.4573}{{\tt 0810.4573}}].

\bibitem{Mizukoshi:2010ky}
J.~K. Mizukoshi, C.~A. de~S.~Pires, F.~S. Queiroz and P.~S. Rodrigues~da Silva,
  {\it {WIMPs in a 3-3-1 model with heavy Sterile neutrinos}},  {\em Phys. Rev.
  D} {\bf 83} (2011) 065024 [\href{http://arXiv.org/abs/1010.4097}{{\tt
  1010.4097}}].

\bibitem{An:2012va}
H.~An, X.~Ji and L.-T. Wang, {\it {Light Dark Matter and $Z'$ Dark Force at
  Colliders}},  {\em JHEP} {\bf 07} (2012) 182
  [\href{http://arXiv.org/abs/1202.2894}{{\tt 1202.2894}}].

\bibitem{An:2012ue}
H.~An, R.~Huo and L.-T. Wang, {\it {Searching for Low Mass Dark Portal at the
  LHC}},  {\em Phys. Dark Univ.} {\bf 2} (2013) 50--57
  [\href{http://arXiv.org/abs/1212.2221}{{\tt 1212.2221}}].

\bibitem{Frandsen:2012rk}
M.~T. Frandsen, F.~Kahlhoefer, A.~Preston, S.~Sarkar and K.~Schmidt-Hoberg,
  {\it {LHC and Tevatron Bounds on the Dark Matter Direct Detection
  Cross-Section for Vector Mediators}},  {\em JHEP} {\bf 07} (2012) 123
  [\href{http://arXiv.org/abs/1204.3839}{{\tt 1204.3839}}].

\bibitem{Barger:2012ey}
V.~Barger, D.~Marfatia and A.~Peterson, {\it {LHC and dark matter signals of Z'
  bosons}},  {\em Phys. Rev. D} {\bf 87} (2013), no.~1 015026
  [\href{http://arXiv.org/abs/1206.6649}{{\tt 1206.6649}}].

\bibitem{Basso:2012gz}
L.~Basso, B.~O'Leary, W.~Porod and F.~Staub, {\it {Dark matter scenarios in the
  minimal SUSY B-L model}},  {\em JHEP} {\bf 09} (2012) 054
  [\href{http://arXiv.org/abs/1207.0507}{{\tt 1207.0507}}].

\bibitem{Arcadi:2013qia}
G.~Arcadi, Y.~Mambrini, M.~H.~G. Tytgat and B.~Zaldivar, {\it {Invisible
  $Z^\prime$ and dark matter: LHC vs LUX constraints}},  {\em JHEP} {\bf 03}
  (2014) 134 [\href{http://arXiv.org/abs/1401.0221}{{\tt 1401.0221}}].

\bibitem{Alves:2013tqa}
A.~Alves, S.~Profumo and F.~S. Queiroz, {\it {The dark $Z^{'}$ portal: direct,
  indirect and collider searches}},  {\em JHEP} {\bf 04} (2014) 063
  [\href{http://arXiv.org/abs/1312.5281}{{\tt 1312.5281}}].

\bibitem{Alves:2015pea}
A.~Alves, A.~Berlin, S.~Profumo and F.~S. Queiroz, {\it {Dark Matter
  Complementarity and the Z$^\prime$ Portal}},  {\em Phys. Rev. D} {\bf 92}
  (2015), no.~8 083004 [\href{http://arXiv.org/abs/1501.03490}{{\tt
  1501.03490}}].

\bibitem{Okada:2016gsh}
N.~Okada and S.~Okada, {\it {$Z^\prime_{BL}$ portal dark matter and LHC Run-2
  results}},  {\em Phys. Rev. D} {\bf 93} (2016), no.~7 075003
  [\href{http://arXiv.org/abs/1601.07526}{{\tt 1601.07526}}].

\bibitem{Fairbairn:2016iuf}
M.~Fairbairn, J.~Heal, F.~Kahlhoefer and P.~Tunney, {\it {Constraints on Z'
  models from LHC dijet searches and implications for dark matter}},  {\em
  JHEP} {\bf 09} (2016) 018 [\href{http://arXiv.org/abs/1605.07940}{{\tt
  1605.07940}}].

\bibitem{Arcadi:2017hfi}
G.~Arcadi, M.~D. Campos, M.~Lindner, A.~Masiero and F.~S. Queiroz, {\it {Dark
  sequential Z' portal: Collider and direct detection experiments}},  {\em
  Phys. Rev. D} {\bf 97} (2018), no.~4 043009
  [\href{http://arXiv.org/abs/1708.00890}{{\tt 1708.00890}}].

\bibitem{Okada:2016tci}
N.~Okada and S.~Okada, {\it {$Z^\prime$-portal right-handed neutrino dark
  matter in the minimal U(1)$_X$ extended Standard Model}},  {\em Phys. Rev. D}
  {\bf 95} (2017), no.~3 035025 [\href{http://arXiv.org/abs/1611.02672}{{\tt
  1611.02672}}].

\bibitem{Belyaev:2017vsx}
A.~S. Belyaev, T.~Flacke, B.~Jain and P.~B. Schaefers, {\it {LHC Dark Matter
  Signals from Vector Resonances and Top Partners}},  {\em Phys. Rev. D} {\bf
  98} (2018), no.~3 035019 [\href{http://arXiv.org/abs/1707.07000}{{\tt
  1707.07000}}].

\bibitem{Okada:2018ktp}
S.~Okada, {\it {$Z'$ Portal Dark Matter in the Minimal $B-L$ Model}},  {\em
  Adv. High Energy Phys.} {\bf 2018} (2018) 5340935
  [\href{http://arXiv.org/abs/1803.06793}{{\tt 1803.06793}}].

\bibitem{Han:2018zcn}
Z.-L. Han and W.~Wang, {\it {$Z'$ Portal Dark Matter in $B-L$ Scotogenic Dirac
  Model}},  {\em Eur. Phys. J. C} {\bf 78} (2018), no.~10 839
  [\href{http://arXiv.org/abs/1805.02025}{{\tt 1805.02025}}].

\bibitem{Cosme:2021baj}
C.~Cosme, M.~Dutra, S.~Godfrey and T.~R. Gray, {\it {Testing freeze-in with
  axial and vector Z' bosons}},  {\em JHEP} {\bf 09} (2021) 056
  [\href{http://arXiv.org/abs/2104.13937}{{\tt 2104.13937}}].

\bibitem{Belyaev:2005ew}
A.~Belyaev, C.~Leroy, R.~Mehdiyev and A.~Pukhov, {\it {Leptoquark single and
  pair production at LHC with CalcHEP/CompHEP in the complete model}},  {\em
  JHEP} {\bf 09} (2005) 005 [\href{http://arXiv.org/abs/hep-ph/0502067}{{\tt
  hep-ph/0502067}}].

\bibitem{Bell:2016fqf}
N.~F. Bell, Y.~Cai and R.~K. Leane, {\it {Dark Forces in the Sky: Signals from
  Z' and the Dark Higgs}},  {\em JCAP} {\bf 08} (2016) 001
  [\href{http://arXiv.org/abs/1605.09382}{{\tt 1605.09382}}].

\bibitem{Zerwekh:2012bf}
A.~R. Zerwekh, {\it {On the Quantum Chromodynamics of a Massive Vector Field in
  the Adjoint Representation}},  {\em Int. J. Mod. Phys. A} {\bf 28} (2013)
  1350054 [\href{http://arXiv.org/abs/1207.5233}{{\tt 1207.5233}}].

\bibitem{Belyaev:2018xpf}
A.~Belyaev, G.~Cacciapaglia, J.~Mckay, D.~Marin and A.~R. Zerwekh, {\it
  {Minimal Spin-one Isotriplet Dark Matter}},  {\em Phys. Rev. D} {\bf 99}
  (2019), no.~11 115003 [\href{http://arXiv.org/abs/1808.10464}{{\tt
  1808.10464}}].

\bibitem{Abe:2020mph}
T.~Abe, M.~Fujiwara, J.~Hisano and K.~Matsushita, {\it {A model of
  electroweakly interacting non-abelian vector dark matter}},  {\em JHEP} {\bf
  07} (2020) 136 [\href{http://arXiv.org/abs/2004.00884}{{\tt 2004.00884}}].

\bibitem{dan-pascos2019}
A.~Belyaev, G.~Cacciapaglia and D.~Locke, ``Minimal consistent fermion dark
  matter.''
  \url{http://indico.hep.manchester.ac.uk/contributionDisplay.py?contribId=403&sessionId=12&confId=5326},
  Presented by Daniel Locke on July 2, 2019.
\newblock PASCOS 2019 conference, July 1-5, 2019.

\bibitem{belyaev-moriond2021}
A.~Belyaev, G.~Cacciapaglia and D.~Locke, ``Minimal consistent dark matter
  models.'' \url{https://moriond.in2p3.fr/2021/Registration/proceedings.html},
  Presented by Alexander Belyaev on 31 March, 2020.
\newblock Moriond QCD 2021 conference, March 29 -- April 4 July, 2021.

\bibitem{DiazSaez:2021pmg}
B.~D\'\i{}az~S\'aez, P.~Escalona, S.~Norero and A.~R. Zerwekh, {\it {Fermion
  singlet dark matter in a pseudoscalar dark matter portal}},  {\em JHEP} {\bf
  10} (2021) 233 [\href{http://arXiv.org/abs/2105.04255}{{\tt 2105.04255}}].

\bibitem{Baek_2017}
S.~Baek, P.~Ko and J.~Li, {\it Minimal renormalizable simplified dark matter
  model with a pseudoscalar mediator},  {\em Physical Review D} {\bf 95} (Apr,
  2017).

\bibitem{Semenov:2008jy}
A.~Semenov, {\it {LanHEP - a package for the automatic generation of Feynman
  rules in field theory. Version 3.0}},  {\em Comput. Phys. Commun.} {\bf 180}
  (2009) 431--454 [\href{http://arXiv.org/abs/0805.0555}{{\tt 0805.0555}}].

\bibitem{hepmdb:FDM+a}
A.~Belyaev, G.~Cacciapaglia and D.~Locke, ``{HEPMDB Model : FDM+a - Dirac
  fermion singlet Dark Matter with psuedoscalar mediator}.''
  \url{https://hepmdb.soton.ac.uk/hepmdb:0222.0334}, Feb, 2022.

\bibitem{Belanger:2014vza}
G.~Bélanger, F.~Boudjema, A.~Pukhov and A.~Semenov, {\it {micrOMEGAs4.1: two
  dark matter candidates}},  {\em Comput. Phys. Commun.} {\bf 192} (2015)
  322--329 [\href{http://arXiv.org/abs/1407.6129}{{\tt 1407.6129}}].

\bibitem{Belanger:2001fz}
G.~Belanger, F.~Boudjema, A.~Pukhov and A.~Semenov, {\it micromegas: A program
  for calculating the relic density in the mssm},  {\em Comput. Phys. Commun.}
  {\bf 149} (2002) 103--120 [\href{http://arXiv.org/abs/hep-ph/0112278}{{\tt
  hep-ph/0112278}}].

\bibitem{Belanger:2004yn}
G.~Belanger, F.~Boudjema, A.~Pukhov and A.~Semenov, ``Micromegas: Version
  1.3.'' hep-ph/0405253, 2004.

\bibitem{ATLAS:2020kdi}
{\bf ATLAS} Collaboration, ``{Combination of searches for invisible Higgs boson
  decays with the ATLAS experiment}.'' ATLAS-CONF-2020-052, 10, 2020.

\bibitem{Atlas:2019qfx}
{\bf ATLAS, CMS} Collaboration, ATLAS and C.~Collaborations, {\it {Report on
  the Physics at the HL-LHC and Perspectives for the HE-LHC}},  in {\em
  {HL/HE-LHC Physics Workshop: final jamboree Geneva, CERN, March 1, 2019}},
  2019.
\newblock \href{http://arXiv.org/abs/1902.10229}{{\tt 1902.10229}}.

\bibitem{Asner:2013psa}
D.~M. Asner {\em et.~al.}, {\it {ILC Higgs White Paper}},  in {\em
  {Proceedings, 2013 Community Summer Study on the Future of U.S. Particle
  Physics: Snowmass on the Mississippi (CSS2013): Minneapolis, MN, USA, July
  29-August 6, 2013}}, 2013.
\newblock \href{http://arXiv.org/abs/1310.0763}{{\tt 1310.0763}}.

\bibitem{tHooft:1978jhc}
G.~'t~Hooft and M.~J.~G. Veltman, {\it {Scalar One Loop Integrals}},  {\em
  Nucl. Phys.} {\bf B153} (1979) 365--401.

\bibitem{Patel:2015tea}
H.~H. Patel, {\it {Package-X: A Mathematica package for the analytic
  calculation of one-loop integrals}},  {\em Comput. Phys. Commun.} {\bf 197}
  (2015) 276--290 [\href{http://arXiv.org/abs/1503.01469}{{\tt 1503.01469}}].

\bibitem{locke_daniel_2022_6308438}
D.~Locke, ``{Supporting material detailing one-loop calculations for paper
  "Minimal Consistent Dark Matter models for Collider an Direct detection
  Characterisation: fermion dark matter"}.''
  [\href{https://zenodo.org/record/6308438}{Zenodo.6308438}], Feb, 2022.

\bibitem{Passarino:1978jh}
G.~Passarino and M.~Veltman, {\it {One Loop Corrections for e+ e- Annihilation
  Into mu+ mu- in the Weinberg Model}},  {\em Nucl. Phys. B} {\bf 160} (1979)
  151--207.

\bibitem{Hahn:2000jm}
T.~Hahn, {\it {Automatic loop calculations with FeynArts, FormCalc, and
  LoopTools}},  {\em Nucl. Phys. Proc. Suppl.} {\bf 89} (2000) 231--236
  [\href{http://arXiv.org/abs/hep-ph/0005029}{{\tt hep-ph/0005029}}].

\end{thebibliography}\endgroup
